\def\ARXIV{}
\def\norevisionmarks{}
\documentclass{article}

\usepackage{iclr2027_conference,times}
\ifdefined\ARXIV\iclrfinalcopy\fi
\newcommand{\releaseurl}{\href{https://huggingface.co/collections/eth-algopt/auxspec}{\nolinkurl{huggingface.co/collections/eth-algopt/auxspec}}}
\newcommand{\weburl}{\href{https://auxspec.org}{\nolinkurl{auxspec.org}}}
\newcommand{\releaselinks}{\par\smallskip{\raggedright\small\textbf{Models and data:} \releaseurl\\ \textbf{Website:} \weburl\par}}
\newcommand{\work}{\texorpdfstring{\textsc{AuXSpec}}{AuXSpec}}
\newcommand{\verina}{Verina}
\newcommand{\clever}{Clever}
\newcommand{\subfig}[2]{\hyperref[#1]{Figure~\ref*{#1}#2}}
\newcommand{\panel}[2]{\hyperref[#1]{\ref*{#1}#2}}

\newcommand{\La}[1]{L#1\textsuperscript{a}}
\newcommand{\Lb}[1]{L#1\textsuperscript{b}}

\usepackage{amsmath,amsfonts,bm}

\def\eqref#1{equation~\ref{#1}}

\def\1{\bm{1}}

\DeclareMathAlphabet{\mathsfit}{\encodingdefault}{\sfdefault}{m}{sl}
\SetMathAlphabet{\mathsfit}{bold}{\encodingdefault}{\sfdefault}{bx}{n}

\usepackage{xcolor}
\usepackage[draft,authormarkup=none]{changes}
\usepackage{hyperref}
\definecolor{linkblue}{HTML}{215CAF}
\hypersetup{colorlinks=true,linkcolor=linkblue,citecolor=linkblue,urlcolor=linkblue}
\usepackage{url}
\usepackage{booktabs}
\usepackage{tabularx}
\usepackage{array}
\newcolumntype{L}[1]{>{\raggedright\arraybackslash}p{#1}}
\usepackage{graphicx}
\usepackage{tikz}
\usetikzlibrary{arrows.meta}
\usepackage{amsmath,amssymb,amsthm}
\usepackage{enumitem}
\usepackage{float}
\usepackage{afterpage}
\usepackage[font=scriptsize,labelsep=period]{caption}
\usepackage{subcaption}
\usepackage{wrapfig}
\usepackage{multirow}

\usepackage{listings}

\usepackage{xcolor}
\definecolor{keywordcolor}{rgb}{0.0, 0.1, 0.6}
\definecolor{tacticcolor}{rgb}{0.0, 0.1, 0.6}
\definecolor{commentcolor}{rgb}{0.35, 0.35, 0.35}
\definecolor{symbolcolor}{rgb}{0.0, 0.1, 0.6}
\definecolor{sortcolor}{rgb}{0.1, 0.0, 0.1}
\definecolor{attributecolor}{rgb}{0.7, 0.1, 0.1}
\definecolor{errorcolor}{rgb}{0.5, 0, 0}
\definecolor{stringcolor}{rgb}{0.5, 0.3, 0.2}

\usepackage[most]{tcolorbox}

\definecolor{promptbar}{HTML}{215CAF}
\definecolor{promptbg}{HTML}{F2F2F2}

\newtcolorbox{promptbox}[1]{
  enhanced,
  colback=promptbg,
  colframe=promptbar,
  colbacktitle=promptbar,
  coltitle=white,
  fonttitle=\bfseries\small,
  boxrule=0.5mm,
  arc=1mm,
  left=5pt, right=5pt, top=4pt, bottom=4pt,
  title={#1}
}

\definecolor{leanbar}{HTML}{215CAF}
\definecolor{leanbg}{RGB}{243,246,250}
\definecolor{leanlabel}{rgb}{0.12,0.12,0.12}

\newtcolorbox{leanbox}{
  enhanced,
  colback=leanbg,
  colframe=leanbar,
  boxrule=0.6pt,
  arc=0pt,
  left=5pt, right=5pt, top=4pt, bottom=4pt,
}

\newif\ifstatusdraft
\statusdrafttrue

\newif\ifshowcomments\showcommentstrue
\ifdefined\ARXIV\showcommentsfalse\fi
\definecolor{authorcommentcolor}{rgb}{0.75,0.1,0.45}
\definecolor{deltaup}{rgb}{0.1,0.5,0.2}
\definecolor{deltadown}{rgb}{0.8,0.1,0.1}

\definecolor{revblue}{HTML}{0000CC}
\ifdefined\norevisionmarks\definecolor{revblue}{HTML}{000000}\fi
\newcommand{\ywx}[1]{\textcolor{revblue}{#1}}
\newenvironment{revised}{\color{revblue}}{}

\theoremstyle{definition}

\title{AuXSpec:\,Learning\,to\,Autoformalize\,Program \\ Specifications via Executable Postconditions}
\ifdefined\ARXIV
\author{Haocheng Wang$^{2,1*}$ \quad Weixuan Yuan$^{1*}$ \quad Florian H\"ubler$^{1}$ \quad Yinya Huang$^{1,4}$ \\
\textbf{Zhijiang Guo$^{2,3}$ \quad Sorrachai Yingchareonthawornchai$^{5}$ \quad Rasmus Kyng$^{1\dagger}$} \\
$^{1}$Department of Computer Science, ETH Zurich \enskip $^{2}$HKUST(GZ) \enskip $^{3}$HKUST \\
$^{4}$ETH AI Center \enskip $^{5}$Institute for Theoretical Studies, ETH Zurich}
\fi

\begin{document}

\maketitle
\ifdefined\ARXIV\lhead{Preprint}
{\renewcommand{\thefootnote}{\fnsymbol{footnote}}%
\footnotetext[1]{Equal contribution.}%
\footnotetext[2]{Corresponding author: kyng@inf.ethz.ch.}}
\fi

\begin{abstract}
Formal specifications define intended program behavior and underpin verification, but writing them in theorem provers such as Lean is slow and expertise-intensive, making specification a major bottleneck to scaling formally correct code. Autoformalization from natural-language problems is promising, yet training specification models requires scalable feedback on whether generated specifications capture intended behavior. Prior approaches obtain this feedback through explicit proofs or separately implemented code with formal correspondence proofs, both expensive.
We introduce \work{}, a Lean framework that pairs declarative postconditions with executable counterparts, obtained through type class resolution when possible and otherwise defined separately with an equivalence proof. Test cases can then evaluate declarative specifications by execution, letting us use agreement as a reinforcement-learning reward. Using \work{}, we synthesize 3{,}037 LeetCode problems and expand training to 19.5K samples through expert iteration on Codeforces.
Experiments show that training Qwen3.5-9B and Qwen3.8-27B with the execution-based reward improves SpecGen pass@4 on \work-Test 0.7\%~$\rightarrow$~82.7\% and 14.3\%~$\rightarrow$~84.3\%, and the 27B model outperforms all frontier models. 
On the out-of-domain \verina{} benchmark with its original prompts, both models improve 7.0\%~$\rightarrow$~90.5\% and 21.9\%~$\rightarrow$~90.5\% and surpass VeriSpecGen-30B-A3B, the prior state of the art among open-source specification models, on pass@1 (theirs 47.6\% vs. ours 73.7\% and 84.5\%) and pass@4 (theirs 63.9\% at pass@10 vs. ours 90.5\%). We release the weights of the trained 9B and 27B models and the 3{,}037 problems with test cases, including the 300-problem \work-Test benchmark with expert review and correctness proofs for future evaluation. 

\ifdefined\ARXIV\releaselinks\fi
\end{abstract}

\section{Introduction}
\label{sec:introduction}

Formal program specifications define the correct behavior of a program over all valid inputs, providing the reference against which implementations can be proved correct or tested. Yet writing such specifications in an interactive theorem prover requires substantial expertise and effort, making specification a fundamental barrier to scaling formal code verification and generation. As AI models make rapid progress in generating code and proving its correctness, this barrier becomes increasingly important: realizing the promise of formally correct code requires automating specification itself. Autoformalization from natural-language programming problems offers a path forward, but immediately raises a central question: \emph{when does a generated formal specification faithfully capture the intended problem?}

This question is particularly important in modern code-generation pipelines. 
Large language models (LLMs) can generate solutions to programming problems (e.g., LeetCode) from natural-language descriptions. These solutions can be evaluated using input-output test cases, which provide scalable reward signals for reinforcement learning \citep{wang2025codecontestsplus,ruan2026evolvecoder}.
However, passing test cases does not guarantee correctness, as a solution may still fail on untested inputs \citep{shi2026codehacker}.
Formal verification offers a stronger guarantee. 
Given a specification, which consists of a precondition defining valid inputs and a postcondition characterizing correct outputs \citep{endres2024nl2postcond,chen2026codespecbench}, a theorem prover like Lean can formally prove that an implementation satisfies its specification for every valid input. 
This guarantee depends on the specification correctly capturing the natural-language problem, yet producing such specifications at scale remains difficult.
Manually curating specifications requires substantial expert effort: \clever{} and \verina{} contain only 161 and 189 problems, respectively \citep{thakur2025clever,ye2026verina}. Automated methods can scale further, but may produce incorrect specifications \citep{fvapps2025,bursuc2025vericoding,fvspec2026}. Scaling specification therefore requires more than generating formal statements. It also requires efficient and scalable feedback on whether generated specifications match the source problem, analogous to the test-case feedback used to train code-generation models.

Training specification generators has been explored in auto-active verification languages, where a verifier, like an SMT solver, provides the reward by checking candidate specifications on test cases \citep{reform2025,specrl2026,baksys2025atlas}.
However, SMT solvers limit the complexity of specifications that can be checked \citep{yang2026coins,ye2026verina}, motivating the use of interactive theorem provers such as Lean.
In Lean, prior work checks generated specifications by turning each test case into a theorem to be proved, a method known as symbolic specification testing \citep{lahiri2024intent,verispecgen2026,yang2026coins}. These theorems can require nontrivial proofs, especially when the specification involves unbounded quantifiers.
If a proof cannot be found, the test case may be left unresolved or evaluated by an LLM judge. Thus, unlike in code generation \citep{wang2025codecontestsplus,he2025hardtests}, test cases do not directly provide cheap rewards for training specification generators in Lean (Table~\ref{tab:intro-comparison-v2}).

\begin{table}[t]
\providecommand{\cmpyes}{\textcolor{green!50!black}{$\checkmark$}}
\providecommand{\cmppart}{\textcolor{orange}{$\sim$}}
\providecommand{\cmpno}{\textcolor{red}{$\times$}}
\centering
\captionsetup{font=scriptsize,skip=3pt}
\caption{Comparison of \work{} with closely related work on specification generation. Specification validation checks whether a specification matches the source problem, and Expert spec. indicates a reference specification written by experts. \textsuperscript{1}$\lor$ indicates that the check falls back to the second method when the first is inconclusive, and $\scriptstyle\rightarrow$ that the second method follows the first. PBT indicates property-based testing. \textsuperscript{2}The postcondition equivalence proof of \work{} shows that the executable postcondition agrees with the postcondition on all valid inputs and outputs. Correctness indicates whether every implementation is proved to satisfy the specification ($\checkmark$), only some are (Partial), or the work has no implementation task (--). For \work{}, the proofs cover the 300 test problems. SFT, EI and RL indicate supervised fine-tuning, expert iteration and reinforcement learning.}
\label{tab:intro-comparison-v2}
\scriptsize
\setlength{\tabcolsep}{0.6pt}
\begin{tabular*}{\linewidth}{@{\extracolsep{\fill}}llllllc@{}}
\toprule
\multirow{2}{*}{Work} & \multirow{2}{*}{Lang.} & \multicolumn{2}{c}{\raisebox{1.5pt}{Specification validation}} & \multirow{2}{*}{Correctness} & \multirow{2}{*}{Manual review} & \multirow{2}{*}{Training} \\[-2pt]
\cmidrule(lr){3-4}
\noalign{\vspace{-2pt}}
 & & \multicolumn{1}{c}{Reference} & \multicolumn{1}{c}{Method\rlap{\textsuperscript{1}}} & & & \\[-1pt]
\midrule
\verina{} \citep{ye2026verina} & Lean & Expert spec. & Proof $\lor$ testing & Partial & Benchmark (189) & None \\
\clever{} \citep{thakur2025clever} & Lean & Expert spec. & Proof & Partial & Benchmark (161) & None \\
VeriSpecGen \citep{verispecgen2026} & Lean & Test cases & Proof $\lor$ LLM judge & -- & Not reported & SFT \\
Coins \citep{yang2026coins} & Rocq & Test cases & Proof, LLM judge & -- & Benchmark (164) & None \\
ATLAS \citep{baksys2025atlas} & Dafny & Test cases & Verifier $\scriptstyle\rightarrow$ LLM judge & $\checkmark$ & Not reported & SFT \\
Verus-SpecGym \citep{agarwal2026specgym} & Verus & Test cases & Verifier $\lor$ execution & -- & Not reported & None \\
LeetProof \citep{feng2026leetproof} & Lean & Test cases & LLM judge $\scriptstyle\rightarrow$ PBT & Partial & Unproved (38) & None \\
\midrule
\textbf{\work{} (ours)} & Lean & Test cases & Equivalence proof\textsuperscript{2}\,$\scriptstyle\rightarrow$\,execution & $\checkmark$ & Test set (300) & SFT, EI, RL \\
\bottomrule
\end{tabular*}
\vspace{-8pt}
\end{table}

To address this issue, we propose \work{}, a framework for training models to generate program specifications in Lean 4.
\work{} retains a declarative postcondition that specifies correctness independently of the implementation and pairs it with an executable postcondition for evaluating test cases\ywx{, a pairing with a long history in formal methods (Section~\ref{sec:related})}. 
For each synthesized problem, we prove a postcondition equivalence theorem, separately from generation (Section~\ref{sec:proofs}), showing that the two postconditions agree for every valid input. This theorem allows us to evaluate any number of test cases by direct execution in Lean, without proving each case separately.

We build a synthesis pipeline to generate Lean specifications and implementations for 3{,}037 LeetCode problems. 
For model training, we combine synthesized data from LeetCode and Codeforces with expert iteration to obtain 19.5K samples for supervised fine-tuning (SFT). 

We fine-tune Qwen3.5-9B and Qwen3.8-27B on these samples and then apply reinforcement learning (RL) with our execution-based reward on 13{,}637 tasks over 8{,}579 further problems. The resulting \work-9B and \work-27B models can generate and repair both specifications and implementations. 
Relative to its base model, \work-27B improves SpecGen pass@4 from 14.3\% to 84.3\% on \work-Test and from 21.9\% to 90.5\% on \verina{} with its original prompts. The corresponding improvements for \work-9B are from 0.7\% to 82.7\% and from 7.0\% to 90.5\%. Moreover, RL with our execution-based reward improves SpecGen pass@4 over SFT alone for both model sizes (Section~\ref{sec:rlanalysis}).

Our contributions are threefold.
\begin{itemize}[leftmargin=*, itemsep=0pt, topsep=2pt, parsep=0pt]
    \item \textbf{Executable specification format.} We introduce a specification format that pairs a declarative postcondition with an executable Boolean postcondition, obtained through type class resolution in Lean when a decidability instance exists and defined separately otherwise. A postcondition equivalence theorem proves that the two agree on valid inputs (Section~\ref{sec:method}).
    \item \textbf{\work{} dataset and \work-Test.} We release the 3{,}037 LeetCode problems with specifications, implementations, frozen test cases and postcondition equivalence proofs, including \work-Test, an expert-reviewed subset of 300 problems that serves as a benchmark for Lean specification generation (Sections~\ref{sec:synthesis} and~\ref{sec:test}).
    \item \textbf{\work-9B-RL and \work-27B-RL.} Using our execution-based reward, we train and release open-weight models that generate and repair Lean specifications and implementations from natural-language descriptions. 
\work-27B-RL surpasses all frontier models in domain, and both models surpass the prior state of the art among open-source specification models on \verina{} (Tables \ref{tab:mainpass} and \ref{tab:verina-notests}).
\end{itemize}

\section{Problem formulation}
\label{sec:method}

In our autoformalization setup, we represent each programming problem with a Lean file. 
This section describes the file format, the generation and repair tasks over its components, and the acceptance criteria used in data synthesis, model training and evaluation.

\textbf{Format.}\label{sec:format}
For each programming problem, we take a natural-language problem description, input constraints, and test cases as input, and aim to produce a Lean file containing a specification, an implementation, two theorem statements, and executable tests (Figures~\ref{fig:example-defs} and \ref{fig:example-tests} in Appendix~\ref{app:example-separate}).
The specification consists of a precondition defining valid inputs, a declarative postcondition characterizing correct outputs independently of how they are computed, and an executable postcondition that expresses the same condition in a form that can be directly evaluated on test cases by Lean.
The implementation is an algorithm that computes an output from a given input.
The file also contains two theorems: the \textit{postcondition equivalence theorem}, asserting that the two postconditions agree on valid inputs, and the \textit{correctness theorem}, asserting that the implementation satisfies the declarative postcondition.
We include the test cases in the Lean file as \texttt{\#guard} commands that check the precondition, the executable postcondition and the implementation.
We illustrate the format with two synthesized problems, \textit{Count Number of Special Subsequences} and \textit{Continuous Subarray Sum} (Appendix~\ref{app:example-typeclass}).

\textbf{Tasks.}\label{sec:tasks-def} We train our models on specification generation (SpecGen) and code generation (CodeGen), as in prior work \citep{ye2026verina}, and their repair variants, SpecGen-Repair and CodeGen-Repair (Table~\ref{tab:tasks}). The synthesis pipeline performs these tasks and assembles the generated components into complete Lean files to build the training data and test set.

\begin{table}[H]
\caption{Input and output of each task. Broken is the rejected output that a repair task revises, and Diagnostics are the compiler errors and failed test cases of that output. Test layers check the output (Table~\ref{tab:layers} in Appendix~\ref{app:layout}).}
\label{tab:tasks}
\centering
\scriptsize
\setlength{\tabcolsep}{4pt}
\begin{tabular}{lccccclc}
\toprule
\multirow{2}{*}{Task} & \multicolumn{5}{c}{Input} & \multicolumn{2}{c}{Output} \\
\cmidrule(l{2pt}r{2pt}){2-6} \cmidrule(l{2pt}r{2pt}){7-8}
 & Description & Constraints & Cases & Broken & Diagnostics & Component & Test layers \\
\midrule
SpecGen & \checkmark & \checkmark & \checkmark &  &  & Specification & \La{1} \Lb{1} \La{2} \Lb{2} \\
SpecGen-Repair & \checkmark & \checkmark & \checkmark & \checkmark & \checkmark & Repaired specification & \La{1} \Lb{1} \La{2} \Lb{2} \\
CodeGen & \checkmark &  & \checkmark &  &  & Implementation & L3 \\
CodeGen-Repair & \checkmark &  & \checkmark & \checkmark & \checkmark & Repaired implementation & L3 \\
\bottomrule
\end{tabular}
\end{table}

\textbf{Acceptance criteria.}\label{sec:acceptance} We apply a common set of acceptance criteria for data filtering, expert iteration, reinforcement learning and evaluation. A generated output is accepted only if the assembled file compiles, the task-specific test layers pass, and LLM judges approve \citep{zheng2023judging}.

Test cases are fixed before synthesis and executed as \texttt{\#guard} commands, which fail at compilation when the expression is false.
The commands are organized into layers by component and type of test (Table~\ref{tab:layers} in Appendix~\ref{app:layout}), and the generated commands must match the frozen test cases.
Beyond the test cases, LLM judges check faithfulness (whether the postcondition captures the problem's requirements), non-tautology (whether the postcondition characterizes correct outputs rather than restating a solution algorithm), completeness (whether the executable postcondition accepts correct outputs on untested inputs) and termination (whether the implementation terminates on valid inputs).

\textbf{Proofs.}\label{sec:proofs}
The Lean file includes two theorem statements, whose proofs are left as \texttt{sorry} throughout data synthesis, training, and evaluation.
We instead adopt executable tests and LLM judges as acceptance criteria, and assess the reliability of these checks separately.
For 35.5\% of the synthesized problems, the executable postcondition is obtained directly through decidability instances and agrees with the declarative postcondition by construction. After synthesis, we attempt to prove postcondition equivalence for all 3{,}037 problems, and 99.0\% are provable without modifying their definitions (Section~\ref{sec:outcomes}), which indicates that the proofs expose errors beyond tests and judges in only 1.0\% of the synthesized problems. The remaining problems are repaired and then proved.
\ywx{For model-generated specifications, we also validate acceptance by proof, and the postcondition equivalence theorem holds for 99.2\% of accepted outputs of both RL models (Appendix~\ref{app:accepted-equiv}).} \ywx{We note that the proved theorem shows only that the two postconditions agree, but both may encode the same misunderstanding: expert review finds 9 such errors among the 300 \work{}-Test problems (Appendix~\ref{app:human-review}). Semantic agreement with the problem therefore rests on the test cases, the judges and expert review (Section~\ref{sec:test}).}

We do not require correctness proofs for generated implementations either, since such proofs are long (a median of about 160 lines on \work{}-Test, Appendix~\ref{app:correct}) and implementation verification is not the focus of this work.
We nevertheless prove correctness for all 300 \work{}-Test problems to provide verified references for evaluation (Section~\ref{sec:test}).

\section{Data synthesis}
\label{sec:synthesis}

To train and evaluate models on the tasks defined in
Section~\ref{sec:method}, we synthesize and release 3{,}037 LeetCode problems in our Lean format.
The dataset provides training data and includes a
held-out, expert-reviewed set of 300 problems,
\work{}-Test, for evaluation.
In this section, we describe the synthesis pipeline, report its results and introduce \work{}-Test.

\begin{figure}[t]
\centering
\definecolor{figCink}{HTML}{1C2740}%
\definecolor{figCmuted}{HTML}{5F6B7E}%
\definecolor{figCarrow}{HTML}{58637A}%
\definecolor{figCioline}{HTML}{C3CAD4}%
\definecolor{figCio}{HTML}{FFFFFF}%
\definecolor{figCin}{HTML}{F5F6F8}%
\definecolor{figCinline}{HTML}{DCE0E6}%
\definecolor{figCpanel}{HTML}{F3F5FA}%
\definecolor{figCpanelline}{HTML}{D6DDEA}%
\definecolor{figClabel}{HTML}{3B5488}%
\definecolor{figClock}{HTML}{34507F}%
\definecolor{figCone}{HTML}{E2E8F4}%
\definecolor{figCtwo}{HTML}{CAD6EC}%
\definecolor{figCthree}{HTML}{B1C3E3}%
\ifdefined\figCflow\else\newsavebox{\figCflow}\fi%
\sbox{\figCflow}{%
  \sffamily\fontsize{7.2}{8.4}\selectfont
  \def\figCarr{\hspace{3.2pt}\begin{tikzpicture}[baseline=-2.55pt]
      \draw[-{Stealth[length=1.2mm,width=1.0mm,inset=0.3mm]},
        draw=figCarrow, line width=0.45pt] (0,0) -- (7.6pt,0);
    \end{tikzpicture}\hspace{3.2pt}}%
  \def\figCloop{\hspace{3.2pt}\begin{tikzpicture}[baseline=-2.55pt]
      \draw[-{Stealth[length=1.2mm,width=1.0mm,inset=0.3mm]},
        draw=figCarrow, line width=0.45pt] (0,1.25pt) -- (9.6pt,1.25pt);
      \draw[-{Stealth[length=1.2mm,width=1.0mm,inset=0.3mm]},
        draw=figCarrow, line width=0.45pt] (9.6pt,-1.25pt) -- (0,-1.25pt);
    \end{tikzpicture}\hspace{3.2pt}}%
  \textcolor{figCmuted}{\itshape Each stage:}\hspace{4.5pt}%
  \textcolor{figCink}{Plan\figCarr Generate\figCarr Validate\figCloop Repair}}%
\begin{tikzpicture}[
  x=1cm, y=1cm,
  font=\sffamily\fontsize{8}{9.2}\selectfont,
  every node/.style={text=figCink},
  io/.style={text height=1.62ex, text depth=0.42ex,
    draw=figCioline, line width=0.45pt, fill=figCio,
    rounded corners=3pt, minimum height=7.4mm, inner xsep=5.5pt,
    font=\sffamily\fontsize{7.7}{8.8}\selectfont,
    inner ysep=1pt, align=center},
  multi/.style={text height=, text depth=},
  flow/.style={-{Stealth[length=1.75mm,width=1.4mm,inset=0.45mm]},
    draw=figCarrow, line width=0.65pt},
  tag/.style={anchor=base west, inner sep=0pt,
    font=\sffamily\bfseries\fontsize{6.2}{7}\selectfont},
  lock/.pic={
    \fill[figClock, rounded corners=0.3pt] (-0.82mm,-0.98mm) rectangle (0.82mm,0.18mm);
    \draw[figClock, line width=0.45pt]
      (-0.50mm,0.18mm) -- (-0.50mm,0.42mm)
      arc[start angle=180, end angle=0, radius=0.50mm] -- (0.50mm,0.18mm);
    \fill[white] (0,-0.36mm) circle[radius=0.17mm];
  },
]
  \def\figCw{2.45}
  \def\figCh{0.46}
  \def\figCp{0.20}
  \def\figCg{0.06}
  \def\figCx{0.16}
  \def\figCtop{0.90}
  \def\figCbot{-1.30}
  \def\figCyB{-0.90}
  \def\figCbase{-0.985}
  \def\figCwi{2.00}
  \def\figCwp{1.56}
  \def\figCwa{1.56}
  \def\figCgap{0.30}
  \pgfmathsetmacro\figCend{\figCx+3*\figCw-2*\figCp+2*\figCg}
  \pgfmathsetmacro\figCpw{\figCend+0.16}
  \pgfmathsetmacro\figCplanE{-\figCgap}
  \pgfmathsetmacro\figCinR{\figCplanE-\figCwp-0.46}
  \pgfmathsetmacro\figCfork{\figCinR+0.16}
  \pgfmathsetmacro\figCinBox{\figCinR-0.14}
  \pgfmathsetmacro\figCinL{\figCinBox-\figCwi-0.14}

  \draw[fill=figCin, draw=figCinline, line width=0.45pt, rounded corners=4pt]
    (\figCinL,\figCtop) rectangle (\figCinR,\figCbot);
  \node[tag, text=figCmuted] at (\figCinL+0.14,0.64)
    {I\kern.6pt N\kern.6pt P\kern.6pt U\kern.6pt T};
  \node[io, multi, anchor=east, minimum width=\figCwi cm] (problem)
    at (\figCinBox,0) {Problem\\description};
  \node[io, anchor=east, minimum width=\figCwi cm, minimum height=5.2mm] (tests)
    at (\figCinBox,\figCyB) {Test cases};
  \node[circle, fill=white, draw=figCinline, line width=0.45pt,
    minimum size=3.7mm, inner sep=0pt] at (\figCinR-0.30,\figCtop) {};
  \pic at (\figCinR-0.30,\figCtop+0.01) {lock};

  \draw[fill=figCpanel, draw=figCpanelline, line width=0.45pt, rounded corners=4pt]
    (0,\figCtop) rectangle (\figCpw,\figCbot);
  \node[tag, text=figClabel] at (\figCx,0.64)
    {G\kern.6pt E\kern.6pt N\kern.6pt E\kern.6pt R\kern.6pt A\kern.6pt T\kern.6pt I\kern.6pt O\kern.6pt N};
  \foreach \i/\c/\t in {0/figCone/Declarative\\Spec,
                        1/figCtwo/Executable\\postcondition,
                        2/figCthree/Implementation}{
    \pgfmathsetmacro\xa{\figCx+\i*(\figCw-\figCp+\figCg)}
    \fill[\c, rounded corners=1.2pt]
      (\xa,-\figCh) -- (\xa+\figCw-\figCp,-\figCh) -- (\xa+\figCw,0)
      -- (\xa+\figCw-\figCp,\figCh) -- (\xa,\figCh) -- (\xa+\figCp,0) -- cycle;
    \node[align=center, inner sep=0pt, font=\sffamily\fontsize{8}{9}\selectfont]
      at (\xa+0.5*\figCw+0.5*\figCp-0.02,0) {\t};
    \pgfmathsetmacro\xt{\xa+\figCw-\figCp+0.5*\figCg}
    \node[circle, fill=white, draw=figCpanelline, line width=0.45pt,
      minimum size=3.7mm, inner sep=0pt] at (\xt,\figCh) {};
    \pic at (\xt,\figCh+0.01) {lock};
  }
  \draw[figCpanelline!70!figCarrow, line width=0.45pt]
    (\figCx+0.04,-0.66) -- ++(0,-0.06) -- (\figCend-0.04,-0.72) -- ++(0,0.06);
  \node[inner sep=0pt, anchor=base] at (0.5*\figCx+0.5*\figCend,\figCbase)
    {\usebox{\figCflow}};

  \node[io, multi, anchor=east, minimum width=\figCwp cm] (plan)
    at (\figCplanE,0) {Problem-\\level plan};
  \node[io, anchor=west, minimum width=\figCwa cm, minimum height=5.6mm] (assembly)
    at (\figCpw+\figCgap,0) {Assembly};
  \node[io, draw=none, fill=none, minimum width=\figCwa cm, minimum height=5.2mm] (file)
    at (assembly |- 0,\figCyB) {Lean file};
  \draw[draw=figCioline, line width=0.45pt, fill=figCio, rounded corners=3pt]
    (file.south west) -- (file.north west) -- ([xshift=-2mm]file.north east)
    [sharp corners] -- ([yshift=-2mm]file.north east)
    [rounded corners=3pt] -- (file.south east) -- cycle;
  \draw[draw=figCioline, line width=0.5pt, fill=figCone, line join=round]
    ([xshift=-2mm]file.north east) -- ([xshift=-2mm,yshift=-2mm]file.north east)
    -- ([yshift=-2mm]file.north east) -- cycle;
  \node[io, draw=none, fill=none, minimum height=5.2mm] at (file) {Lean file};

  \draw[figCarrow, line width=0.65pt] (\figCinR,-0.45) -- (\figCfork,-0.45);
  \draw[flow, rounded corners=2.5pt] (\figCfork,-0.45) |- (plan.west);
  \draw[flow, rounded corners=2.5pt] (\figCfork,-0.45) |- (0,\figCyB);
  \draw[flow] (plan.east) -- (0,0);
  \draw[flow] (\figCend,0) -- (assembly.west);
  \draw[flow] (assembly) -- (file);
\end{tikzpicture}%
\caption{\ywx{Data synthesis pipeline for \work{}. A problem-level plan guides three sequential generation stages. Each stage plans, generates, validates, and repairs its component before the next stage. Locks mark what stays frozen during synthesis: the inputs and every accepted component. Assembly combines the accepted components with the two theorem statements into a Lean file. Figure~\ref{fig:pipeline-example} in Appendix~\ref{app:pipeline} shows each stage on a running example.}}
\label{fig:pipeline}
\end{figure}
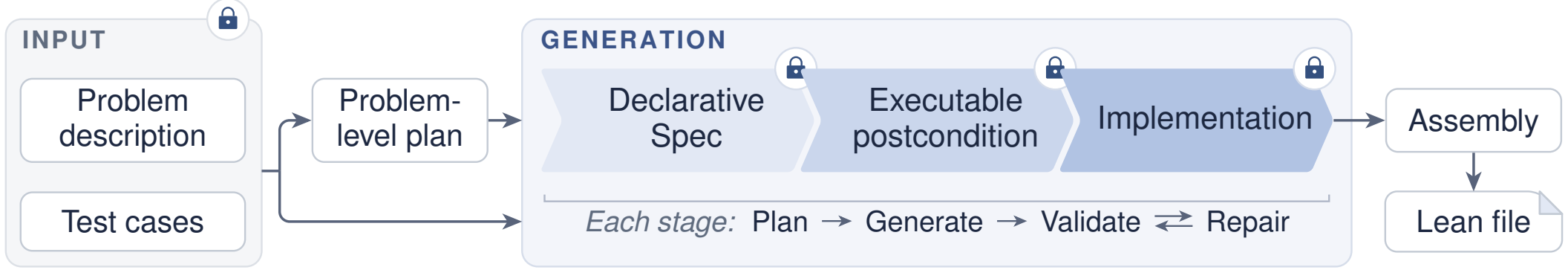

\subsection{Synthesis pipeline}
\label{sec:pipeline}
Our pipeline takes a problem description and test cases as input
(Figure~\ref{fig:pipeline}).
We first make a problem-level plan, then generate the declarative
specification (the precondition and the declarative postcondition), the executable postcondition, and the implementation in
three stages.
In each stage, we plan and generate a component, validate it against the acceptance criteria of Section~\ref{sec:acceptance} and repair it if validation fails. We separate the stages to prevent the generation of a later component from overfitting the test cases or weakening an accepted specification \citep{baksys2025atlas}. Finally, we assemble the accepted components into a Lean file
in the format of Section~\ref{sec:method}.

\textbf{Inputs.}
We collect 3{,}109 LeetCode problems \citep{xia2025leetcodedataset}
with typed function signatures, reference solutions,
and existing test cases (Table~\ref{tab:sources}).
Before synthesis, we construct and freeze positive and
negative test cases for each problem (Appendix~\ref{sec:testcorpus}).
A positive test case pairs a valid input with the output of the reference solution, which is computed independently of the specifications under synthesis \citep{wang2025codecontestsplus,fu2025klearcodetest,he2025hardtests}.
A negative test case is either an invalid input or a valid
input paired with an incorrect output.

\begin{table}[!htb]
\caption{Number of problems and samples for each source. Problems are those used for SFT or RL, and SFT, RL and \work-Test share no problem. Difficulty follows the labels of each source, and an LLM judge assigns the 1{,}484 problems without a label, including all Aizu problems (Appendix~\ref{app:data}). RL tasks are pairs of a problem and a task.}
\label{tab:sources}
\centering
\scriptsize
\setlength{\tabcolsep}{4pt}
\begin{tabular}{lrrrrrrrrr}
\toprule
\multirow{2}{*}{Source} & \multirow{2}{*}{Problems} & \multicolumn{3}{c}{Difficulty} & \multicolumn{2}{c}{SFT} & \multicolumn{2}{c}{RL} & \work-Test \\
\cmidrule(l{2pt}r{2pt}){3-5} \cmidrule(l{2pt}r{2pt}){6-7} \cmidrule(l{2pt}r{2pt}){8-9} \cmidrule(l{2pt}r{2pt}){10-10}
 & & Easy & Medium & Hard & Problems & Samples & Problems & Tasks & Problems \\
\midrule
LeetCode & 2{,}761 & 629 & 1{,}442 & 690 & 2{,}701 & 14{,}928 & 60 & 105 & 300 \\
Codeforces & 6{,}155 & 1{,}898 & 1{,}568 & 2{,}689 & 1{,}292 & 4{,}564 & 4{,}863 & 7{,}790 & 0 \\
AtCoder & 900 & 329 & 385 & 186 & 0 & 0 & 900 & 1{,}452 & 0 \\
Aizu & 988 & 382 & 391 & 215 & 0 & 0 & 988 & 1{,}545 & 0 \\
CodeChef & 536 & 187 & 214 & 135 & 0 & 0 & 536 & 920 & 0 \\
GeeksforGeeks & 1{,}232 & 749 & 483 & 0 & 0 & 0 & 1{,}232 & 1{,}825 & 0 \\
\midrule
Total & 12{,}572 & 4{,}174 & 4{,}483 & 3{,}915 & 3{,}993 & 19{,}492 & 8{,}579 & 13{,}637 & 300 \\
\bottomrule
\end{tabular}
\end{table}

\textbf{Specification.}
We first generate the precondition and the declarative
postcondition. To make the declarative postcondition executable, we apply \texttt{decide} when a decidability instance is available and otherwise attempt to synthesize such an instance (Appendix~\ref{app:route}). The result, called the decidable form, agrees with the declarative postcondition by construction. When instance synthesis fails, the declarative postcondition is reformulated into a separate executable postcondition, and the postcondition equivalence theorem, proved after synthesis (Section~\ref{sec:outcomes}), states that the executable postcondition agrees with the declarative postcondition.

\textbf{Implementation.}
Given the accepted specification, we plan an algorithm and generate the implementation, where LLM judges check that the implementation terminates on all valid inputs and is fast enough for the stated constraints (Appendix~\ref{app:stages}).

\textbf{Repair.}
If a component fails validation, the model revises it from compiler errors, failed test cases and judge feedback for up to 18 rounds. Each repair round modifies only the faulty code block, leaving other blocks and accepted components unchanged, which prevents repairs from weakening a specification to pass the test cases. The recorded repairs provide the repair samples of Section~\ref{sec:tasks}.

\textbf{Models.}
Based on a pilot comparison of frontier models (Appendix~\ref{app:models}), we select Kimi-K3 \citep{kimik3} for generation and repair, DeepSeek-V4-Pro \citep{deepseekv4} for planning, and DeepSeek-V4-Flash for the judges.

\subsection{Results}
\label{sec:outcomes}

The pipeline synthesizes 3{,}037 of the 3{,}109 problems (97.7\%, Appendix~\ref{app:accounts}).
Only 34.8\% of the problems pass all stages at the first attempt, and repair recovers most of the others within 2 rounds (Appendix~\ref{app:repairanalysis}).
For 35.5\% of the synthesized problems, the declarative postcondition can be made executable by \texttt{decide}.
These executable postconditions pass validation on the first attempt in 98.1\% of cases, compared with 52.4\% for those generated separately (Appendix~\ref{app:specanalysis}).
Since tests on small inputs do not detect nontermination or inefficiency, the implementation judge rejects 19.6\% of the generated implementations that pass all test cases (Appendix~\ref{app:stages}).

After synthesis, we attempt to prove the postcondition equivalence theorem for all 3{,}037 problems (Appendix~\ref{app:proofs}), and the attempts expose statement errors in 31 problems (1.0\%), all of which had passed the test layers and the judges.
Most of these errors (19) are in the declarative postcondition, which execution never evaluates (Table~\ref{tab:stmterrors} in Appendix~\ref{app:proofs}).
We repair the 31 problems and then prove the theorem.
For the remaining 99.0\% of the problems, the theorem is provable as synthesized.

\subsection{\work-Test}
\label{sec:test}

\work{}-Test is a held-out evaluation benchmark consisting of 300 problems from the synthesized dataset, with 72 easy, 155 medium and 73 hard problems, yielding 834 samples across four tasks (Table~\ref{tab:stagetasks} in Appendix~\ref{app:databreakdown}).
We manually review every component against the problem description and identify errors in 21 problems (Appendix~\ref{app:human-review}).
Attempts to prove the correctness theorems reveal one additional error (Appendix~\ref{app:correct}).
We correct these 22 problems and prove correctness for all 300; the remaining 278 problems (92.7\%) require no changes to their definitions.

These reviewed problems serve as references for evaluating model-generated specifications and implementations.
For SpecGen, we pair a generated specification with the reference implementation, and for CodeGen, a generated implementation with the reference specification. The repair tasks are evaluated in the same way, and we check the resulting Lean files against the acceptance criteria of Section~\ref{sec:acceptance}.
Since evaluation relies on test cases and judges rather than comparison with the reference specification, it can accept alternative correct formulations.

\section{Training}
\label{sec:train}

\subsection{Training data}
\label{sec:tasks}

Each synthesized problem contributes samples to the tasks of Section~\ref{sec:tasks-def}. The generation tasks take accepted components as targets, and the repair tasks take components rejected during synthesis as inputs. To obtain a starting point for RL that reasons concisely, we rephrase the long and often redundant reasoning recorded during synthesis into a short trace of the planning behind a generated component or the diagnosis behind a repair (Appendix~\ref{app:traces}).

Increasing the number of problems improves SpecGen performance more than continued pretraining on a Mathlib corpus (Appendix~\ref{app:cpt}). 
We therefore source further problems for supervised fine-tuning (SFT) and reinforcement learning (RL) from Codeforces, AtCoder, Aizu, CodeChef and GeeksforGeeks through TACO \citep{li2023taco} and open-r1/codeforces \citep{penedo2025codeforces} (Table~\ref{tab:sources}). To eliminate the risk of data contamination, the final SFT and RL data are disjoint from each other and from the evaluation benchmarks (Appendix~\ref{app:data}).

\subsection{Supervised fine-tuning}
\label{sec:training}

We expand the SFT data to 19.5K training samples over 3 rounds of expert iteration \citep{polu2023expertiteration}, each fine-tuning Qwen3.8-27B from the base model on the accumulated samples and adding the outputs that meet the acceptance criteria of Section~\ref{sec:acceptance}. SpecGen performance rises as the data grow, with no sign of saturation (Appendix~\ref{app:ei}). Fine-tuning Qwen3.5-9B \citep{qwen3.5} and Qwen3.8-27B \citep{qwen38} on these samples yields \work-9B-SFT and \work-27B-SFT, with the settings of Appendix~\ref{app:training}.

\subsection{Reinforcement learning}
\label{sec:rl}

We further apply RL to the SFT models for 1 epoch on 13{,}637 SpecGen and CodeGen tasks over 8{,}579 problems disjoint from the SFT data and \work-Test (Table~\ref{tab:sources}), yielding \work-9B-RL and \work-27B-RL. \ywx{We ablate the components of our recipe (Appendix~\ref{app:ablation-components}) and compare with VeriSpecGen on the same backbone (Appendix~\ref{app:shared-backbone}).}

\textbf{Reward design.}
Since the SFT models solve only 10.1\% and 18.7\% of SpecGen problems at pass@1 (Section~\ref{sec:main}), a binary reward that requires all acceptance criteria induces low reward variance, which slows optimization even when the reward is faithful to the acceptance criteria \citep{razin2025goodteacher}. We therefore score each rollout with a 6-stage progressive reward that follows the order of the acceptance criteria of Section~\ref{sec:acceptance} \citep{le2022coderl,specrl2026}. Failing a stage assigns the rollout the value of that stage, from format (0), compilation (0.1), test execution (0.2), test case alignment (0.1), non-tautology (0.1) to overall quality (0.4), and passing all stages yields 1. With $v_k$ the value of stage $k$, the reward of rollout $y$ is
\begin{equation}
r(y) = \begin{cases}
1 & \text{if } y \text{ passes all stages,} \\
v_k & \text{if } k \text{ is the first stage that } y \text{ fails.}
\end{cases}
\label{eq:reward}
\end{equation}
We value non-tautology at 0.1 to rank specifications that restate the computation without specifying the problem below failed tests. Replaying the training rollouts shows that at this value tautological specifications have a negative mean advantage in every group, whereas at a higher value such as 0.3 the advantage turns positive in groups where no rollout passes (Appendix~\ref{app:rldynamics}). Two kinds of rollouts that pass all stages receive partial credit instead of 1, enumerative specifications (0.4) and slow implementations (0.5), which evaluation accepts. To prevent reward hacking, a test counts toward the required number of tests only if it calls a declaration of the output, which excludes tests that hold trivially. Falling short of this number reduces the reward to 0.1. Hard-coding answers, by special-casing the test inputs or by copying the given test cases into the implementation, sets the reward to 0, and so does returning a repair unchanged (Appendix~\ref{app:rlreward}).

\textbf{Self-repair.}
Besides generation, we train on repair tasks built from failed rollouts of the current policy, which avoids distribution shift \citep{kumar2025score}.
During training, these failed rollouts are paired with compiler messages or judge feedback in a buffer, from which 40\% of each batch is drawn as single-turn repair tasks (Table~\ref{tab:tasks}). Repairs continue for up to 3 rounds, and the reward of a repair rollout is scaled down linearly from 1 in the first round to 0.8 in the last.

\textbf{Policy optimization.}
Our RL algorithm extends Decoupled Clip and Dynamic Sampling Policy Optimization \citep[DAPO;][]{yu2025dapo} by normalizing the advantage by the standard deviation over the batch instead of the group \citep{hu2025reinforcepp} and masking the clipped surrogate $\ell_{i,t}$ of token $t$ in rollout $i$ with $m_{i,t} \in \{0,1\}$. We maximize
\begin{equation}
\mathcal{J}(\theta) = \mathbb{E}\Big[\frac{\sum_{i,t} m_{i,t}\, \ell_{i,t}(\theta)}{\sum_{i,t} m_{i,t}}\Big],
\label{eq:objective}
\end{equation}
which reduces to the original algorithm with $m_{i,t} = 1$ and group normalization. We classify a rollout as repetitive if a 40-character n-gram occurs at least 20 times and over 40\% of n-grams are repeats \citep{welleck2020unlikelihood}. Repetitive rollouts that reach the length limit receive reward 0, and the mask removes the gradient of such a rollout beyond 200 tokens past the repetition onset. For the 27B model, the loss adds a penalty with weight 0.1 on the probability of the end-of-sequence token inside the thinking segment where it exceeds 1e-6. Appendix~\ref{app:training} lists the other hyperparameters.

\section{Experiments}
\label{sec:experiments}

\subsection{Setup}
\label{sec:evalsetup}

Training settings, including hyperparameters and the Lean environment, are listed in Appendix~\ref{app:training}, and the sampling settings in Appendix~\ref{app:inference}. We report pass@$k$ for $k \in \{1, 4\}$, the fraction of problems with at least one of $k$ samples meeting the acceptance criteria of Section~\ref{sec:acceptance} (Appendix~\ref{app:metrics}). All evaluated models receive the same prompts (Appendix~\ref{app:prompts}). The baselines are our backbones, frontier models and prior work on specification generation.
\begin{itemize}[leftmargin=*, itemsep=0pt, topsep=2pt, parsep=0pt]
\item \textbf{Instruct Models:} To isolate the effect of fine-tuning, we evaluate the backbones Qwen3.5-9B \citep{qwen3.5} and Qwen3.8-27B \citep{qwen38}.
\item \textbf{Frontier Models:} Claude-Opus-5 \citep{opus5}, Claude-Sonnet-5 \citep{sonnet5}, GPT-5.6-Sol and GPT-5.6-Luna \citep{gpt56}, Kimi-K3 \citep{kimik3}, and DeepSeek-V4-Pro and DeepSeek-V4-Flash \citep{deepseekv4}.
\item \textbf{Prior Work:} We compare against VeriSpecGen \citep{verispecgen2026}, which fine-tunes Qwen3-4B and Qwen3-Coder-30B-A3B on refinement trajectories distilled from a closed teacher, using the reported results as the models are not publicly available.
\end{itemize}

\subsection{Main results}
\label{sec:main}

We report the performance on the in-domain test set in Table~\ref{tab:mainpass}. Training improves SpecGen pass@4 from 0.7\% to 82.7\% for the 9B backbone and from 14.3\% to 84.3\% for the 27B backbone. \work-27B-RL surpasses every frontier model in both pass@1 and pass@4 on SpecGen and CodeGen, with 65.7\% SpecGen pass@1 (vs.\ 57.7\% for Claude-Opus-5, the strongest frontier model). The 9B RL model also outperforms every frontier model in SpecGen pass@1 (58.6\%). On the repair tasks, the 27B RL model attains the best CodeGen-Repair pass@1 (88.1\%) and stays within 2 points of the best result on the other metrics, while the 9B RL model surpasses most frontier models in SpecGen-Repair pass@1. \ywx{To assess false positives of SpecGen acceptance, we prove the postcondition equivalence theorem for 99.2\% of accepted RL outputs (Appendix~\ref{app:accepted-equiv}), and expert review confirms 90 of 100 sampled outputs, with all errors in the validity predicate (Appendix~\ref{app:specgen-audit}).}

\begin{table}[htbp]
\providecommand{\gain}[1]{\rlap{\,{\color{green!50!black}\tiny(+#1)}}}
\providecommand{\drop}[1]{\rlap{\,{\color{red!70!black}\tiny(\textminus#1)}}}
\centering
\captionsetup{font=scriptsize}
\caption{Main results on \work-Test. \textbf{Bold} indicates the best result for each metric and \underline{underline} the second best. Green numbers in parentheses show the gain in points over the backbone of the same size.}
\label{tab:mainpass}
\scriptsize
\setlength{\tabcolsep}{6.2pt}
\begin{tabular}{@{\hspace{6pt}}l *{8}{c}@{\hspace{18pt}}}
\toprule
\multicolumn{1}{c}{\multirow{2}{*}{Model}} & \multicolumn{2}{c}{SpecGen} & \multicolumn{2}{c}{CodeGen} & \multicolumn{2}{c}{SpecGen-Repair} & \multicolumn{2}{c}{CodeGen-Repair} \\
\cmidrule(l{2pt}r{2pt}){2-3}\cmidrule(l{2pt}r{2pt}){4-5}\cmidrule(l{2pt}r{2pt}){6-7}\cmidrule(l{2pt}r{2pt}){8-9}
 & pass@1 & pass@4 & pass@1 & pass@4 & pass@1 & pass@4 & pass@1 & pass@4 \\
\specialrule{\lightrulewidth}{\aboverulesep}{2.5pt}
\multicolumn{9}{@{\hspace{6pt}}l}{\textit{Frontier Models}} \\
\specialrule{\lightrulewidth}{2pt}{\belowrulesep}
Claude-Opus-5 & 57.7 & 81.3 & \underline{93.0} & \underline{98.3} & \textbf{83.0} & \textbf{90.1} & \underline{86.9} & \underline{95.1} \\
Claude-Sonnet-5 & 13.8 & 39.3 & 54.8 & 73.0 & 66.4 & 81.7 & 77.2 & 88.3 \\
GPT-5.6-Sol & 29.8 & 59.7 & 80.0 & 96.3 & 61.8 & 77.1 & 82.3 & 89.3 \\
GPT-5.6-Luna & 9.5 & 28.7 & 39.9 & 71.3 & 49.4 & 69.5 & 71.8 & 87.4 \\
Kimi-K3 & 43.8 & \underline{83.7} & 67.2 & 94.3 & 73.5 & 87.8 & 81.3 & \textbf{97.1} \\
DeepSeek-V4-Pro & 21.2 & 54.3 & 48.6 & 72.3 & 57.6 & 82.4 & 76.9 & 94.2 \\
DeepSeek-V4-Flash & 38.4 & 76.7 & 45.0 & 69.7 & 67.0 & 88.5 & 73.3 & \underline{95.1} \\
\specialrule{\lightrulewidth}{\aboverulesep}{2.5pt}
\multicolumn{9}{@{\hspace{6pt}}l}{\textit{Instruct Models}} \\
\specialrule{\lightrulewidth}{2pt}{\belowrulesep}
Qwen3.5-9B & 0.2 & 0.7 & 8.2 & 18.0 & 8.4 & 26.0 & 4.9 & 14.6 \\
Qwen3.8-27B & 4.5 & 14.3 & 8.9 & 22.0 & 44.7 & 73.3 & 35.7 & 64.1 \\
\specialrule{\lightrulewidth}{\aboverulesep}{2.5pt}
\multicolumn{9}{@{\hspace{6pt}}l}{\textit{Ours}} \\
\specialrule{\lightrulewidth}{2pt}{\belowrulesep}
\work-9B-SFT & 10.1\gain{9.9} & 24.3\gain{23.6} & 13.1\gain{4.9} & 28.3\gain{10.3} & 56.7\gain{48.3} & 76.3\gain{50.3} & 59.5\gain{54.6} & 80.6\gain{66.0} \\
\work-9B-RL & \underline{58.6}\gain{58.4} & 82.7\gain{82.0} & 77.1\gain{68.9} & 96.0\gain{78.0} & 73.5\gain{65.1} & 84.7\gain{58.7} & 74.5\gain{69.6} & 94.2\gain{79.6} \\
\work-27B-SFT & 18.7\gain{14.2} & 40.7\gain{26.4} & 27.6\gain{18.7} & 52.7\gain{30.7} & 73.1\gain{28.4} & 84.7\gain{11.4} & 72.3\gain{36.6} & 87.4\gain{23.3} \\
\work-27B-RL & \textbf{65.7}\gain{61.2} & \textbf{84.3}\gain{70.0} & \textbf{94.8}\gain{85.9} & \textbf{99.7}\gain{77.7} & \underline{81.9}\gain{37.2} & \underline{89.3}\gain{16.0} & \textbf{88.1}\gain{52.4} & \underline{95.1}\gain{31.0} \\
\bottomrule
\end{tabular}
\end{table}

\subsection{Out-of-domain evaluation}
\label{sec:ood}
To assess generalization beyond the synthesized data, we evaluate on \verina{} \citep{ye2026verina}, a 189-problem manually curated benchmark with reference specifications, implementations and test cases.
Following the original evaluation, generated specifications must be sound and complete with respect to the reference specification on the test cases, and generated implementations must pass all positive test cases.
\work-27B-RL outperforms all baselines in SpecGen pass@1 with 91.9\% and trails the best baseline in pass@4 by 1.1 points (Table~\ref{tab:verina}). On CodeGen, both RL models outperform all baselines, with pass@1 of 92.6\% for the 9B model and 91.8\% for the 27B model (vs.\ 56.6\% for the best baseline). The 9B model also attains the highest pass@4 (98.4\%).
Including the test cases in the prompt improves SpecGen pass@4 of both RL models from 90.5\% to 96.8\%, and the same test cases improve the pass@4 of frontier models by 5 to 24 points (Table~\ref{tab:verina}).

\begin{table}[H]
\providecommand{\gain}[1]{\rlap{\,{\color{green!50!black}\tiny(+#1)}}}
\providecommand{\drop}[1]{\rlap{\,{\color{red!70!black}\tiny(\textminus#1)}}}
\centering
\captionsetup{font=scriptsize}
\caption{Out-of-domain evaluation on \verina{} without and with test cases in the prompt. The setting without test cases follows the default of \verina{}. Bold, underlines and green numbers as in Table~\ref{tab:mainpass}. \textsuperscript{\dag} indicates results reported in the original paper \citep{verispecgen2026} and scored by the same \verina{} evaluator, which are available only without test cases in the prompt. VeriSpecGen-4B and VeriSpecGen-30B-A3B are the SFT No-Test models of VeriSpecGen fine-tuned from Qwen3-4B-Instruct-2507 and Qwen3-Coder-30B-A3B, respectively. Italic values are pass@10 in place of pass@4 and are not underlined.}
\label{tab:verina}\label{tab:verina-notests}
\scriptsize
\setlength{\tabcolsep}{6.2pt}
\begin{tabular}{@{\hspace{6pt}}l *{8}{c}@{\hspace{18pt}}}
\toprule
\multicolumn{1}{c}{\multirow{3}{*}{Model}} & \multicolumn{4}{c}{w/o tests} & \multicolumn{4}{c}{w/ tests} \\
\cmidrule(l{2pt}r{2pt}){2-5}\cmidrule(l{2pt}r{2pt}){6-9}
 & \multicolumn{2}{c}{SpecGen} & \multicolumn{2}{c}{CodeGen} & \multicolumn{2}{c}{SpecGen} & \multicolumn{2}{c}{CodeGen} \\
\cmidrule(l{2pt}r{2pt}){2-3}\cmidrule(l{2pt}r{2pt}){4-5}\cmidrule(l{2pt}r{2pt}){6-7}\cmidrule(l{2pt}r{2pt}){8-9}
 & pass@1 & pass@4 & pass@1 & pass@4 & pass@1 & pass@4 & pass@1 & pass@4 \\
\specialrule{\lightrulewidth}{\aboverulesep}{2.5pt}
\multicolumn{9}{@{\hspace{6pt}}l}{\textit{Frontier Models}} \\
\specialrule{\lightrulewidth}{2pt}{\belowrulesep}
Claude-Opus-5 & \underline{77.8} & \underline{89.9} & 57.0 & 78.8 & 86.2 & 95.2 & 56.6 & 79.4 \\
Claude-Sonnet-5 & 57.8 & 85.7 & 20.2 & 46.6 & 64.8 & 90.5 & 17.9 & 41.8 \\
GPT-5.6-Sol & 42.1 & 73.5 & 16.3 & 35.4 & 85.1 & \textbf{97.9} & 20.8 & 40.7 \\
GPT-5.6-Luna & 50.8 & 86.2 & 20.2 & 51.9 & 67.3 & 94.7 & 24.5 & 58.2 \\
Kimi-K3 & 69.2 & 89.4 & 34.0 & 69.8 & 77.6 & 96.3 & 34.9 & 72.5 \\
DeepSeek-V4-Pro & 44.7 & 76.7 & 32.3 & 61.4 & 53.8 & 82.5 & 37.6 & 70.4 \\
DeepSeek-V4-Flash & 51.9 & 84.7 & 6.5 & 21.2 & 60.2 & 92.6 & 5.8 & 18.5 \\
\specialrule{\lightrulewidth}{\aboverulesep}{2.5pt}
\multicolumn{9}{@{\hspace{6pt}}l}{\textit{Instruct Models}} \\
\specialrule{\lightrulewidth}{2pt}{\belowrulesep}
Qwen3.5-9B & 2.0 & 7.0 & 25.3 & 48.7 & 7.0 & 18.5 & 23.4 & 49.2 \\
Qwen3.8-27B & 7.2 & 21.9 & 4.2 & 15.9 & 28.8 & 55.0 & 3.3 & 13.2 \\
\specialrule{\lightrulewidth}{\aboverulesep}{2.5pt}
\multicolumn{9}{@{\hspace{6pt}}l}{\textit{Prior Work}} \\
\specialrule{\lightrulewidth}{2pt}{\belowrulesep}
VeriSpecGen-4B & 30.3\textsuperscript{\dag} & \textit{45.2}\textsuperscript{\dag} & 35.4\textsuperscript{\dag} & \textit{66.8}\textsuperscript{\dag} & -- & -- & -- & -- \\
VeriSpecGen-30B-A3B & 47.6\textsuperscript{\dag} & \textit{63.9}\textsuperscript{\dag} & 58.7\textsuperscript{\dag} & \textit{84.0}\textsuperscript{\dag} & -- & -- & -- & -- \\
\specialrule{\lightrulewidth}{\aboverulesep}{2.5pt}
\multicolumn{9}{@{\hspace{6pt}}l}{\textit{Ours}} \\
\specialrule{\lightrulewidth}{2pt}{\belowrulesep}
\work-9B-SFT & 34.9\gain{32.9} & 65.6\gain{58.6} & 36.5\gain{11.2} & 71.4\gain{22.7} & 37.3\gain{30.3} & 69.8\gain{51.3} & 38.0\gain{14.6} & 70.4\gain{21.2} \\
\work-9B-RL & 73.7\gain{71.7} & \textbf{90.5}\gain{83.5} & \textbf{89.8}\gain{64.5} & \textbf{97.4}\gain{48.7} & \underline{87.0}\gain{80.0} & \underline{96.8}\gain{78.3} & \textbf{92.6}\gain{69.2} & \textbf{98.4}\gain{49.2} \\
\work-27B-SFT & 52.9\gain{45.7} & 77.2\gain{55.3} & 58.7\gain{54.5} & 85.2\gain{69.3} & 57.3\gain{28.5} & 82.5\gain{27.5} & 57.8\gain{54.5} & 86.8\gain{73.6} \\
\work-27B-RL & \textbf{84.5}\gain{77.3} & \textbf{90.5}\gain{68.6} & \underline{89.7}\gain{85.5} & \underline{93.7}\gain{77.8} & \textbf{91.9}\gain{63.1} & \underline{96.8}\gain{41.8} & \underline{91.8}\gain{88.5} & \underline{93.7}\gain{80.5} \\
\bottomrule
\end{tabular}
\end{table}

We further evaluate general capabilities on AIME 25 \citep{aime25}, GPQA-Diamond \citep{rein2024gpqa}, and HumanEval+ and MBPP+ \citep{liu2023evalplus} with the sampling settings of DeepSeek-R1 \citep{guo2025deepseekr1}. Supervised fine-tuning lowers every score, and RL recovers about half or more of each drop, with \work-9B-RL surpassing its backbone on both code benchmarks (Appendix~\ref{app:general}).

\ywx{Besides providing additional learning signal during training, the repair tasks enable the model to iteratively refine its specifications based on previous outputs and feedback from the compiler and judges at test time. On ICPC-Eval \citep{xu2025icpceval}, a benchmark of problems from 11 recent contests of the International Collegiate Programming Contest (ICPC), \work-27B-RL solves a higher percentage of problems with refine@16 (82.7\%) than with pass@16 (61.3\%, Appendix~\ref{app:icpc}).}

\subsection{RL analysis}
\label{sec:rlanalysis}

\begin{figure}[t]
\centering
\captionsetup{font=scriptsize,skip=4pt}
\includegraphics[width=\linewidth]{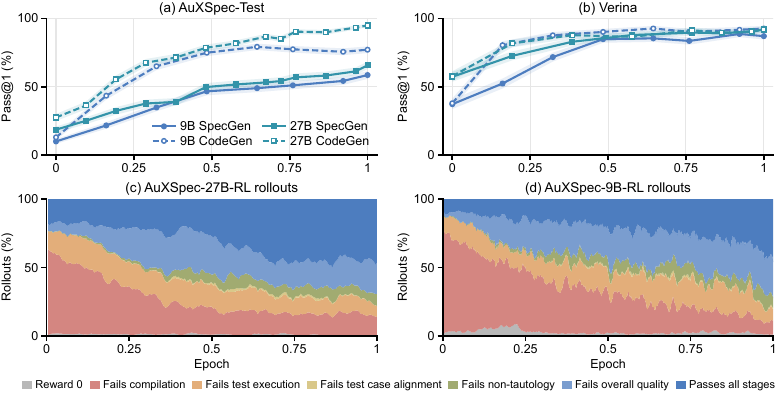}
\caption{Training dynamics of RL over 1 epoch. (a, b) SpecGen (solid) and CodeGen (dashed) pass@1 of the 9B and 27B models on \work-Test and \verina{}, where epoch 0 is the SFT model and bands show 1 standard error. (c, d) Fraction of rollouts by reward outcome, from the first failed stage to passing all stages, over all 4 tasks with a 5-step moving average.}
\label{fig:rltrain}
\end{figure}

Reinforcement learning from the SFT models improves SpecGen pass@1 on \work-Test from 10.1\% to 58.6\% for the 9B model and from 18.7\% to 65.7\% for the 27B model (Table~\ref{tab:mainpass} and \subfig{fig:rltrain}{a}), and the gains transfer to \verina{}, where SpecGen pass@1 rises from 37.3\% to 87.0\% and from 57.3\% to 91.9\% (Table~\ref{tab:verina} and \subfig{fig:rltrain}{b}). The reward is computed on the synthesized problems only, yet \verina{} scores each specification against the reference specification on test cases absent from the prompt, so the policy learns faithfulness to the description beyond the frozen test cases.

Gains at pass@1 can also come from sharpening the sampling distribution, in which case the SFT model would catch up at large $k$ \citep{yue2025rlreasoning}. SpecGen pass@256 of \work-27B-SFT (85.7\%) stays below pass@8 of \work-27B-RL (88.2\%), and the SFT model therefore needs more than 32 times the samples of the RL model to reach this pass rate. Within 256 samples, 33 of the 300 problems are solved only by the RL model and 1 only by the SFT model (Appendix~\ref{app:passk}). The RL model is also token-efficient, requiring 20.3\% of the output tokens per solved SpecGen sample of the most economical frontier model (Appendix~\ref{app:tokens}).

We utilize a progressive reward to assign fine-grained credit to partial progress. A binary reward assigns 0 to every rollout that fails any stage and provides no signal in groups where no sample passes. With fine-grained values, the group-relative advantage can distinguish better samples from worse ones within a group, which raises the probability of the better samples until the group saturates. Replaying the rollouts of the final runs confirms the denser signal. Groups without a passing rollout make up 37\% of SpecGen groups for the 27B model and 47\% for the 9B model and would be discarded by dynamic sampling under a binary reward \citep{yu2025dapo}, whereas the progressive reward yields nonzero variance in 87.9\% and 86.2\% of them (Appendix~\ref{app:rlreward}). The rollouts shift accordingly and follow this group-relative credit. In the early stage, few CodeGen groups contain a fast and correct program, where slow but correct programs receive a positive advantage and grow (up to 67.4\% of correct CodeGen rollouts of the 27B model). These slow programs then decrease gradually as fast programs appear (0.4\% at the end), whereas the 9B model, which solves fewer CodeGen problems, ends with more of them (30.4\%). Enumerative specifications show a similar pattern (17.0\% to 2.8\% of otherwise accepted specifications for the 27B model, \subfig{fig:rltrain}{c}).

The selection of the intermediate values relies on the reward dynamics, which we measure by replaying the training rollouts and recording the advantage of each stage in groups of different strength. For instance, correct but slow implementations are valued at 0.5, at which their advantage is positive only in groups with fewer than two passing rollouts, whereas at 0.8 it would stay positive with six passing rollouts of 16 (Appendix~\ref{app:rldynamics}). As CodeGen groups without a passing rollout fall from 44\% to 0\% over training (\subfig{fig:rldyn}{b2}), the mean advantage of slow implementations turns from $+0.10$ to $-0.17$ (steps 1--40 and 161--208). In SpecGen, 29\% to 52\% of groups remain without a passing rollout in each period (\subfig{fig:rldyn}{b1}), so hard SpecGen problems contribute a learning signal throughout training.

\section{Related work}
\label{sec:related}

\textbf{Specification generation.}
Specification autoformalization translates requirements into formal specifications. Prior work studies postcondition generation from natural language \citep{endres2024nl2postcond}, executable Python specifications \citep{chen2026codespecbench} and user intent in Dafny \citep{lahiri2024intent}.

\textbf{Training for specification generation.}
Training for program specification generation has concentrated on auto-active verification languages (Section~\ref{sec:introduction}). In Lean and Rocq, prior work checks a specification by proving each test case separately or by searching for counterexamples \citep{feng2026leetproof}, and reinforcement learning from verifier feedback has targeted correctness proofs \citep{li2026goedelcodeprover} but remains underexplored for specification generation \citep{baksys2025atlas}. Specification generators there are trained instead on synthesized and repaired examples, where a reference implementation labels tests, a proof attempt checks each labeled test, and LLM judgments guide refinement when proof automation is inconclusive \citep{verispecgen2026}. \work{} also labels tests with reference solutions, and relies on the test results of the executable postcondition to filter the training data and to define the reinforcement-learning reward (Section~\ref{sec:acceptance}).

\ywx{\textbf{Executable specifications.}
The executability of specifications has long been debated in formal methods. Executable specifications can be validated against intended behavior by execution \citep{kemmerer1985testing,fuchs1992executable}, whereas declarative specifications are more expressive and independent of any algorithm \citep{hayes1989executable}. Runtime assertion checking and property-based testing run contracts and properties as tests \citep{meyer1992contract,cheon2002jml,claessen2000quickcheck}, and proof assistants test conjectures before proving them \citep{berghofer2004random,plausible}. Boolean reflection pairs a proposition with a Boolean function and a proof that the two agree \citep{gonthier2010ssreflect,paraskevopoulou2015foundational}. Executable counterparts can also be derived automatically from inductive relations in proof assistants, but the derivation covers a restricted class of relations, and users still write executable variants by hand and prove that the two agree \citep{berghofer2009inductive,paraskevopoulou2022computing}. \work{} utilizes both sides of this trade-off, where we let models generate the executable variants, a step that previously required manual curation.}

\textbf{Adjacent settings.}
Generating specifications for existing implementations utilizes a verifier to check the specification against the code but assumes the code is correct \citep{safe2025,reform2025}. 
In this setting, reinforcement learning has included executable checks on mutated-output tests in the reward \citep{specrl2026}. Our SpecGen task instead starts from the problem description, input constraints and tests, without access to an implementation. Vericoding benchmarks evaluate the reverse task, generating implementations and proofs from a given formal specification \citep{algoveri2026,bursuc2025vericoding}.

\section{Conclusion}
\label{sec:conclusion}

We presented \work{}, a framework for training models to generate program specifications in Lean with execution-based feedback.
\work{} pairs each declarative postcondition with an executable postcondition, and the postcondition equivalence theorem states that both agree on all valid inputs.
This allows us to evaluate specifications by directly executing test cases, without proving each case separately.
At the same time, we retain the declarative postcondition for formal correctness proofs.
With this format, we synthesize and release 3{,}037 LeetCode problems, each with a proved postcondition equivalence theorem.
The dataset includes \work{}-Test, a held-out benchmark of 300 expert-reviewed problems, all with proved correctness theorems.

Training on the synthesized data with an execution-based reinforcement-learning reward yields 9B and 27B models, and the 27B model surpasses all frontier models on \work{}-Test and outperforms all baselines in SpecGen pass@1 on the out-of-domain \verina{} benchmark.
Since the models are trained only on synthesized problems, the gains on \verina{} suggest that training with the reward improves faithfulness to the problem description beyond the test cases used in the reward.
We discuss the limitations of this work in Appendix~\ref{app:limitations}.

\ifdefined\ARXIV
\subsubsection*{Acknowledgments}
This research was supported by the starting grant ``A New Paradigm for Flow and Cut Algorithms'' (no. TMSGI2 218022) of the Swiss National Science Foundation and the Swiss AI small project grant ``Formal Reasoning in Sparse Data Regimes'' (a0093). Sorrachai Yingchareonthawornchai is supported by Dr.~Max R\"ossler, the Walter Haefner Foundation, and the ETH Z\"urich Foundation.
\fi
\subsection*{AI use statement}

We used generative AI tools to synthesize the dataset and to produce proofs. The pipeline uses LLMs to synthesize, repair, judge and rewrite specifications and implementations (Section~\ref{sec:synthesis}, Appendices~\ref{app:pipeline} and~\ref{app:prompts}), and proof agents built on LLMs produce the postcondition equivalence and correctness proofs, which the Lean kernel checks (Appendix~\ref{app:proofs}). We did not use generative AI tools to develop the conceptual framework or to interpret results, and translation and qualitative data analysis do not apply to this work. We also used generative AI tools to edit the text, create figures, search the literature and format references. We reviewed all AI-assisted work, ran the AI-generated code, checked every number against the script that computed it and every citation against the original publication, and take responsibility for the final content of this work, including text, claims or artifacts produced with the aid of generative AI.

\subsection*{Ethics statement}

One of the authors carried out the expert review of \work-Test, and 2 of the authors reviewed the accepted outputs (Appendix~\ref{app:human}), and the work involves no other human participants and no personal data. The problems come from LeetCodeDataset \citep{xia2025leetcodedataset} and TACO \citep{li2023taco}, both released under the Apache-2.0 license, with the web-crawled problem statements of TACO under CC BY 4.0. We use them for research with attribution and will release our data and models under compatible licenses. The released models generate specifications and implementations for programming problems, and every accepted implementation remains checkable by proof against the postcondition.

\subsection*{Reproducibility statement}

The synthesis pipeline is described in Section~\ref{sec:synthesis}, with every stage in Appendix~\ref{app:pipeline}, the counts at each step in Appendix~\ref{app:accounts}, the models used in Appendix~\ref{app:models}, condensed prompts of the stages and judges in Appendix~\ref{app:prompts}, and the proving of the postcondition equivalence and correctness theorems in Appendix~\ref{app:proofs}. Training is described in Section~\ref{sec:train}, with the training and sampling settings in Appendices~\ref{app:training} and~\ref{app:inference}, expert iteration in Appendix~\ref{app:ei}, the metrics in Appendix~\ref{app:metrics} and the reinforcement-learning reward and ablations in Appendix~\ref{app:rl}. We will release the synthesized problems with test cases and postcondition equivalence proofs, \work-Test with correctness proofs, the evaluation code and the model weights.

\clearpage
\bibliography{references}
\bibliographystyle{iclr2027_conference}

\clearpage
\appendix

\section{File format}
\label{app:format}

\subsection{Declarations}
\label{app:terms}

Each term of the paper is implemented by a declaration (Table~\ref{tab:declarations}).

\begin{table}[h]
\caption{Declarations per \work{} problem.}
\label{tab:declarations}
\centering
\scriptsize
\setlength{\tabcolsep}{4pt}
\begin{tabular}{@{}lll@{}}
\toprule
Term & Declaration & Description \\
\midrule
Precondition & {\scriptsize\texttt{isValidInput}} & Executable predicate encoding the input constraints \\
Postcondition & {\scriptsize\texttt{\char`\<algo\char`\>\_spec}} & \texttt{Prop} relating input and output \\
Executable postcondition & {\scriptsize\texttt{\char`\<algo\char`\>\_check}} & \texttt{Bool} function of input and output \\
Postcondition equivalence theorem & {\scriptsize\texttt{\char`\<algo\char`\>\_check\_iff\_spec}} & \texttt{\_check} decides \texttt{\_spec} on valid inputs \\
Implementation & {\scriptsize\texttt{\char`\<algo\char`\>}} & Total function computing the output \\
Correctness theorem & {\scriptsize\texttt{\char`\<algo\char`\>\_correct}} & \texttt{\char`\<algo\char`\>} satisfies \texttt{\_spec} on valid inputs \\
\bottomrule
\end{tabular}
\end{table}

\subsection{Format}
\label{app:layout}

Each file includes precondition, postcondition and executable postcondition with postcondition equivalence theorem, implementation with correctness theorem, and test cases. Helper definitions follow a topological order of dependency. The namespace, the type abbreviations, the printed expressions of the frozen test cases, and both theorem statements are assembled. The test cases are organized into test layers (Table~\ref{tab:layers}).

\begin{table}[htbp]
\caption{Test layers with the component under test and the error detected.}
\label{tab:layers}
\centering
\scriptsize
\setlength{\tabcolsep}{4pt}
\begin{tabular}{@{}>{\raggedright\arraybackslash}m{0.7cm}>{\raggedright\arraybackslash}m{2.9cm}>{\raggedright\arraybackslash}m{3.4cm}>{\raggedright\arraybackslash}m{1.1cm}>{\raggedright\arraybackslash}m{3.2cm}@{}}
\toprule
Layer & Component Under Test & Test Case & Assertion & Error Detected \\
\midrule
\La{1} & \multirow{2}{=}{Precondition} & Valid input & Accepts & Overly restrictive \\
\Lb{1} & & Invalid input & Rejects & Overly permissive \\
\midrule
\La{2} & \multirow{2}{=}{Executable postcondition} & Valid input, correct output & Accepts & Incomplete \\
\Lb{2} & & Valid input, incorrect output & Rejects & Unsound \\
\midrule
L3 & Implementation & Valid input, expected output & Matches & Incorrect \\
\midrule
L4 & Executable postcondition on\newline implementation output & Valid input & Accepts & Implementation output rejected \\
\bottomrule
\end{tabular}
\end{table}

\section{Configuration}
\label{app:training}

\subsection{Training}

Both model sizes are fine-tuned with the settings in Table~\ref{tab:sft}. \work-9B-SFT is fine-tuned on an earlier version of the SFT data with 18{,}489 samples over 4{,}026 problems. In a round of expert iteration, the model is fine-tuned for 3 epochs on the training samples of that round. Checkpoints are evaluated after each epoch, and the models reported in the main text are the third-epoch checkpoints. Loss is taken on assistant tokens only.

\begin{table}[H]
\centering
\scriptsize
\setlength{\tabcolsep}{3pt}
\begin{minipage}[t]{0.43\linewidth}
\centering
\caption{Supervised fine-tuning settings.}
\label{tab:sft}
\begin{tabular}{lcc}
\toprule
Setting & 9B & 27B \\
\midrule
Base model & Qwen3.5-9B & Qwen3.8-27B \\
Hardware & 4 GH200 & 8 GH200 \\
Batch (sequences) & \multicolumn{2}{c}{64} \\
Sequence length (tokens) & \multicolumn{2}{c}{16{,}384} \\
Learning rate & \multicolumn{2}{c}{1e-5, cosine} \\
Warmup & \multicolumn{2}{c}{5\%} \\
Epochs & \multicolumn{2}{c}{3} \\
\bottomrule
\end{tabular}
\end{minipage}%
\hfill
\begin{minipage}[t]{0.55\linewidth}
\centering
\caption{Reinforcement learning settings.}
\label{tab:rl}
\begin{tabular}{lcc}
\toprule
Setting & 9B & 27B \\
\midrule
Start model & \work-9B-SFT & \work-27B-SFT \\
Hardware & 16 GH200 & 64 GH200 \\
Prompts per step & 32 & 64 \\
Rollouts per prompt & \multicolumn{2}{c}{16} \\
Prompts per mini-batch & \multicolumn{2}{c}{8} \\
Learning rate & \multicolumn{2}{c}{1e-6, constant} \\
Clip range $\epsilon_{\text{low}}$ / $\epsilon_{\text{high}}$ & \multicolumn{2}{c}{0.2 / 0.28} \\
Temperature / top-$p$ & \multicolumn{2}{c}{0.7 / 1.0} \\
Max new tokens & \multicolumn{2}{c}{16{,}384} \\
Training steps (1 epoch) & 434 & 208 \\
\bottomrule
\end{tabular}
\end{minipage}
\end{table}

Reinforcement learning follows the settings of Table~\ref{tab:rl} for 1 epoch of the 13{,}637 tasks, which takes 434 steps for the 9B model and 208 steps for the 27B model. Rollout and training share the same GPUs, the Lean compilation pool runs on the CPUs of the same nodes, and the LLM judges are called through an external API. Rollouts longer than 12{,}288 tokens lose reward linearly, up to 1.0 at the limit of 16{,}384 tokens, following the soft overlong punishment of DAPO \citep{yu2025dapo}. The optimizer is AdamW with weight decay 0.01 and gradient clipping at 1.0, and the clipped surrogate adds dual clipping with constant 3.0 for negative advantages. The maximum prompt length is 8{,}192 tokens. A SpecGen rollout needs 12 tests and a CodeGen rollout 3, fewer for problems with fewer test rows. A rollout fails compilation when a declaration does not elaborate, a test cannot be evaluated, compilation exceeds the time limit, the compiler reports an error without a position, or the rollout contains \texttt{sorry} outside the two theorem statements, \texttt{admit}, \texttt{axiom} or \texttt{native\_decide}. A rollout receives 0.3 when a judge call fails (Table~\ref{tab:rlstages}). A rollout is repetitive when a 40-character n-gram, taken at a stride of 4 characters, occurs at least 20 times and fewer than 60\% of the n-grams are distinct.

\subsection{Inference}
\label{app:inference}

Expert iteration, the evaluation of our models and the analyses of Appendix~\ref{app:additional} sample with the settings of Table~\ref{tab:inference}. Baselines follow the sampling settings recommended for each model with the same maximum output length, except the long-reasoning models DeepSeek-V4-Flash, DeepSeek-V4-Pro and Kimi-K3, which run without a length limit.

\begin{table}[H]
\caption{Inference settings.}
\label{tab:inference}
\centering
\scriptsize
\setlength{\tabcolsep}{4pt}
\begin{tabular}{lc}
\toprule
Inference engine & vLLM \\
Temperature & 0.3 (evaluation), 0.7 (expert iteration and analyses) \\
Top-$p$ & 0.95 \\
Max new tokens & 16{,}384 \\
Samples per problem, compile@$k$ and pass@$k$ & 4 \\
\bottomrule
\end{tabular}
\end{table}

\section{Training data}
\label{app:data}

\subsection{Source selection}

The LeetCode problems combine 2{,}869 problems of LeetCodeDataset \citep{xia2025leetcodedataset} with 746 free problems collected from the site after the dataset snapshot. Of the 746 problems, 263 pass a buildability filter that requires at least 3 examples and a Python function signature and excludes database, shell, concurrency and design problems. The seeds are 3{,}109 problems, 2{,}864 from LeetCodeDataset and 245 from the site. We take the Codeforces and AtCoder problems from the train split of TACO \citep{li2023taco}. For RL, we add problems from the train and test splits of TACO, including Aizu, CodeChef and GeeksforGeeks problems, and 989 Codeforces problems from open-r1/codeforces \citep{penedo2025codeforces}. Problems are admitted with at least 3 tests, a reference solution, and a single accepted answer, since one reference output cannot label a problem with several accepted answers. Stored tests whose inputs violate the constraints of the problem description are removed before use, which drops 1{,}028 of 90{,}885 test rows on Codeforces and 2{,}119 of 22{,}115 on AtCoder. Difficulty in Table~\ref{tab:sources} follows the labels of LeetCode and TACO and the ratings of open-r1/codeforces. The 1{,}484 problems without a label, including all Aizu problems, take the tier that an LLM judge assigns in the conversion audit, where easy problems need direct simulation or a single formula, medium problems a standard technique and hard problems a non-obvious idea or a combination of techniques.

\subsection{Problem format}

Each Codeforces and AtCoder problem, whose input is given from standard input, is rephrased into a typed function interface and admitted only when the reference solution still passes the stored tests through that interface. Conversion admits 3{,}490 of the 4{,}314 Codeforces problems (80.9\%). The rejected problems have stored tests that no reference solution reproduces (361), inputs that change under parsing and unparsing (259), several accepted answers (184), or other harness errors (20). Freezing the test cases removes 205 more problems, 201 of which lack 3 invalid inputs for \Lb{1}, and leaves 3{,}285. Of the 1{,}440 AtCoder problems, conversion admits 892, an audit for descriptions that drop a stated condition and for examples that violate the stated constraints leaves 677, and freezing leaves 623. Most AtCoder rejections at conversion are inputs that change under parsing and unparsing (352 of 548), which trace to faulty stored tests such as leading zeros or counts that do not match the input.

\subsection{Data breakdown}
\label{app:databreakdown}

We count the samples of each task for supervised fine-tuning, reinforcement learning, and \work-Test (Table~\ref{tab:stagetasks}). A reinforcement learning sample is a generation prompt, and repair prompts are built during training from failed rollouts (Section~\ref{sec:rl}).

\begin{table}[htbp]
\caption{Number of samples of each task in each stage.}
\label{tab:stagetasks}
\centering
\scriptsize
\setlength{\tabcolsep}{4pt}
\begin{tabular}{lrrr}
\toprule
Task & SFT & RL & \work-Test \\
\midrule
SpecGen & 6{,}528 & 8{,}281 & 300 \\
SpecGen-Repair & 3{,}807 & 0 & 131 \\
CodeGen & 6{,}645 & 5{,}356 & 300 \\
CodeGen-Repair & 2{,}512 & 0 & 103 \\
\midrule
Total & 19{,}492 & 13{,}637 & 834 \\
\bottomrule
\end{tabular}
\end{table}

The 19{,}492 supervised fine-tuning samples are drawn from the synthesis pipeline of Section~\ref{sec:synthesis} and expert iteration (Table~\ref{tab:sftbreakdown}). An expert-iteration repair chain accepted after more than 1 round becomes 1 generation sample with a reasoning trace over the first attempt, feedback and repair.

\begin{table}[htbp]
\caption{Supervised fine-tuning samples by source.}
\label{tab:sftbreakdown}
\centering
\scriptsize
\setlength{\tabcolsep}{4pt}
\begin{tabular}{lrrrrr}
\toprule
Source & SpecGen & SpecGen-Repair & CodeGen & CodeGen-Repair & Total \\
\midrule
Synthesis pipeline & 4{,}786 & 2{,}479 & 4{,}475 & 1{,}100 & 12{,}840 \\
Expert iteration & 373 & 1{,}328 & 739 & 1{,}412 & 3{,}852 \\
Expert iteration, rewritten & 1{,}369 & 0 & 1{,}431 & 0 & 2{,}800 \\
\midrule
Total & 6{,}528 & 3{,}807 & 6{,}645 & 2{,}512 & 19{,}492 \\
\bottomrule
\end{tabular}
\end{table}

The corpus builder also removes samples longer than 16{,}384 tokens (42), duplicate samples (57), samples beyond 3 per problem and task (11), and 1 sample that is not a whole file.

\subsection{Expert iteration details}

Each round samples 1 refinement chain per problem and task from a fine-tuned Qwen3.8-27B. A chain generates an output, compiles the output, and repairs the output for up to 8 rounds, or 16 rounds in the last round of expert iteration. A chain is accepted under the criteria of Section~\ref{sec:acceptance}, and the reasons of each judge rejection enter the next repair round as review text. Expert iteration runs 3 rounds (Table~\ref{tab:ei}). The first 2 rounds sample from the model fine-tuned on the LeetCode data, and the last round from the model fine-tuned on the data of the first 2 rounds and the Codeforces pipeline data. Accepted chains that pass at the first round become generation samples, and later rounds become repair samples or are rewritten into generation samples (Appendix~\ref{app:databreakdown}). A scan over all refinement records adds 384 samples from accepted chains that no round collected. The chains on reinforcement learning problems in the last round add no supervised fine-tuning samples.

\begin{table}[htbp]
\caption{Rounds of expert iteration. A chain is 1 refinement run on 1 problem and task.}
\label{tab:ei}
\centering
\scriptsize
\setlength{\tabcolsep}{4pt}
\begin{tabular}{lrrrr}
\toprule
Round & Maximum repair rounds & Chains & Accepted chains & SFT samples \\
\midrule
LeetCode & 8 & 5{,}239 & 2{,}268 & 3{,}610 \\
Codeforces & 8 & 3{,}778 & 1{,}248 & 2{,}137 \\
Last round & 16 & 3{,}208 & 1{,}444 & 521 \\
Scan over all records & -- & -- & -- & 384 \\
\midrule
Total & & 12{,}225 & 4{,}960 & 6{,}652 \\
\bottomrule
\end{tabular}
\end{table}

Expert iteration adds repairs more than first attempts. Of the 2{,}268 accepted chains of the LeetCode round, 685 pass at the first round, and of the 298 accepted SpecGen chains on reinforcement learning problems in the last round, 290 need at least 1 repair round.

\subsection{Decontamination}

The training problems pass filters that remove the 300 \work-Test problems, the problems that \verina{} links to, and the reinforcement learning problems from supervised fine-tuning. \verina{} problems are matched through the LeetCode links in the \verina{} descriptions, where 43 of the 189 problems link to 36 LeetCode problems. The filters remove 21 samples of \work-Test problems and 166 samples of 32 LeetCode problems that \verina{} links to, and a recount of the final training data finds no problem of these sets.

\section{Evaluation}
\label{app:evaluation}

\subsection{Metrics}
\label{app:metrics}

\textbf{Compile@$k$ and Pass@$k$.} We follow the standard definition of pass@$k$ \citep{chen2021codex} and define compile@$k$ by analogy. For a problem with $k$ samples, compile@$k$ is 1 when at least one sample compiles, and pass@$k$ is 1 when at least one sample meets the acceptance criteria of Section~\ref{sec:acceptance}. The reported value is the mean over problems, which equals the unbiased estimator of pass@$k$ \citep{chen2021codex} since exactly $k$ samples are drawn per problem. Test case alignment compares the test lines of a sample with the frozen test cases regardless of layer headings, and Lean code blocks are extracted regardless of the capitalization of the language tag.

\subsection{Token efficiency}
\label{app:tokens}

We measure output tokens per response and per solved sample (Table~\ref{tab:tokens}). \work-27B-RL requires the fewest output tokens per solved sample of all 13 models in the 4 generation settings. On SpecGen of \work-Test, it requires 80\% fewer tokens per solved sample than GPT-5.6-Sol, the most economical frontier model (2.7K vs.\ 13.2K), while its pass@1 is 35.9 points higher (65.7\% vs.\ 29.8\%; Table~\ref{tab:mainpass}). RL lowers the cost per solved SpecGen sample at both model sizes, as pass@1 grows faster than the response length (19.3K to 2.7K tokens for the 27B model and 36.0K to 11.0K for the 9B model).

RL changes the length of the reasoning trace in opposite directions at the 2 model sizes. Reasoning traces of the 27B model become shorter after RL, whereas reasoning traces of the 9B model become longer (1.6K to 0.4K and 1.7K to 4.6K thinking tokens per SpecGen response). A plausible reading is that the larger model needs less explicit reasoning to state a specification, whereas the smaller model relies on longer reasoning to reach a correct output. Thinking tokens are the tokens of the thinking segment. Tokens of our models and the backbones are counted with the tokenizer of the backbone, and tokens of the frontier models are the output tokens reported by the API, thinking included.

\begin{table}[htbp]
\centering
\captionsetup{font=scriptsize}
\caption{Average output tokens per response (Resp.) and output tokens per solved sample (Solved), the total output tokens of a task divided by the samples that meet the acceptance criteria, both in thousands. Responses are capped at 16{,}384 tokens for our models, the backbones and the Claude and GPT models, at 32{,}768 for Kimi-K3 and at 65{,}536 for the DeepSeek models. \underline{Underline} indicates the lowest value of each column within each group, and \textbf{bold} the lowest overall.}
\label{tab:tokens}
\scriptsize
\setlength{\tabcolsep}{4pt}
\begin{tabular}{@{\hspace{6pt}}l *{8}{r}}
\toprule
 & \multicolumn{4}{c}{\work-Test} & \multicolumn{4}{c}{\verina{}} \\
\cmidrule(l{2pt}r{2pt}){2-5}\cmidrule(l{2pt}r{2pt}){6-9}
 & \multicolumn{2}{c}{SpecGen} & \multicolumn{2}{c}{CodeGen} & \multicolumn{2}{c}{SpecGen} & \multicolumn{2}{c}{CodeGen} \\
\cmidrule(l{2pt}r{2pt}){2-3}\cmidrule(l{2pt}r{2pt}){4-5}\cmidrule(l{2pt}r{2pt}){6-7}\cmidrule(l{2pt}r{2pt}){8-9}
\multicolumn{1}{c}{Model} & Resp. & Solved & Resp. & Solved & Resp. & Solved & Resp. & Solved \\
\specialrule{\lightrulewidth}{\aboverulesep}{2.5pt}
\multicolumn{9}{@{\hspace{6pt}}l}{\textit{Frontier Models}} \\
\specialrule{\lightrulewidth}{2pt}{\belowrulesep}
Claude-Opus-5 & 9.7 & 16.8 & 4.8 & 5.2 & 5.2 & 6.0 & 3.9 & \underline{6.9} \\
Claude-Sonnet-5 & 10.2 & 74.4 & 7.3 & 13.4 & 4.4 & 6.9 & 9.6 & 53.7 \\
GPT-5.6-Sol & 3.9 & \underline{13.2} & 2.8 & \underline{3.5} & \underline{2.6} & \underline{3.0} & 3.1 & 14.7 \\
GPT-5.6-Luna & \underline{3.4} & 36.0 & \underline{2.8} & 7.0 & 2.7 & 4.0 & \underline{2.5} & 10.1 \\
Kimi-K3 & 10.7 & 24.4 & 6.7 & 10.0 & 5.2 & 6.7 & 9.0 & 25.7 \\
DeepSeek-V4-Pro & 27.0 & 126.9 & 24.1 & 49.5 & 17.1 & 31.8 & 20.8 & 55.2 \\
DeepSeek-V4-Flash & 49.6 & 129.2 & 49.0 & 109.0 & 37.8 & 62.9 & 43.2 & 741.4 \\
\specialrule{\lightrulewidth}{\aboverulesep}{2.5pt}
\multicolumn{9}{@{\hspace{6pt}}l}{\textit{Instruct Models}} \\
\specialrule{\lightrulewidth}{2pt}{\belowrulesep}
Qwen3.5-9B & \underline{3.2} & 1{,}912.1 & \underline{6.8} & \underline{82.9} & \underline{2.6} & 37.6 & \underline{2.3} & \underline{9.7} \\
Qwen3.8-27B & 13.4 & \underline{296.8} & 15.9 & 178.6 & 8.1 & \underline{28.0} & 15.5 & 469.8 \\
\specialrule{\lightrulewidth}{\aboverulesep}{2.5pt}
\multicolumn{9}{@{\hspace{6pt}}l}{\textit{Ours}} \\
\specialrule{\lightrulewidth}{2pt}{\belowrulesep}
\work-9B-SFT & 3.6 & 36.0 & 2.2 & 16.6 & 4.0 & 10.6 & 1.5 & 4.0 \\
\work-9B-RL & 6.4 & 11.0 & 4.1 & 5.3 & 3.9 & 4.5 & 1.9 & 2.0 \\
\work-27B-SFT & 3.6 & 19.3 & 1.9 & 7.0 & 3.8 & 6.7 & 1.4 & 2.5 \\
\work-27B-RL & \textbf{1.8} & \textbf{2.7} & \textbf{1.7} & \textbf{1.8} & \textbf{1.6} & \textbf{1.7} & \textbf{1.1} & \textbf{1.2} \\
\bottomrule
\end{tabular}
\end{table}

\subsection{General capabilities}
\label{app:general}

To assess general capabilities after specialization, we evaluate the backbones, SFT models and RL models on AIME 25 \citep{aime25}, GPQA-Diamond \citep{rein2024gpqa}, and HumanEval+ and MBPP+ of EvalPlus \citep{liu2023evalplus}. EvalPlus extends HumanEval \citep{chen2021codex} and MBPP \citep{mbpp2021} with more test cases per problem. Sampling follows the settings of DeepSeek-R1 \citep{guo2025deepseekr1} with temperature 0.6, top-$p$ 0.95 and at most 32K output tokens, and pass@1 averages 64 samples per problem on AIME 25, 8 on GPQA-Diamond and 4 on HumanEval+ and MBPP+.

Supervised fine-tuning lowers all 4 scores of both backbones, most on AIME 25, by 36.7 points for the 9B model and 53.1 for the 27B model (Table~\ref{tab:general}). RL recovers 72\% of this drop on AIME 25 at both sizes and about half of the drop on GPQA-Diamond (47\% and 52\%). On the 2 code benchmarks, \work-9B-RL surpasses its backbone (88.6\% vs.\ 86.6\% on HumanEval+ and 70.8\% vs.\ 65.3\% on MBPP+), and \work-27B-RL trails its backbone by 3.2 and 1.5 points.

\begin{table}[htbp]
\providecommand{\gain}[1]{\rlap{\,{\color{green!50!black}\tiny(+#1)}}}
\providecommand{\drop}[1]{\rlap{\,{\color{red!70!black}\tiny(\textminus#1)}}}
\centering
\captionsetup{font=scriptsize}
\caption{General capabilities of the backbones, SFT models and RL models (pass@1, \%). Numbers in parentheses show the change in points from the backbone of the same size.}
\label{tab:general}
\scriptsize
\setlength{\tabcolsep}{10pt}
\begin{tabular}{l *{4}{c}@{\hspace{24pt}}}
\toprule
Model & AIME 25 & GPQA-Diamond & HumanEval+ & MBPP+ \\
\midrule
Qwen3.5-9B & 62.3 & 80.2 & 86.6 & 65.3 \\
\work-9B-SFT & 25.6\drop{36.7} & 62.2\drop{18.0} & 63.9\drop{22.7} & 59.9\drop{5.4} \\
\work-9B-RL & 52.2\drop{10.1} & 70.7\drop{9.5} & 88.6\gain{2.0} & 70.8\gain{5.5} \\
\midrule
Qwen3.8-27B & 84.1 & 84.5 & 92.5 & 80.4 \\
\work-27B-SFT & 31.0\drop{53.1} & 62.9\drop{21.6} & 79.0\drop{13.5} & 71.1\drop{9.3} \\
\work-27B-RL & 69.3\drop{14.8} & 74.2\drop{10.3} & 89.3\drop{3.2} & 78.9\drop{1.5} \\
\bottomrule
\end{tabular}
\end{table}

\subsection{Self-repair on ICPC-Eval}
\label{app:icpc}

Besides generation, \work{} models are trained on the repair tasks (Section~\ref{sec:tasks-def}), which revise a rejected output from compiler and judge feedback. To assess the value of this ability, we evaluate \work-27B-RL on ICPC-Eval \citep{xu2025icpceval}, a benchmark of competitive programming problems whose specifications require careful definitions. Of its 118 problems, we exclude 28 problems with multiple correct outputs, which a fixed expected output cannot judge, and 15 problems whose test cases exceed the length range of \work-Test, which leaves 75 problems. Following Refine@K of ICPC-Eval and the comparison of self-repair with i.i.d.\ sampling at an equivalent budget \citep{olausson2024selfrepair}, we compare 3 settings within a budget of $k$ model calls per problem:
\begin{itemize}[leftmargin=*, itemsep=0pt, topsep=2pt, parsep=0pt]
\item \textbf{Pass@$k$} draws $k$ independent samples and is estimated from 16 samples.
\item \textbf{Refine@$k$ with compiler and judge feedback} draws 1 sample and returns each rejected output with the compiler messages or the judge findings, in the format of the repair tasks, until an output is accepted or $k$ calls are made.
\item \textbf{Refine@$k$ with compiler feedback only} follows the same chain and stops at the first output that compiles.
\end{itemize}
All settings sample at temperature 0.6 and top-p 0.95 with at most 32{,}768 generated tokens, and the acceptance criteria (Section~\ref{sec:acceptance}) score every output.

\begin{figure}[H]
\centering
\begin{minipage}[c]{0.46\linewidth}
\includegraphics[width=\linewidth]{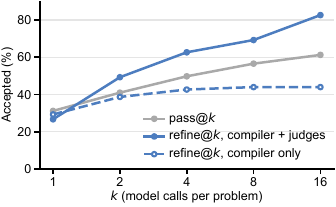}
\end{minipage}\hfill
\begin{minipage}[c]{0.52\linewidth}
\centering
\scriptsize
\setlength{\tabcolsep}{3pt}
\begin{tabular}{@{}l *{5}{c}@{}}
\toprule
$k$ & 1 & 2 & 4 & 8 & 16 \\
\midrule
Pass@$k$ & 31.2 & 41.0 & 49.8 & 56.6 & 61.3 \\
Refine@$k$, compiler + judges & 26.7 & \textbf{49.3} & \textbf{62.7} & \textbf{69.3} & \textbf{82.7} \\
Refine@$k$, compiler only & 29.3 & 38.7 & 42.7 & 44.0 & 44.0 \\
\bottomrule
\end{tabular}

\end{minipage}
\vspace{-4pt}
\caption{SpecGen results of \work-27B-RL on 75 problems of ICPC-Eval, as the percentage of problems with an accepted output within $k$ model calls per problem. Refine@$k$ returns compiler and judge feedback (compiler + judges) or compiler messages only (compiler only). Bold indicates refine@$k$ values above pass@$k$.}
\label{fig:icpc-repair}
\end{figure}

With compiler and judge feedback, refine@16 reaches 82.7\%, compared with 61.3\% for pass@16 at an equivalent budget (Figure~\ref{fig:icpc-repair}). With compiler feedback only, more chains compile within 2 rounds than independent samples do (88.0\% vs.\ 79.3\% for compile@2), but refine@16 reaches 44.0\%, as the chain stops at the first output that compiles and the 42 outputs that the judges reject are never revised.

\raggedbottom
\setlength{\intextsep}{6pt}
\section{Training method analysis}
\label{app:additional}

\subsection{Expert iteration}
\label{app:ei}

SpecGen pass@4 rises at each stage of the SFT data (Table~\ref{tab:eiablation}), from 30.7\% on the LeetCode data to 39.3\% after adding the first round of expert iteration and 47.3\% after further adding the Codeforces problems and 2 more rounds.

\begin{table}[H]
\centering
\captionsetup{font=scriptsize}
\caption{Pass@4 on the in-domain test set after each stage of the SFT data, sampled with the analysis settings of Table~\ref{tab:inference}. Each stage adds data to the previous stage and is fine-tuned from Qwen3.8-27B for 3 epochs.}
\label{tab:eiablation}
\scriptsize
\setlength{\tabcolsep}{3.2pt}
\begin{tabular}{l *{4}{c}}
\toprule
Training data & Problems & Samples & SpecGen & CodeGen \\
\midrule
LeetCode & 2{,}687 & 7{,}993 & 30.7 & 38.0 \\
+ expert iteration round 1 & 2{,}733 & 11{,}811 & 39.3 & 52.0 \\
+ Codeforces, expert iteration rounds 2 and 3 (\work-27B-SFT) & 3{,}993 & 19{,}492 & 47.3 & 48.7 \\
\bottomrule
\end{tabular}
\end{table}

\subsection{Continued pretraining}
\label{app:cpt}

\textbf{Motivation.} An early supervised fine-tuning run of Qwen3.5-9B compiles 5.05\% of the 1,208 SpecGen samples and 3.8\% of all 3,596 samples on the held-out problems. Classifying the first compiler error of the failed samples shows that 40.8\% misuse Mathlib declarations, through application type mismatches (17.3\%), invalid fields or projections (14.7\%) and type mismatches (8.7\%), and 65.2\% once unknown identifiers (11.6\%) and missing instances (12.8\%) are counted. With retrieved declaration names in 58\% of the answers, the failures come from the use of Mathlib.

\textbf{Corpus.} Synthesis augments prompts with Mathlib declarations retrieved for the problem (Appendix~\ref{sec:retrieval}), and across the 3,109 LeetCode problems the retrievals cover 1,381 distinct declarations, of which 1,298 are found in the pinned Mathlib source. These declarations come from 411 files, and we take all declarations of those files rather than the retrieved ones alone, which yields 26,936 declaration heads. The heads are rewritten with DeepSeek-V4-Flash \citep{deepseekv4} into English lessons that state the content, the application, the occasion and the contrast of a declaration, which follows the rewriting of raw corpora into reading-comprehension text \citep{cheng2024adaptllm} and textbook-style synthetic corpora \citep{gunasekar2023textbooks}. Together with 8,312 rows of Mathlib source, 2,336 lessons on decidability routes and 3,790 pairs of compiler error and repair, the corpus $\mathcal{M}$ holds 41,374 rows and about 43M tokens.

\textbf{Training and result.} Adding problems to supervised fine-tuning improves pass@4 more than continued pretraining does (Table~\ref{tab:cpt}). Continued pretraining minimizes the next-token loss on the corpus,
\begin{equation}
\mathcal{L}_{\text{CPT}}(\theta) = -\,\mathbb{E}_{u \sim \mathcal{M}} \sum_{t} \log \pi_\theta(u_t \mid u_{<t}),
\label{eq:cpt}
\end{equation}
for 2 epochs, 92 steps, on Qwen3.5-9B base with learning rate 5e-6 and sequence length 16,384, and supervised fine-tuning follows on 4,826 rows from 1,631 problems. Domain corpora of this kind help on downstream tasks \citep{gururangan2020dont} and formal provers are pretrained on Lean and mathematics before fine-tuning \citep{azerbayev2024llemma,dsprover2024}, yet the benefit of continued pretraining is not universal \citep{shi2023dontstop}. Here continued pretraining improves SpecGen pass@4 from 8.2 to 9.8 and pass@4 over the 4 tasks from 26.4 to 27.4, whereas adding 1,056 problems, 3,167 rows, to supervised fine-tuning improves the same 2 numbers to 13.7 and 30.2. The corpus loss falls from 1.03 to 0.69 with held-out perplexity 2.03. The corpus is therefore fit while the downstream gain stays at 1.0 point, the distance between document perplexity and downstream use reported for knowledge injection through continued pretraining \citep{jiang2024knowledge}. Content memorized from a corpus becomes usable after many representations of the same content \citep{allenzhu2024physics,yang2025synthetic}, and each of the 26,936 declarations appears in 1 lesson here. Lessons state the meaning of declarations, whereas a compiling specification requires declarations composed into terms that the elaborator accepts, and added problems carry test cases and judges that label samples of that composition. Later rounds therefore put the effort into data.

\begin{table}[H]
\providecommand{\gain}[1]{\rlap{\,{\color{green!50!black}\tiny(+#1)}}}
\centering
\captionsetup{font=scriptsize}
\caption{Continued pretraining compared with more fine-tuning data for Qwen3.5-9B, pass@$k$ with 4 samples on an earlier version of the training data (4,826 rows from 1,631 problems). SFT, more data adds 3,167 rows from 1,056 problems. Parentheses show the change from SFT.}
\label{tab:cpt}
\scriptsize
\setlength{\tabcolsep}{12pt}
\begin{tabular}{@{\hspace{6pt}}l *{8}{c}@{\hspace{16pt}}}
\toprule
\multirow{3}{*}{Model} & \multicolumn{6}{c}{\work-Test} & \multicolumn{2}{c}{\verina{} SpecGen} \\
\cmidrule(l{2pt}r{2pt}){2-7}\cmidrule(l{2pt}r{2pt}){8-9}
 & \multicolumn{2}{c}{SpecGen} & \multicolumn{2}{c}{CodeGen} & \multicolumn{2}{c}{All 4 tasks} & \multicolumn{2}{c}{Basic + Advanced} \\
\cmidrule(l{2pt}r{2pt}){2-3}\cmidrule(l{2pt}r{2pt}){4-5}\cmidrule(l{2pt}r{2pt}){6-7}\cmidrule(l{2pt}r{2pt}){8-9}
 & p@1 & p@4 & p@1 & p@4 & p@1 & p@4 & p@1 & p@4 \\
\midrule
Qwen3.5-9B & 0.0 & 0.0 & 0.0 & 0.0 & 2.4 & 7.3 & 2.0 & 7.0 \\
SFT & 2.0 & 8.2 & 4.3 & 15.6 & 13.6 & 26.4 & 16.0 & 35.4 \\
CPT + SFT & 2.3\gain{0.3} & 9.8\gain{1.6} & 5.9\gain{1.6} & 16.0\gain{0.4} & 14.1\gain{0.5} & 27.4\gain{1.0} & 16.0\rlap{\,{\tiny(0.0)}} & 37.6\gain{2.2} \\
SFT, more data & 7.0\gain{5.0} & 13.7\gain{5.5} & 7.0\gain{2.7} & 18.8\gain{3.2} & 18.3\gain{4.7} & 30.2\gain{3.8} & 18.3\gain{2.3} & 43.9\gain{8.5} \\
\bottomrule
\end{tabular}
\end{table}

\subsection{Reinforcement learning}
\label{app:rl}

The final recipe for \work-9B-RL and \work-27B-RL combines the progressive reward of Eq.~(\ref{eq:reward}), the self-repair buffer, the overlong penalty and the mask on repetitive rollouts on the clipped surrogate of DAPO with batch-normalized advantages (Section~\ref{sec:rl}), and adds the end-of-sequence penalty for the 27B model. The runs on the 9B model develop the recipe, and both models train for 1 epoch, 434 steps for the 9B model and 208 steps for the 27B model, with the settings of Table~\ref{tab:rl}. Both runs remain stable through the epoch, with 0.12\% and 0.60\% of rollouts repetitive and 0.13\% and 0.11\% malformed for the 27B and 9B models. The reported models are the final checkpoints, and SpecGen pass@1 on \work-Test still rises over the last evaluated checkpoints, from 54.3\% at step 400 to 58.6\% for the 9B model and from 61.3\% at step 200 to 65.7\% for the 27B model. The subsections below analyze the reward stages, the reward dynamics and pass@$k$.

\subsubsection{Reward stages}
\label{app:rlreward}

The progressive reward keeps most groups without a passing rollout in the gradient. Under a binary reward these groups have equal rewards and are discarded by dynamic sampling \citep{yu2025dapo}. They make up 37\% of SpecGen groups for the 27B model and 47\% for the 9B model, and the progressive reward keeps nonzero variance in 87.9\% and 86.2\% of them (Table~\ref{tab:rlspread}).

\begin{table}[H]
\caption{Generation groups with nonzero reward variance under the progressive reward and under a binary reward that requires all stages, scored on the rollouts of the whole run of each model. A binary reward has zero variance in every group without a passing rollout.}
\label{tab:rlspread}
\centering
\scriptsize
\begin{tabular}{lrrrr}
\toprule
 & \multicolumn{2}{c}{27B} & \multicolumn{2}{c}{9B} \\
\cmidrule(l{2pt}r{2pt}){2-3}\cmidrule(l{2pt}r{2pt}){4-5}
 & SpecGen & CodeGen & SpecGen & CodeGen \\
\midrule
Groups & 8{,}346 & 5{,}384 & 7{,}461 & 5{,}279 \\
Nonzero variance, progressive (\%) & 91.6 & 59.9 & 91.3 & 73.0 \\
Nonzero variance, binary (\%) & 59.4 & 41.9 & 50.9 & 49.3 \\
Groups without a passing rollout & 3{,}048 & 1{,}309 & 3{,}497 & 1{,}743 \\
\quad nonzero variance, progressive (\%) & 87.9 & 74.3 & 86.2 & 71.8 \\
\bottomrule
\end{tabular}
\end{table}

Table~\ref{tab:rlstages} lists the reward and the fraction of rollouts of each outcome. Each rule against reward hacking closes a shortcut that the stages would otherwise reward. A test that calls no declaration of the output, such as \texttt{\#guard true}, passes without testing the output, and counting only tests that call a declaration keeps the test stages meaningful. An escape hatch, the \texttt{sorry} and \texttt{admit} tactics, an \texttt{axiom} declaration or the \texttt{native\_decide} tactic, lets an output compile without the checks that later stages assume. Special-casing the test inputs, excluding a test value from the validity predicate and copying the given test cases into the implementation all pass the test stages without solving the problem and receive 0. Altering a given test case lets a rollout pass the test stages against a different expected output and receives 0.1. A test that cannot run would otherwise earn the test execution credit of a test that ran and failed. An unchanged repair would keep the credit of the stages the failed input already passes. The non-tautology stage precedes the overall quality stage, the order in which evaluation runs the judges.

The partial credit separates accepted outputs by form. The non-tautology judge classifies a specification as declarative, enumerative or tautological. A declarative specification states the condition of the problem and is sound and complete: every answer that satisfies the specification is correct, and every correct answer satisfies the specification. An enumerative specification states the same condition over a candidate list that the specification constructs, such as all sublists of the input, instead of quantifying over the candidate type. A tautological specification replaces the condition with an algorithm for the answer, which is the role of the implementation. The implementation judge marks a correct CodeGen implementation as slow when the implementation takes more than about $10^7$ steps on the largest admitted input and a method faster by more than a logarithmic factor exists, known as a slow positive \citep{li2022alphacode}. Appendix~\ref{app:evaljudges} lists both prompts.

\begin{table}[H]
\caption{Reward and fraction of rollouts (\%) for each outcome of the progressive reward in SpecGen (left) and CodeGen (right), each with the repair task, over the whole run of each model. $^{1}$Only in the repair tasks, which otherwise share the outcomes and values, with the reward of a repair rollout scaled down from 1 in the first round to 0.8 in the last.}
\label{tab:rlstages}
\centering
\scriptsize
\setlength{\tabcolsep}{2.5pt}
\begin{tabular}{@{}lrrr|lrrr@{}}
\toprule
\multicolumn{4}{c|}{SpecGen} & \multicolumn{4}{c}{CodeGen} \\
Outcome & Reward & 27B & 9B & Outcome & Reward & 27B & 9B \\
\midrule
Malformed format or no code block & 0 & 0.30 & 1.70 & Malformed format or no code block & 0 & 0.22 & 1.17 \\
Special-cased or copied test cases & 0 & 0.17 & 0.08 & Special-cased or copied test cases & 0 & 0.66 & 1.41 \\
Unchanged repair$^{1}$ & 0 & 0.36 & 0.67 & Unchanged repair$^{1}$ & 0 & 0.73 & 0.59 \\
Fewer tests than required & 0.1 & 0.61 & 0.52 & Fewer tests than required & 0.1 & 0.19 & 0.18 \\
Fails compilation & 0.1 & 29.16 & 38.70 & Fails compilation & 0.1 & 12.42 & 17.37 \\
Fails test case alignment & 0.1 & 0.96 & 0.68 & Fails test case alignment & 0.1 & 0.21 & 0.35 \\
Fails test execution & 0.2 & 17.01 & 18.12 & Fails test execution & 0.2 & 10.72 & 12.04 \\
Fails non-tautology & 0.1 & 7.67 & 8.46 & No verdict from a judge after retries & 0.3 & 0.00 & 0.06 \\
No verdict from a judge after retries & 0.3 & 0.00 & 0.02 & Fails overall quality & 0.4 & 0.31 & 0.57 \\
Fails overall quality & 0.4 & 17.18 & 10.95 & Correct but slow implementation & 0.5 & 21.94 & 37.67 \\
Enumerative specification & 0.4 & 1.89 & 2.14 & Passes all stages & 1.0 & 52.58 & 28.58 \\
Passes all stages & 1.0 & 24.68 & 17.96 &  &  &  &  \\
\midrule
Rollouts &  & 251{,}232 & 224{,}528 & Rollouts &  & 112{,}000 & 111{,}792 \\
\bottomrule
\end{tabular}
\end{table}

As training proceeds, compilation failures fall from 49.0\% of rollouts in the first 40 steps to 15.5\% in the last 40 steps for the 27B model and from 66.0\% to 10.7\% for the 9B model, while passing rollouts rise from 18.4\% to 45.0\% and from 9.9\% to 38.5\% (\subfig{fig:rltrain}{c} and~\panel{fig:rltrain}{d}).

\subsubsection{Reward dynamics}
\label{app:rldynamics}

\begin{figure}[H]
\centering
\includegraphics[width=\linewidth]{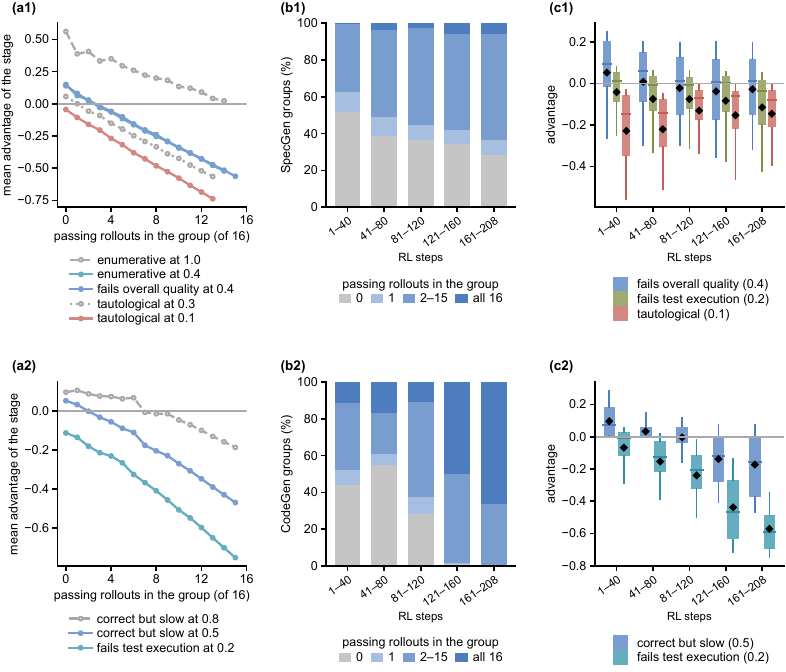}
\caption{Advantage of stages in the 27B run, SpecGen in the top row (a1 to c1) and CodeGen in the bottom row (a2 to c2). (a) Mean advantage against the number of passing rollouts in a group of 16, replayed on the same rollouts with the value of one stage changed. (b) Groups by the number of passing rollouts in each period of training. (c) Advantage of each stage per period under the final values, with boxes at the 25th and 75th percentiles, whiskers at the 10th and 90th percentiles and diamonds at the mean.}
\label{fig:rldyn}
\end{figure}

We analyze the group-relative advantage \citep{shao2024deepseekmath} before batch normalization. Replaying the training rollouts with the value of one outcome changed shows how the value sets this advantage (\subfig{fig:rldyn}{a1} and~\panel{fig:rldyn}{a2}). Failures that attempt to score unfaithfully are shortcuts rather than progress. For instance, tautological specifications restate algorithms instead of specifying problems, and failed test case alignments usually rewrite the given test cases. Such unfaithful attempts should be discouraged in every group. Higher values, though well below the full reward of 1, still assign these failures positive advantages in groups without correct rollouts, whereas replaying at 0.1 shows advantages that are negative and below those of faithful attempts in almost every group (Table~\ref{tab:shortcut}).

\begin{table}[H]
\caption{Advantage of shortcut failures at the final value and at an alternative higher value, replayed on the training rollouts. For failed test case alignments, the advantage is given as the fraction of SpecGen-Repair rollouts with a positive advantage.}
\label{tab:shortcut}
\centering
\scriptsize
\begin{tabular}{lclcl}
\toprule
Shortcut Failure & \begin{tabular}[c]{@{}c@{}}Final\\Value\end{tabular} & Advantage & \begin{tabular}[c]{@{}c@{}}Alternative\\Value\end{tabular} & Advantage \\
\midrule
Tautological specification & 0.1 & Negative in every group & 0.3 & Positive in groups without a full-score rollout \\
Failed test case alignment & 0.1 & Positive for 0.8\% of rollouts & 0.2 & Positive for 58.3\% of rollouts \\
\bottomrule
\end{tabular}
\end{table}

Based on observation, some attempts make partial progress that is correct in substance but not yet in the perfect form. For instance, 1) enumerative specifications already state the correct condition and only need to be rewritten declaratively, and 2) slow implementations are already correct and only need faster algorithms. Binary judges also classify enumerative specifications poorly. An intermediate reward for partial progress refines sparse signals into fine-grained rewards. When rewards are low, like SpecGen in the early steps, partial progress rewards small improvements of failing attempts, while once correct rollouts are common, declarative specifications and fast implementations are assigned higher advantages.

By replaying the rollouts, we fix intermediate values at which partial progress receives positive advantages in groups with few correct rollouts and negative ones in groups with many. A slow implementation at 0.5 turns negative from 2 passing rollouts of 16, compared with 6 at 0.8, and an enumerative specification at 0.4 turns negative from 3, compared with 14 when valued like a full-score answer. As the policy improves, the advantage of partial progress turns from positive to negative (\subfig{fig:rldyn}{b2} and~\panel{fig:rldyn}{c2}). CodeGen groups without a passing rollout fall from 44\% in steps 1 to 40 to 0\% in steps 161 to 208, and the mean advantage of slow implementations falls from $+0.10$ to $-0.17$ over the same periods. SpecGen keeps 29\% to 52\% of groups without a passing rollout in each period, where quality rejections keep a mean advantage near 0 with both signs (\subfig{fig:rldyn}{c1}). Partial progress thus provides intermediate supervision that keeps the reward informative throughout training (Table~\ref{tab:partial}).

\begin{table}[H]
\caption{Partial progress during RL. The advantage of partial progress turns negative from the listed fraction, which refers to full-score rollouts in a whole group of 16, replayed on the 27B rollouts. The effect of slow implementations comes from the 27B run, with fractions over 20-step windows of CodeGen rollouts. A repair round from an enumerative specification succeeds when the repaired output is a declarative specification that passes all stages, counted over all repair rounds started from enumerative specifications (336 for the 27B model and 480 for the 9B model).}
\label{tab:partial}
\centering
\scriptsize
\renewcommand{\tabularxcolumn}[1]{m{#1}}
\begin{tabularx}{\linewidth}{lcc>{\arraybackslash}X}
\toprule
Partial Progress & Reward & \begin{tabular}[c]{@{}c@{}}Negative\\Advantage From\end{tabular} & Effect \\
\midrule
Enumerative specification & 0.4 & 3/16 & Repair rounds turn 74.1\% (27B) and 54.6\% (9B) of enumerative specifications into declarative ones that pass all stages. \\
\addlinespace
Slow implementation & 0.5 & 2/16 & Slow implementations rise from 14.5\% of correct CodeGen rollouts in steps 1 to 20 to 67.4\% in steps 81 to 100 and fall to 0.4\% by the end of the 27B run. \\
\bottomrule
\end{tabularx}
\end{table}

Whether a partial value acts as credit depends heavily on the capability of the policy to sample better rollouts in the same group. For the 9B model, 71.7\% of correct but slow CodeGen rollouts fall in groups without a fast correct rollout, where the slow value keeps a positive advantage, compared with 48.4\% for the 27B model, and the 9B run does not start repair rounds from slow rollouts. CodeGen pass@1 of \work-9B-RL accordingly drops from 77.1\% to 44.7\% when slow programs count as failures, compared with 94.8\% to 88.0\% for \work-27B-RL.

\subsubsection{Pass@$k$}
\label{app:passk}

A key question for RL with verifiable rewards is whether it expands the set of problems a model can solve or only raises the probability of solutions that the starting model already samples. Pass@$k$ at large $k$ separates the two, since a starting model that catches up at large $k$ solves the same problems with more samples \citep{yue2025rlreasoning}. Prolonged RL has been reported to expand the set most where the starting model is weak \citep{liu2025prorl}.

We compare \work-27B-RL with \work-27B-SFT, the starting model of RL, at temperature 0.6 and top-$p$ 0.95. Pass@$k$ follows the unbiased estimator \citep{chen2021codex} on 64 samples per problem for the SFT model and 32 for the RL model. Pass@64, pass@128 and pass@256 count a problem as solved when any of 64, 128 or 256 samples passes, and additional samples are drawn only for problems not yet solved.

\begin{figure}[H]
\centering
\includegraphics[width=0.6\linewidth]{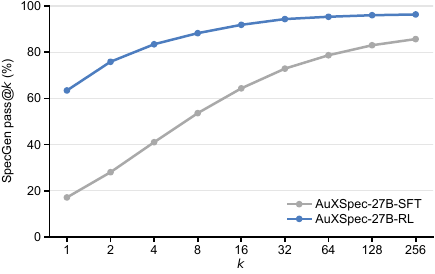}
\caption{SpecGen pass@$k$ on \work-Test of \work-27B-SFT and \work-27B-RL.}
\label{fig:passk}
\end{figure}

The RL model stays above the SFT model at every $k$ and reaches 96.3\%, compared with 85.7\% at $k = 256$, where the SFT model remains below the pass@8 of the RL model (88.2\%, Figure~\ref{fig:passk}). The SFT model still gains at $k = 256$, while the gain per doubling of $k$ falls from 5.8 points at $k = 64$ to 4.3 and 2.7 points at $k = 128$ and $k = 256$. The gap narrows from 13.0 to 10.6 points between $k = 128$ and $k = 256$ as the RL model approaches the ceiling, with 11 unsolved problems of which 10 are solved by neither model.

As a result, 33 problems are solved only by the RL model and 1 only by the SFT model within 256 samples (Table~\ref{tab:passkcover}).

\begin{table}[H]
\caption{SpecGen problems of \work-Test solved within 256 samples by \work-27B-SFT and \work-27B-RL.}
\label{tab:passkcover}
\centering
\scriptsize
\begin{tabular}{lrr}
\toprule
 & Problems & Fraction (\%) \\
\midrule
Solved by both models & 256 & 85.3 \\
Solved only by \work-27B-SFT & 1 & 0.3 \\
Solved only by \work-27B-RL & 33 & 11.0 \\
Solved by neither model & 10 & 3.3 \\
\bottomrule
\end{tabular}
\end{table}

Appendix~\ref{app:opd} reports RL with a distillation term from \work-27B-SFT.

\subsection{On-policy distillation}
\label{app:opd}

To transfer the capability of \work-27B-SFT, the strongest of our models on \work-Test, to the smaller \work-9B-SFT, we apply on-policy distillation (OPD) and OPD within reinforcement learning \citep{agarwal2024gkd,kdrl}, but neither improves the student model. OPD, with the k3 estimator in the advantage, suffers from mode collapse, where the student model concentrates on the decidable form of the executable postcondition, leading to a compile rate of 2.0\% without compiler feedback. SpecGen pass@4 drops from 24.3 to 5.3 while CodeGen changes by at most 3 points. Distillation within reinforcement learning suffers from teacher redundancy, where a teacher model fine-tuned on the same data carries 4\% to 7\% of the advantage with little information about which samples succeed. We next describe the setup, the failure patterns and the attribution.

\textbf{Setup.} On-policy distillation trains student models on self-generated responses with token-level supervision from teacher models \citep{agarwal2024gkd}. We select \work-27B-SFT as the teacher model $\pi_T$, the strongest of our models on \work-Test (Table~\ref{tab:mainpass}), and \work-9B-SFT as the student model $\pi_\theta$, which has the same vocabulary and chat template as the teacher model for a locally served teacher model to score student tokens. For a prompt $x$ and a response $y \sim \pi_\theta(\cdot \mid x)$, the per-token log-ratio
\begin{equation}
\ell_t = \log \pi_\theta(y_t \mid x, y_{<t}) - \log \pi_T(y_t \mid x, y_{<t})
\label{eq:logratio}
\end{equation}
is the k1 estimate of the reverse KL divergence $\mathrm{KL}(\pi_\theta \,\|\, \pi_T)$, and $k_3(t) = e^{-\ell_t} - 1 + \ell_t$ is a nonnegative estimate of the same divergence \citep{schulman2020kl}. Both settings enter the policy gradient through the per-token advantage $A_t$.
\begin{itemize}[leftmargin=*, itemsep=0pt, topsep=2pt, parsep=0pt]
\item \textbf{OPD}: the per-token advantage is $A_t = -k_3(t)$ with no task reward, as in on-policy distillation \citep{agarwal2024gkd,thinkingmachines2025opd}, over 200 steps on 12,791 prompts (1 epoch), with student responses and teacher log-probabilities at temperature 1.0.
\item \textbf{OPD within RL}: the objective adds a distillation term with per-token advantage $-\ell_t$, weighted by $\beta = 0.1$, to the reinforcement learning objective of Appendix~\ref{app:rl}, as in joint distillation and reinforcement learning \citep{kdrl} with the k1 estimator in place of k2. Variants keep the distillation term only where the sign agrees with the task advantage \citep{opdvr2026} or drop the term and reweight the task advantage by clipped teacher ratios \citep{distilledrl2026}. Student responses are sampled at temperature 0.7 and teacher log-probabilities at 1.0.
\end{itemize}

\textbf{Diagnostics.} Besides pass@$k$, we track token entropy, the fraction of generated text inside repetition loops, the fraction of SpecGen samples in the decidable form of Section~\ref{sec:pipeline}, and teacher informativeness, the mean of $-\ell_t$ on sequences with positive task advantage minus the same mean on sequences with negative advantage \citep{demystifyingopd2026}.

\textbf{Results.} OPD degrades SpecGen and changes CodeGen by at most 3 points (Table~\ref{tab:opd}). SpecGen pass@4 drops from 24.3 to 5.3 on \work-Test and from 42.0 to 18.5 on \verina{} advanced problems, and CodeGen pass@4 from 30.7 to 27.7. In OPD within RL, teacher informativeness stays at +0.004 or below and the teacher term carries 4\% to 7\% of the total advantage over the first steps. The 3 runs ended at steps 4 to 9 when the judge service ran out of budget, and the measured informativeness led us to stop this line rather than resume.

\begin{table}[htbp]
\providecommand{\gain}[1]{\rlap{\,{\color{green!50!black}\tiny(+#1)}}}
\providecommand{\drop}[1]{\rlap{\,{\color{red!70!black}\tiny(\textminus#1)}}}
\centering
\captionsetup{font=scriptsize}
\caption{On-policy distillation of \work-9B-SFT from \work-27B-SFT, pass@$k$ with 4 samples. Parentheses show the change from \work-9B-SFT, and the teacher row is a reference. \textsuperscript{*}An earlier checkpoint of \work-27B-SFT, used as the teacher.}
\label{tab:opd}
\scriptsize
\setlength{\tabcolsep}{10.5pt}
\begin{tabular}{@{\hspace{6pt}}l *{8}{c}@{\hspace{22pt}}}
\toprule
\multirow{3}{*}{Model} & \multicolumn{4}{c}{\work-Test} & \multicolumn{4}{c}{\verina{} SpecGen} \\
\cmidrule(l{2pt}r{2pt}){2-5}\cmidrule(l{2pt}r{2pt}){6-9}
 & \multicolumn{2}{c}{SpecGen} & \multicolumn{2}{c}{CodeGen} & \multicolumn{2}{c}{All} & \multicolumn{2}{c}{Advanced} \\
\cmidrule(l{2pt}r{2pt}){2-3}\cmidrule(l{2pt}r{2pt}){4-5}\cmidrule(l{2pt}r{2pt}){6-7}\cmidrule(l{2pt}r{2pt}){8-9}
 & p@1 & p@4 & p@1 & p@4 & p@1 & p@4 & p@1 & p@4 \\
\midrule
\work-9B-SFT & 9.6 & 24.3 & 13.5 & 30.7 & 30.3 & 61.4 & 17.6 & 42.0 \\
\quad + OPD & 2.3\drop{7.3} & 5.3\drop{19.0} & 15.2\gain{1.7} & 27.7\drop{3.0} & 29.1\drop{1.2} & 43.9\drop{17.5} & 8.3\drop{9.3} & 18.5\drop{23.5} \\
\midrule
\work-27B-SFT\textsuperscript{*} (teacher) & 20.4 & 47.0 & 23.1 & 44.3 & 49.1 & 82.5 & 41.4 & 79.0 \\
\bottomrule
\end{tabular}
\end{table}

\begin{figure}[H]
\centering
\captionsetup{font=footnotesize,skip=3pt}
\includegraphics[width=\linewidth]{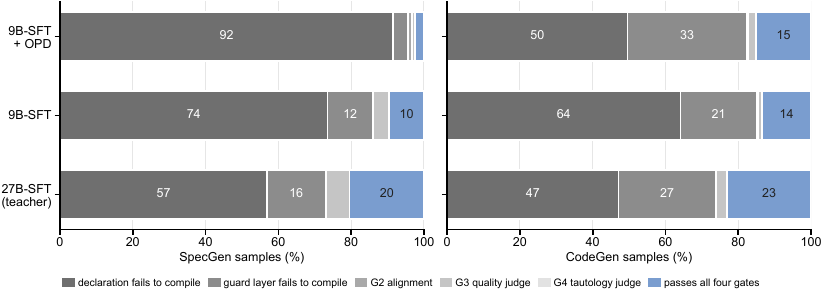}
\caption{Distribution of SpecGen and CodeGen samples by failure stage, 1,200 samples per model and task. After OPD, 92\% of SpecGen samples stop at a declaration that does not compile, while CodeGen samples shift from declaration errors to guard-layer errors with an unchanged pass rate.}
\label{fig:opd_funnel}
\end{figure}

\textbf{Phenomena.} After OPD, 1,147 of 1,200 SpecGen samples (95.6\%) fail at compilation, compared with 1,032 (86.0\%) before OPD and 875 (72.9\%) for the teacher (Figure~\ref{fig:opd_funnel}), and OPD also narrows the sample distribution and lets repetition loops grow unpenalized.

\begin{figure}[H]
\centering
\captionsetup{font=footnotesize,skip=3pt}
\includegraphics[width=0.8\linewidth]{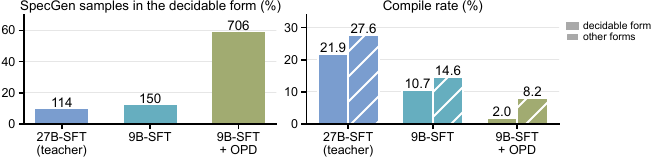}
\caption{Use of the decidable form in SpecGen samples, with counts out of 1,200 above the bars (left), and compile rates of the decidable form against other forms (right).}
\label{fig:opd_template}
\end{figure}

\emph{Template concentration.} The number of SpecGen samples stating the executable postcondition in the decidable form increases from 150 (12.5\%) before OPD to 706 of 1,200 (58.8\%) after OPD, compared with 114 (9.5\%) for the teacher (Figure~\ref{fig:opd_template}). Within this form, the compile rate falls from 10.7\% to 2.0\%, compared with 21.9\% for the teacher, and failed samples report a missing decidability instance.

\begin{figure}[H]
\centering
\captionsetup{font=footnotesize,skip=3pt}
\includegraphics[width=0.8\linewidth]{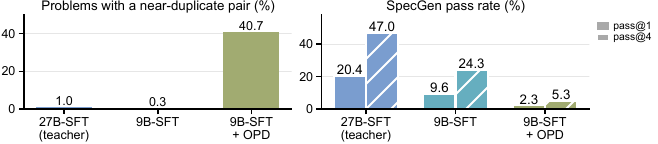}
\caption{Near-duplicate SpecGen samples per problem (left) and SpecGen pass@1 against pass@4 (right).}
\label{fig:opd_diversity}
\end{figure}

\begin{figure}[H]
\centering
\captionsetup{font=footnotesize,skip=3pt}
\includegraphics[width=\linewidth]{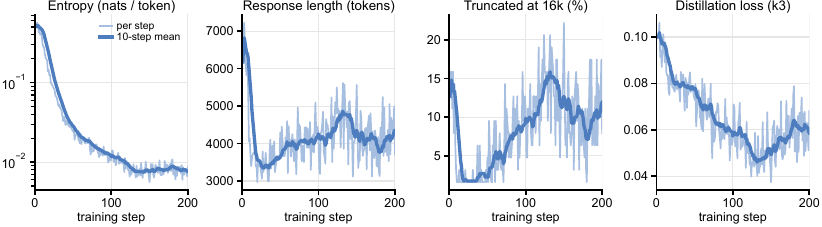}
\caption{Training dynamics of OPD: token entropy, mean response length, fraction of responses truncated at 16,384 tokens, and distillation loss. Light lines show single steps and dark lines a moving average over 10 steps.}
\label{fig:opd_dynamics}
\end{figure}

\emph{Diversity collapse.} Token entropy falls from 0.451 to 0.012 ($-$97.3\%) within the first 100 steps (Figure~\ref{fig:opd_dynamics}), the entropy collapse also reported for reinforcement learning \citep{yu2025dapo,he2025skyworkor1}. Near-duplicate sample pairs occur on 40.7\% of SpecGen problems, compared with 0.3\% before OPD (Figure~\ref{fig:opd_diversity}). On \verina{} basic problems, pass@1 rises from 39.8 to 44.7 while pass@4 falls from 75.9 to 63.0, the pattern of diversity collapse in which gains at small $k$ are paired with losses at larger $k$ \citep{dang2025diversity,yue2025rlreasoning}.

\begin{figure}[H]
\centering
\captionsetup{font=footnotesize,skip=3pt}
\includegraphics[width=\linewidth]{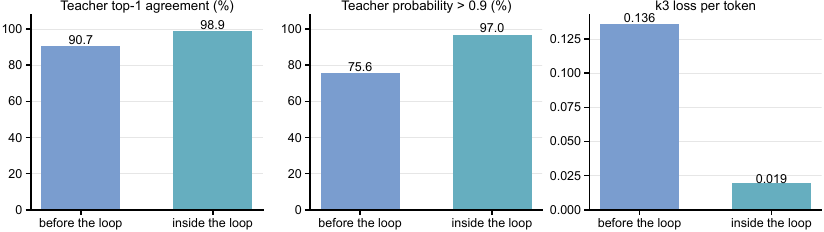}
\caption{Degenerate agreement inside repetition loops: teacher top-1 agreement, fraction of tokens with teacher probability above 0.9, and k3 loss per token, on 26 looping SpecGen outputs.}
\label{fig:opd_agreement}
\end{figure}

\emph{Degenerate agreement.} Repetition loops \citep{holtzman2020degeneration} grow during training, and from step 51 on, looping responses hold 43\% to 51\% of the generated text. Inside a loop, the teacher ranks the next student token first on 98.9\% of tokens, compared with 90.7\% before the loop, and the k3 loss per token therefore drops from 0.136 to 0.019 ($-$86.0\%, Figure~\ref{fig:opd_agreement}). The same degenerate agreement is reported for on-policy distillation \citep{tide2026}. Loops account for 1 of the 60 SpecGen problems lost to OPD, where all 4 samples are truncated or loop.

\textbf{Attribution.} We attribute the SpecGen drop to estimator bias, which lowers entropy throughout training, to mode collapse, which moves probability onto a form that often fails to compile, and to teacher redundancy, which makes the teacher term uninformative about sample success.
\begin{enumerate}[leftmargin=*, itemsep=0pt, topsep=2pt, parsep=0pt]
\item \emph{Estimator bias.} With $k_3$ inside the advantage, the policy gradient is a biased estimate of the reverse KL gradient that also raises the forward KL, and this configuration is reported to collapse training \citep{shah2025comedy}. The entropy decline of Figure~\ref{fig:opd_dynamics} matches this bias.
\item \emph{Mode collapse.} The pipeline sets the decidable form as the first executable postcondition and replaces the form through instance synthesis or a separate definition when compilation fails (Section~\ref{sec:pipeline}), and the form covers 35.5\% of synthesized problems as a result. Evaluation scores single samples without compiler feedback. We conjecture that the mode-seeking reverse KL \citep{agarwal2024gkd,jin2026eopd} moves probability onto this frequent form, including on problems where the form does not compile. The student adopts the form in 54.6\% of samples on problems where the teacher never adopts it. OPD therefore amplifies the form beyond teacher usage.
\item \emph{Teacher redundancy.} Teacher models with higher scores help only when offering capabilities beyond the experience of the student \citep{rethinkingopd2026}. Teacher and student are fine-tuned on the same data, and although OPD within RL adds the fine-grained reward of reinforcement learning, the teacher term stays uninformative about which samples succeed.
\end{enumerate}
Identical tokenizers and chat templates rule out a format mismatch, and on problems the teacher fails, the student stops at the failing stage of the teacher no more often after OPD than before.

\section{Ablation study}
\label{app:ablation}
\begin{revised}

\subsection{Shared backbone with VeriSpecGen}
\label{app:shared-backbone}

To compare with VeriSpecGen on a shared backbone, we fix the backbone to Qwen3-4B-Instruct-2507, the backbone of VeriSpecGen-4B, and keep the rest of our pipeline: the same SFT data and configuration (19{,}492 samples, 3 epochs), the RL recipe of \work-27B (Section~\ref{sec:rl}; 64 problems and 16 samples per step), and the same evaluation protocol (Section~\ref{sec:ood}), with the official \verina{} evaluator at temperature 0.3.

With our prompts and acceptance criteria, the backbone before training scores 0.0\% SpecGen pass@1 on both \work-Test and \verina{} (Tables~\ref{tab:shared-backbone-indomain} and~\ref{tab:shared-backbone}). None of its 1{,}200 SpecGen samples on \work-Test compiles: the outputs either reach the generation length limit or contain Lean errors such as Lean~3 syntax. VeriSpecGen reports 14.7\% for the same backbone with its own prompt and evaluation harness, so the 2 base rows are not directly comparable.

SFT on our data improves generation only slightly (6.9\% SpecGen pass@1 on \verina{} without test cases, close to 9.5\% for VeriSpecGen-4B SFT Spec-Only), while the repair tasks improve the most (from 6.5\% to 49.0\% SpecGen-Repair pass@1 on \work-Test).

\begin{table}[H]
\centering
\captionsetup{font=scriptsize}
\caption{In-domain evaluation of the shared-backbone model on \work-Test, with the protocol of Table~\ref{tab:mainpass}. The model starts from Qwen3-4B-Instruct-2507, the backbone of VeriSpecGen-4B, and is trained with our SFT data and RL recipe. Base Model indicates Qwen3-4B-Instruct-2507 before training, evaluated with our prompts. VeriSpecGen does not release its models, so the table has no VeriSpecGen rows. Bold indicates the best value of each column.}
\label{tab:shared-backbone-indomain}
\scriptsize
\setlength{\tabcolsep}{5.5pt}
\begin{tabular}{@{\hspace{6pt}}l *{8}{c}@{\hspace{6pt}}}
\toprule
\multicolumn{1}{c}{\multirow{2}{*}{Checkpoint}} & \multicolumn{2}{c}{SpecGen} & \multicolumn{2}{c}{CodeGen} & \multicolumn{2}{c}{SpecGen-Repair} & \multicolumn{2}{c}{CodeGen-Repair} \\
\cmidrule(l{2pt}r{2pt}){2-3}\cmidrule(l{2pt}r{2pt}){4-5}\cmidrule(l{2pt}r{2pt}){6-7}\cmidrule(l{2pt}r{2pt}){8-9}
 & pass@1 & pass@4 & pass@1 & pass@4 & pass@1 & pass@4 & pass@1 & pass@4 \\
\specialrule{\lightrulewidth}{\aboverulesep}{2.5pt}
\multicolumn{9}{@{\hspace{6pt}}l}{\textit{Ours, Qwen3-4B-Instruct-2507}} \\
\specialrule{\lightrulewidth}{2pt}{\belowrulesep}
Base Model & 0.0 & 0.0 & 2.8 & 4.0 & 6.5 & 8.4 & 4.1 & 6.8 \\
SFT & 1.2 & 4.0 & 7.2 & 15.3 & 49.0 & 72.5 & 48.5 & 66.0 \\
RL step 20 & 3.8 & 11.3 & 15.2 & 34.3 & 52.1 & 69.5 & 51.2 & 72.8 \\
RL step 40 & 6.8 & 17.7 & 23.8 & 45.7 & 54.6 & 72.5 & 54.6 & 74.8 \\
RL step 60 & 9.3 & 23.7 & 34.7 & 65.0 & \textbf{56.1} & \textbf{77.9} & 54.1 & 76.7 \\
RL step 80 & 13.7 & 32.3 & 42.8 & 72.0 & 54.2 & 75.6 & 56.8 & 81.6 \\
RL step 100 & 17.9 & 38.3 & 51.3 & 75.7 & 46.6 & 71.0 & 56.1 & 81.6 \\
RL step 120 & \textbf{24.0} & \textbf{43.7} & \textbf{57.8} & \textbf{81.7} & 45.6 & 76.3 & \textbf{59.0} & \textbf{84.5} \\
\bottomrule
\end{tabular}
\end{table}

RL accounts for most of the gain. On \verina{} without test cases, SpecGen pass@1 improves by about 6 points every 20 steps, from 23.0\% at step 20 to 54.2\% at step 120, and still improves by 3.5 points between steps 100 and 120 (Table~\ref{tab:shared-backbone}). From step 40, all 4 scores surpass every VeriSpecGen-4B model, and from step 100 every VeriSpecGen model including the 30B-A3B models (54.2\% and 77.2\% SpecGen and CodeGen pass@1 at step 120, compared with at most 47.6\% and 61.2\%), where our pass@4 is compared with their pass@10. On \work-Test, SpecGen pass@1 improves from 1.2\% after SFT to 24.0\% at step 120 and CodeGen pass@1 from 7.2\% to 57.8\%, and both still rise (Table~\ref{tab:shared-backbone-indomain}). Our SFT data contain 19{,}492 samples, compared with 121{,}531 (No-Test) or 343{,}827 (Full) examples for the VeriSpecGen models.

\begin{table}[H]
\centering
\captionsetup{font=scriptsize}
\caption{Out-of-domain evaluation on \verina{} with a shared backbone. The setting without test cases follows the default of \verina{}. We also report the setting with test cases in the prompt, since concrete input-output pairs specify behavior that the description may state ambiguously (Section~\ref{sec:ood}). Our model starts from Qwen3-4B-Instruct-2507, the backbone of VeriSpecGen-4B, and is trained with our SFT data and RL recipe; Base Model indicates the backbone before training, evaluated with our prompts. \textsuperscript{\dag} indicates results reported in the original paper \citep{verispecgen2026} without test cases in the prompt, and italic values are pass@10 in place of pass@4. VeriSpecGen does not release its models, so its results with test cases are not available (--). Among our values without test cases, \underline{underline} indicates values above every VeriSpecGen-4B model but not above every VeriSpecGen-30B-A3B model, and \textbf{bold} indicates values above every VeriSpecGen model, including VeriSpecGen-30B-A3B. Our pass@4 is compared with their pass@10.}
\label{tab:shared-backbone}
\scriptsize
\setlength{\tabcolsep}{5.5pt}
\begin{tabular}{@{\hspace{6pt}}l *{8}{c}@{\hspace{6pt}}}
\toprule
\multicolumn{1}{c}{\multirow{3}{*}{Model}} & \multicolumn{4}{c}{w/o tests} & \multicolumn{4}{c}{w/ tests} \\
\cmidrule(l{2pt}r{2pt}){2-5}\cmidrule(l{2pt}r{2pt}){6-9}
 & \multicolumn{2}{c}{SpecGen} & \multicolumn{2}{c}{CodeGen} & \multicolumn{2}{c}{SpecGen} & \multicolumn{2}{c}{CodeGen} \\
\cmidrule(l{2pt}r{2pt}){2-3}\cmidrule(l{2pt}r{2pt}){4-5}\cmidrule(l{2pt}r{2pt}){6-7}\cmidrule(l{2pt}r{2pt}){8-9}
 & pass@1 & pass@4 & pass@1 & pass@4 & pass@1 & pass@4 & pass@1 & pass@4 \\
\specialrule{\lightrulewidth}{\aboverulesep}{2.5pt}
\multicolumn{9}{@{\hspace{6pt}}l}{\textit{Ours, Qwen3-4B-Instruct-2507}} \\
\specialrule{\lightrulewidth}{2pt}{\belowrulesep}
Base Model & 0.0 & 0.0 & 0.4 & 0.5 & 0.0 & 0.0 & 0.0 & 0.0 \\
SFT & 6.9 & 24.9 & 21.6 & 54.5 & 12.4 & 35.4 & 21.8 & 53.4 \\
RL step 20 & 23.0 & \underline{54.5} & \underline{50.3} & \underline{79.4} & 25.3 & 57.7 & 51.3 & 76.2 \\
RL step 40 & \underline{32.9} & \underline{59.8} & \underline{57.9} & \textbf{85.2} & 42.1 & 68.3 & 61.9 & 86.8 \\
RL step 60 & \underline{39.3} & \textbf{68.8} & \textbf{64.8} & \textbf{86.8} & 48.8 & 79.4 & 70.6 & 92.1 \\
RL step 80 & \underline{46.6} & \textbf{75.7} & \textbf{67.3} & \textbf{91.5} & 59.8 & 84.1 & 73.7 & 92.6 \\
RL step 100 & \textbf{50.7} & \textbf{78.8} & \textbf{74.3} & \textbf{92.6} & 66.7 & 87.3 & 81.5 & 94.7 \\
RL step 120 & \textbf{54.2} & \textbf{84.1} & \textbf{77.2} & \textbf{95.2} & 73.8 & 89.9 & 84.0 & 96.8 \\
\specialrule{\lightrulewidth}{\aboverulesep}{2.5pt}
\multicolumn{9}{@{\hspace{6pt}}l}{\textit{VeriSpecGen-4B, Qwen3-4B-Instruct-2507}} \\
\specialrule{\lightrulewidth}{2pt}{\belowrulesep}
Base Model & 14.7\textsuperscript{\dag} & \textit{16.3}\textsuperscript{\dag} & 25.3\textsuperscript{\dag} & \textit{28.3}\textsuperscript{\dag} & -- & -- & -- & -- \\
SFT Spec-Only & 9.5\textsuperscript{\dag} & \textit{16.6}\textsuperscript{\dag} & 18.0\textsuperscript{\dag} & \textit{26.7}\textsuperscript{\dag} & -- & -- & -- & -- \\
SFT Full & 22.8\textsuperscript{\dag} & \textit{38.8}\textsuperscript{\dag} & 43.5\textsuperscript{\dag} & \textit{72.7}\textsuperscript{\dag} & -- & -- & -- & -- \\
SFT No-Test & 30.3\textsuperscript{\dag} & \textit{45.2}\textsuperscript{\dag} & 35.4\textsuperscript{\dag} & \textit{66.8}\textsuperscript{\dag} & -- & -- & -- & -- \\
\specialrule{\lightrulewidth}{\aboverulesep}{2.5pt}
\multicolumn{9}{@{\hspace{6pt}}l}{\textit{VeriSpecGen-30B-A3B, Qwen3-Coder-30B-A3B}} \\
\specialrule{\lightrulewidth}{2pt}{\belowrulesep}
Base Model & 29.4\textsuperscript{\dag} & \textit{41.7}\textsuperscript{\dag} & 39.7\textsuperscript{\dag} & \textit{53.5}\textsuperscript{\dag} & -- & -- & -- & -- \\
SFT Spec-Only & 30.4\textsuperscript{\dag} & \textit{50.0}\textsuperscript{\dag} & 41.5\textsuperscript{\dag} & \textit{68.4}\textsuperscript{\dag} & -- & -- & -- & -- \\
SFT Full & 43.5\textsuperscript{\dag} & \textit{58.0}\textsuperscript{\dag} & 61.2\textsuperscript{\dag} & \textit{79.7}\textsuperscript{\dag} & -- & -- & -- & -- \\
SFT No-Test & 47.6\textsuperscript{\dag} & \textit{63.9}\textsuperscript{\dag} & 58.7\textsuperscript{\dag} & \textit{84.0}\textsuperscript{\dag} & -- & -- & -- & -- \\
\bottomrule
\end{tabular}
\end{table}

\subsection{RL recipe ablation}
\label{app:ablation-components}

To isolate the contribution of each component, we construct 3 variants of the RL recipe of \work-9B, each removing one component, and train them from \work-9B-SFT with the same training prompts, order and seed:
\begin{itemize}[leftmargin=*, itemsep=0pt, topsep=2pt, parsep=0pt]
\item \textbf{w/o test execution} compiles each rollout without its test cases, so no test case is executed, and keeps all judge stages.
\item \textbf{w/o LLM judges} assigns the top reward to every rollout that passes the test cases.
\item \textbf{w/o self-repair} fills every batch slot with a new training prompt.
\end{itemize}
The variant without self-repair therefore consumes training prompts faster, so each variant is compared with \work-9B-RL after approximately the same number of training prompts rather than at the same step (Table~\ref{tab:ablation-indomain}). Evaluation follows Table~\ref{tab:mainpass}.

\begin{table}[H]
\centering
\captionsetup{font=scriptsize}
\caption{Component ablations of the 9B recipe on \work-Test, with the protocol of Table~\ref{tab:mainpass}. Each ablation run starts from \work-9B-SFT and removes one component, and \work-9B-RL is the reference run. Training prompts indicates the number of problems drawn from the RL pool up to each checkpoint. Checkpoints are evaluated every 25 steps up to step 150. Bold indicates the best value of each column and underline the second best.}
\label{tab:ablation-indomain}
\scriptsize
\setlength{\tabcolsep}{4pt}
\begin{tabular}{@{}l r r *{8}{c}@{}}
\toprule
\multirow{2}{*}{Model} & \multirow{2}{*}{Step} & \multirow{2}{*}{\shortstack{Training\\prompts}} & \multicolumn{2}{c}{SpecGen} & \multicolumn{2}{c}{CodeGen} & \multicolumn{2}{c}{SpecGen-Repair} & \multicolumn{2}{c}{CodeGen-Repair} \\
\cmidrule(l{2pt}r{2pt}){4-5}\cmidrule(l{2pt}r{2pt}){6-7}\cmidrule(l{2pt}r{2pt}){8-9}\cmidrule(l{2pt}r{2pt}){10-11}
 & & & pass@1 & pass@4 & pass@1 & pass@4 & pass@1 & pass@4 & pass@1 & pass@4 \\
\midrule
\work-9B-SFT & 0 & 0 & 10.1 & 24.3 & 13.1 & 28.3 & 56.7 & 76.3 & 59.5 & 80.6 \\
\midrule
\work-9B-RL & 70 & 2,149 & 21.8 & 51.3 & 43.5 & 80.7 & 54.3 & 84.7 & 55.6 & 86.4 \\
 & 140 & 4,162 & 34.9 & 62.0 & 65.0 & 89.0 & 49.6 & 84.7 & 55.3 & \underline{92.2} \\
 & 210 & 6,186 & \textbf{46.7} & \textbf{71.7} & \textbf{74.9} & \textbf{92.0} & 67.4 & \textbf{87.0} & \underline{71.1} & 90.3 \\
\midrule
w/o test execution & 25 & 961 & 12.2 & 28.0 & 21.8 & 40.0 & 59.4 & 80.9 & 60.0 & 77.7 \\
 & 50 & 1,833 & 13.7 & 30.7 & 26.0 & 48.0 & 62.2 & 80.2 & 60.2 & 82.5 \\
 & 75 & 2,604 & 16.8 & 35.3 & 24.0 & 45.0 & 63.0 & 85.5 & 62.4 & 85.4 \\
 & 100 & 3,362 & 18.1 & 37.0 & 28.9 & 52.0 & 62.6 & 80.2 & 67.5 & 88.3 \\
 & 125 & 4,111 & 24.3 & 45.3 & 34.4 & 61.7 & 66.8 & 85.5 & 69.7 & 90.3 \\
 & 150 & 4,796 & 31.4 & 53.7 & 37.6 & 62.3 & 65.8 & 84.7 & 63.1 & 86.4 \\
\midrule
w/o LLM judges & 25 & 830 & 11.7 & 29.0 & 26.0 & 48.0 & 62.4 & 80.9 & 60.2 & 76.7 \\
 & 50 & 1,626 & 15.1 & 32.3 & 40.2 & 66.3 & 62.4 & 83.2 & 64.3 & 90.3 \\
 & 75 & 2,346 & 26.3 & 50.0 & 58.3 & 86.3 & 68.3 & \textbf{87.0} & 67.0 & 86.4 \\
 & 100 & 3,030 & 35.8 & 62.0 & 67.0 & 90.0 & \underline{69.1} & \textbf{87.0} & 68.2 & 88.3 \\
 & 125 & 3,711 & 31.6 & 51.0 & 64.6 & 84.7 & 63.4 & 82.4 & 69.2 & \textbf{93.2} \\
 & 150 & 4,556 & 20.0 & 35.7 & 54.8 & 75.3 & 43.3 & 67.9 & 60.4 & 84.5 \\
\midrule
w/o self-repair & 25 & 1,300 & 13.8 & 30.3 & 21.3 & 42.3 & 57.1 & 77.1 & 61.4 & 81.6 \\
 & 50 & 2,396 & 14.7 & 32.3 & 26.2 & 53.3 & 45.8 & 79.4 & 51.7 & 76.7 \\
 & 75 & 3,378 & 20.6 & 44.3 & 46.8 & 75.0 & 66.4 & 80.9 & 63.6 & 80.6 \\
 & 100 & 4,346 & 23.0 & 44.3 & 53.7 & 81.3 & 65.8 & 81.7 & 64.8 & 84.5 \\
 & 125 & 5,332 & 33.2 & 57.3 & 60.5 & 88.3 & 66.6 & 80.2 & \textbf{71.4} & 90.3 \\
 & 150 & 6,334 & \underline{43.1} & \underline{67.3} & \underline{70.2} & \underline{91.7} & \textbf{70.8} & \underline{86.3} & 68.9 & 88.3 \\
\bottomrule
\end{tabular}
\end{table}

\textbf{LLM judges.} Without LLM judges in the reward, the model exploits the reward by restating the given test cases as the specification. We note that the training reward keeps increasing, while SpecGen pass@1 on \work-Test reaches its highest value around step 100 and then decreases (\subfig{fig:ablation-dynamics}{c1}). The gap between the 2 therefore widens over training (84.7\% of SpecGen rollouts accepted in steps 141--150 vs.\ 20.0\% pass@1 at step 150, down from 35.8\% at step 100). Tautological specifications account for a growing part of the evaluation samples (1.9\% at step 50 vs.\ 58.8\% at step 150). On \verina{}, these specifications pass the test cases given in the prompt but fail without them (88.4\% vs.\ 52.0\% SpecGen pass@1 at step 150).

\textbf{Test execution.} Without test execution in the reward, the reward accepts implementations that the test cases would reject. About half of the CodeGen rollouts pass the remaining stages early in training (49.7\% vs.\ 19.7\% of CodeGen rollouts accepted in steps 41--50 without and with test execution), while CodeGen pass@1 on \work-Test improves slowly (\subfig{fig:ablation-dynamics}{b2}). After approximately the same number of training prompts the variant falls behind \work-9B-RL in both tasks (34.4\% vs.\ 65.0\% CodeGen and 24.3\% vs.\ 34.9\% SpecGen pass@1 after 4{,}111 and 4{,}162 prompts). The gap in CodeGen persists at step 150 even though the variant has drawn more training prompts (37.6\% vs.\ 65.0\% after 4{,}796 vs.\ 4{,}162 prompts; Table~\ref{tab:ablation-indomain}). Repair remains comparable to the other runs (65.8\% SpecGen-Repair pass@1 at step 150), so the variant falls behind in generating a correct output from scratch. The judges alone cannot check the behavior of an implementation, and test execution keeps the reward accurate.

\begin{figure}[tbp]
\centering
\includegraphics[width=\linewidth]{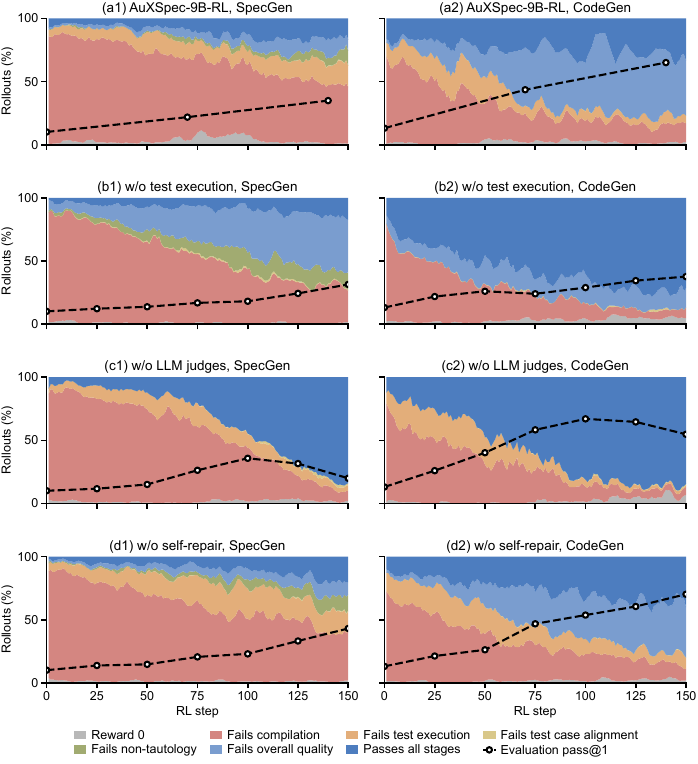}
\caption{Training rollouts by reward outcome for \work-9B-RL and the 3 variants, one row per run, on SpecGen (left) and CodeGen (right), from the first failed stage to passing all stages, with a 5-step moving average, up to step 150. The variant without self-repair draws more training prompts per step than the other runs, and Table~\ref{tab:ablation-indomain} compares the runs after approximately the same number of training prompts. \subfig{fig:rltrain}{d} shows the full run of \work-9B-RL. Each variant drops the bands of its removed stages, so rollouts that those stages would reject count as passing all stages. Dashed lines indicate pass@1 of the same task on \work-Test at the evaluated checkpoints, and the first point of each line indicates \work-9B-SFT. A widening gap between the top band and the dashed line indicates rewarded outputs that evaluation rejects.}
\label{fig:ablation-dynamics}
\end{figure}

\textbf{Self-repair.} Without self-repair every batch slot holds a new training prompt. The variant therefore consumes training prompts faster and outperforms \work-9B-RL at the same step (43.1\% vs.\ 34.9\% SpecGen pass@1 at steps 150 and 140 after 6{,}334 vs.\ 4{,}162 prompts). \work-9B-RL outperforms it in both generation and repair even after fewer training prompts (21.8\% vs.\ 14.7\% SpecGen and 54.3\% vs.\ 45.8\% SpecGen-Repair pass@1 after 2{,}149 and 2{,}396 prompts). After about 6.2K prompts it still outperforms the variant in generation (46.7\% vs.\ 43.1\% SpecGen pass@1; Table~\ref{tab:ablation-indomain}). Training without repair tasks is thus a viable recipe when new problems are plentiful.

The repair tasks are designed for settings in which a model revises its output from feedback over several rounds, such as long-horizon verification and collaboration between agents. Refinement from compiler and judge feedback is part of the training objective, and the model also refines its specifications at test time: on ICPC-Eval, refine@16 reaches 82.7\% of the problems, compared with 61.3\% for pass@16 (Appendix~\ref{app:icpc}).

\clearpage

We also report \verina{} at the same checkpoints, where the variants follow similar trends (Table~\ref{tab:ablation-verina}).

\begin{table}[H]
\centering
\captionsetup{font=scriptsize}
\caption{Component ablations of the 9B recipe on \verina{}, with and without test cases in the prompt. The setting without test cases follows the default of \verina{}. We also report the setting with test cases in the prompt, since concrete input-output pairs specify behavior that the description may state ambiguously (Section~\ref{sec:ood}). Rows as in Table~\ref{tab:ablation-indomain}.}
\label{tab:ablation-verina}
\scriptsize
\setlength{\tabcolsep}{4pt}
\begin{tabular}{@{}l r r *{8}{c}@{}}
\toprule
\multirow{2}{*}{Model} & \multirow{2}{*}{Step} & \multirow{2}{*}{\shortstack{Training\\prompts}} & \multicolumn{2}{c}{SpecGen, w/o tests} & \multicolumn{2}{c}{CodeGen, w/o tests} & \multicolumn{2}{c}{SpecGen, w/ tests} & \multicolumn{2}{c}{CodeGen, w/ tests} \\
\cmidrule(l{2pt}r{2pt}){4-5}\cmidrule(l{2pt}r{2pt}){6-7}\cmidrule(l{2pt}r{2pt}){8-9}\cmidrule(l{2pt}r{2pt}){10-11}
 & & & pass@1 & pass@4 & pass@1 & pass@4 & pass@1 & pass@4 & pass@1 & pass@4 \\
\midrule
\work-9B-SFT & 0 & 0 & 34.9 & 65.6 & 36.5 & 71.4 & 37.3 & 69.8 & 38.0 & 70.4 \\
\midrule
\work-9B-RL & 70 & 2,149 & 52.1 & 82.0 & 75.8 & 93.7 & 52.4 & 86.2 & 80.4 & 97.4 \\
 & 140 & 4,162 & 67.3 & \underline{91.5} & \underline{86.4} & \textbf{97.4} & 71.7 & 92.1 & 87.6 & \textbf{98.4} \\
 & 210 & 6,186 & \underline{73.4} & \textbf{92.1} & \textbf{89.2} & \underline{96.8} & \underline{84.8} & \textbf{97.4} & \textbf{90.1} & 96.8 \\
\midrule
w/o test execution & 25 & 961 & 47.6 & 70.9 & 63.9 & 84.7 & 48.8 & 75.1 & 63.9 & 86.2 \\
 & 50 & 1,833 & 52.0 & 78.8 & 67.9 & 85.7 & 56.6 & 82.0 & 66.5 & 85.7 \\
 & 75 & 2,604 & 55.2 & 78.8 & 69.3 & 87.3 & 60.3 & 82.0 & 71.2 & 89.4 \\
 & 100 & 3,362 & 58.6 & 81.0 & 73.5 & 91.0 & 65.5 & 83.1 & 73.0 & 89.9 \\
 & 125 & 4,111 & 65.1 & 85.7 & 76.5 & 91.5 & 72.8 & 88.9 & 77.6 & 92.6 \\
 & 150 & 4,796 & 69.0 & 85.7 & 80.0 & 92.6 & 73.9 & 90.5 & 80.8 & 96.3 \\
\midrule
w/o LLM judges & 25 & 830 & 41.1 & 73.0 & 63.1 & 83.6 & 45.2 & 71.4 & 64.2 & 85.2 \\
 & 50 & 1,626 & 48.9 & 77.8 & 68.8 & 91.0 & 46.7 & 76.7 & 70.1 & 91.0 \\
 & 75 & 2,346 & 55.3 & 84.1 & 76.1 & 94.7 & 63.0 & 89.4 & 76.6 & 96.3 \\
 & 100 & 3,030 & 63.4 & 88.4 & 84.7 & 95.8 & 75.0 & 95.2 & 83.2 & \textbf{98.4} \\
 & 125 & 3,711 & 63.9 & 87.3 & 84.0 & \underline{96.8} & 82.1 & 95.8 & \underline{89.0} & 97.4 \\
 & 150 & 4,556 & 52.0 & 79.4 & 81.5 & 94.7 & \textbf{88.4} & \underline{96.3} & 87.7 & \underline{97.9} \\
\midrule
w/o self-repair & 25 & 1,300 & 43.3 & 73.5 & 57.1 & 81.0 & 50.0 & 78.8 & 61.5 & 83.1 \\
 & 50 & 2,396 & 57.1 & 84.7 & 68.5 & 86.8 & 58.3 & 84.1 & 70.5 & 90.5 \\
 & 75 & 3,378 & 61.1 & 84.7 & 70.9 & 89.9 & 63.2 & 88.4 & 74.2 & 92.1 \\
 & 100 & 4,346 & 64.4 & 86.2 & 77.1 & 93.1 & 65.1 & 87.8 & 79.8 & 93.1 \\
 & 125 & 5,332 & 67.3 & 85.7 & 81.0 & 96.3 & 78.0 & 92.6 & 81.0 & 92.6 \\
 & 150 & 6,334 & \textbf{73.8} & 89.9 & 83.1 & 95.8 & 82.8 & 94.7 & 83.3 & 96.8 \\
\bottomrule
\end{tabular}
\end{table}

\end{revised}

\clearpage

\section{Pipeline details}
\label{app:pipeline}

\begin{figure}[H]
\centering
\includegraphics[width=\linewidth]{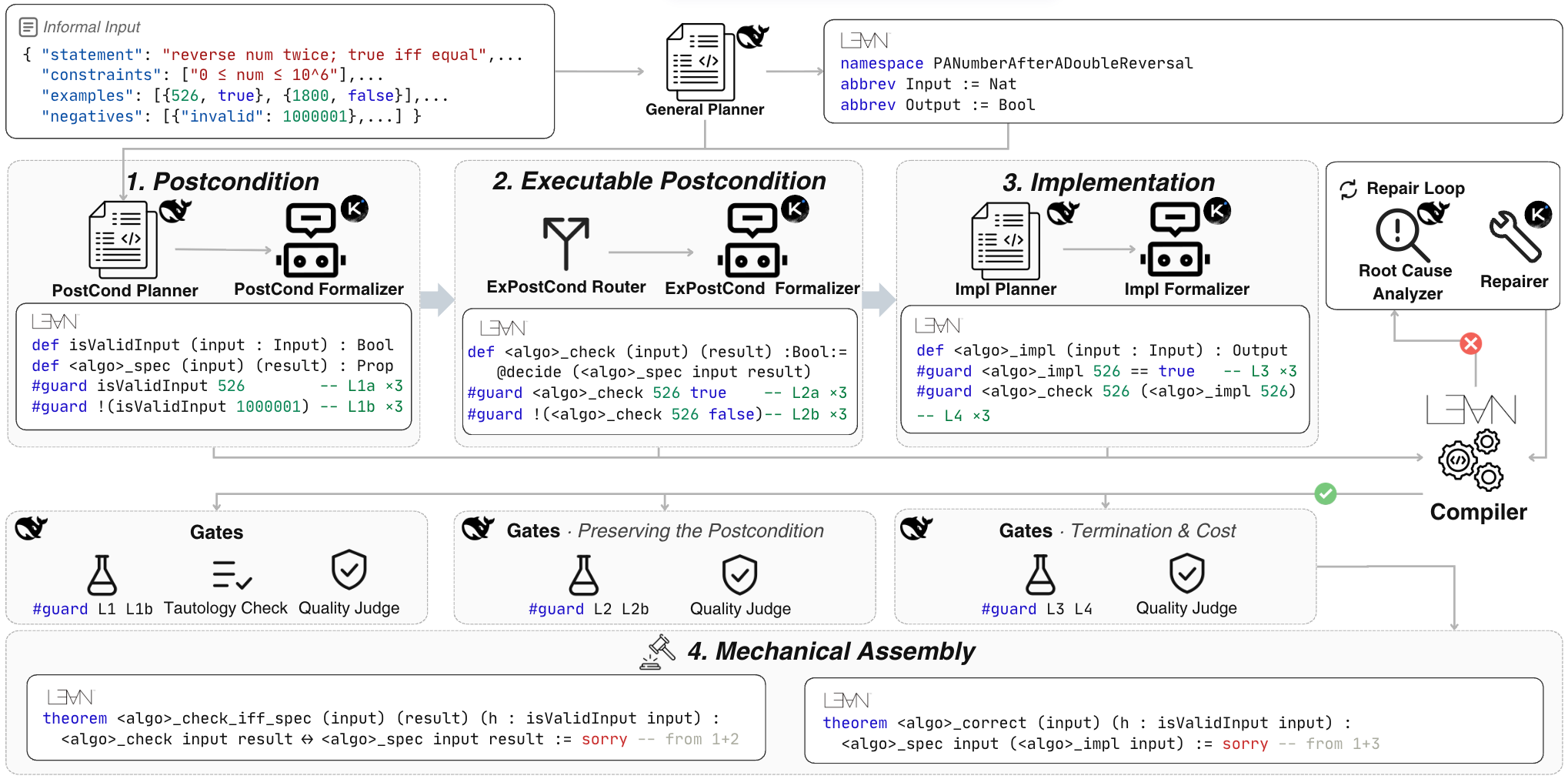}
\caption{Data synthesis pipeline for \work{} on a running example. Stages (1) to (3) synthesize the postcondition, the executable postcondition, and the implementation under separate plans, and stage (4) assembles the components with the theorem statements.}
\label{fig:pipeline-example}
\end{figure}

Figure~\ref{fig:pipeline-example} shows the pipeline of Figure~\ref{fig:pipeline} on a running example, with the input of each stage and the Lean code that the stage produces. Each stage below ends with an analysis of the stage on the 3{,}037 synthesized problems.

\subsection{Preprocessing}

\subsubsection{Test cases}
\label{sec:testcorpus}

For each of 3{,}109 problems (719 easy, 1{,}622 medium, 768 hard), we screen the stored examples and propose negative test cases, invalid inputs and incorrect outputs beside the positive test cases. Invalid inputs are proposed by category (a value just outside a stated bound, a size outside the stated range, a container of the wrong shape, a broken structural invariant, parameters invalid only together, a value outside the stated domain), and incorrect outputs by category (a near miss, a suboptimal feasible answer, an unattainable answer, the answer under a relaxed constraint, an output of the wrong shape, a degenerate answer). Benchmarks store negative test cases beside positive ones \citep{ye2026verina,verispecgen2026}, since a specification that accepts correct outputs on the given examples can still accept incorrect outputs \citep{endres2024nl2postcond}.

A proposed incorrect output is admitted when the reference solution, run on the proposed valid input, returns an output that is not a reordering of the proposed output. Invalid inputs are proposed together with the violated constraint and are dropped on coinciding with a valid example of the problem. Cases too large for compiled evaluation and cases whose stored answer is null are refused before any model call, with the measured thresholds in Appendix~\ref{app:accounts}.

The 27{,}981 test cases, 9 per problem, are assembled into 18 \texttt{\#guard} commands per problem, one per test case and layer of Table~\ref{tab:layers}, with each positive test case reused at \La{1}, \La{2}, L3, and L4. A \texttt{\#guard} command runs compiled code, and a passed test case relies on the compiler and runtime as trusted base, the same as \texttt{native\_decide}. For the test cases, we store the provenance (public example or synthesized) and the admission round. A test case judged wrong is replaced by a held-back test case at most once per problem.

\textbf{Analysis.} The frozen test cases, negative test cases included, are shown to the generation model, and the negative test cases act as part of the problem statement rather than as a hidden filter. Of 3{,}276 specification candidates tested at \La{1} and \Lb{1}, 8 (0.24\%) fail only at \Lb{1}, and of 3{,}195 executable postconditions tested at \La{2} and \Lb{2}, 4 (0.13\%) fail only at \Lb{2}. The 4 failures at \Lb{2} have one cause, a lossy output decoder in the specification, such as a tree decoder that drops children below null nodes. The 8 failures at \Lb{1} fall on 5 problems, and in 4 of the 5 problems the frozen invalid input is in fact valid. Repair narrowed the precondition to reject that input in all 5 problems. Frozen test values keep test cases from adapting to a wrong specification, and these 5 preconditions are the only adaptation in the opposite direction.

\textbf{Screening.} Before any model call, test cases are filtered by size and by null answers. A test case containing a list or string longer than 128 items is refused, since a case is frozen as a literal that compiled evaluation builds item by item. We cap compilation at 600 seconds on the Lean server and observe that a test case exceeding the cap terminates the serving REPL. The threshold was measured over the stored test cases, and any threshold between 102 and 185 items refuses the same test cases. A case whose stored answer is null is refused in the layers that read an answer, since the source dataset stores null wherever the answer is a mutated argument, and a null true answer makes each negative built from the null answer meaningless.

\subsubsection{Library retrieval}
\label{sec:retrieval}

Models reference library declarations absent from the pinned library version \citep{chakraborty2025fstar}. Typed queries against Mathlib, for a definition, a theorem, or an instance, are generated together with one sentence stating the use the specification has for the answer, and relevance is judged against that sentence. Search candidates are deduplicated, elaborated with \texttt{\#check} against the pinned toolchain, and judged for relevance by an LLM given the problem description. Specification synthesis is augmented with the declarations admitted by the relevance judge, 4.2 per problem, each with a checked signature and a one-sentence reason from the LLM. Appendix~\ref{app:accounts} reports the counts at each step.

\textbf{Analysis.} Elaboration keeps declarations that exist only in newer Mathlib versions out of specification synthesis (Table~\ref{tab:retrieval}). 23{,}608 queries (14{,}888 definition, 5{,}052 theorem, 3{,}668 instance) produced 92{,}104 candidates. Of the elaborated declarations, 5.4\% were absent from the pinned Mathlib, and review dropped 10.0\% of the reviewed candidates. Of the 13{,}191 stored declarations, specification synthesis of the 3{,}037 problems received 12{,}876, 4.2 per problem, with 131 problems receiving none. The absent declarations originate in a search index built on the latest Mathlib, while synthesis compiles against the pinned version. Retrieved declarations do not change the first compilation. Specifications that reference a retrieved declaration by the full name compile at the first attempt in 83.2\% of 1{,}871 problems, and the other specifications in 83.9\% of 1{,}032.

\begin{table}[htbp]
\caption{Library retrieval by step over 3{,}109 problems and 3{,}715 rounds. The first 5 rows indicate work across all rounds, and the last row indicates the number of declarations passed to specification synthesis, so a problem retrieved more than once counts only the last round there.}
\label{tab:retrieval}
\centering
\scriptsize
\begin{tabular}{lrl}
\toprule
Step & Count & Note \\
\midrule
Queries generated & 23{,}608 & Definition, theorem, and instance queries \\
Candidates returned & 92{,}104 & Search hits \\
Declarations elaborated & 83{,}679 & Deduplicated, \#check against the pinned toolchain \\
Declarations present & 79{,}126 & 5.4\% of elaborated declarations were absent \\
Reviewed & 17{,}372 & 1{,}729 dropped as irrelevant \\
Stored & 13{,}191 & Last record per problem, 4.2 per problem \\
\bottomrule
\end{tabular}
\end{table}

\subsection{Specifications}

\textbf{Planning.}
A type plan fixes the declaration names and one type per parameter, with a conversion function wherever the stored value has a different type. The postcondition is then planned regardless of decidability, which keeps computation out of the postcondition. A frozen test value is printed once as the expression used by all test layers to prevent models from altering the value.

\textbf{Formalization.}

Each component is generated separately and compiled with the accepted components. The \texttt{\#guard} commands of the layers are generated from the expressions the plan fixed, and each command is accepted at the first passing check in the order shape, verbatim expression, token-level expression, and judge review of equivalence. The shape check admits one command per test case, attached to the declaration of that layer and with the polarity of the test case. 
The expression check compares the tokens of the printed expression, which keeps whitespace from merging 2 values.

\textbf{Executable postcondition.}
\label{app:route}
The executable postcondition is first set to the decidable form of the postcondition, and the file is compiled. When no decidability instance is derived, a model supplies instances from the compiler diagnostics for up to 3 rounds. When the supplied instances still fail to elaborate, a separate definition is generated. Only a separate definition can disagree with the postcondition, which the \La{2} and \Lb{2} test cases detect on the tested outputs and the postcondition equivalence theorem excludes on all valid inputs. Appendix~\ref{app:accounts} reports the counts by form.

\textbf{Judges.}
\label{app:gates}
The postcondition judge is given the problem description, the constraints, and the postcondition source. A rejection quotes a sentence of the description and names an output that the postcondition admits and the problem excludes. The executable postcondition judge is given the constraints, the accepted precondition and postcondition, and the executable postcondition, and a rejection names a valid input with a witness or counterexample that the executable postcondition misses or wrongly admits. The tautology judge rejects a postcondition that computes the output. Problems whose natural statement is the transformation itself are labeled in advance by a majority vote of 3 classifier calls, and the label exempts the problem from the tautology judge. The rejected component is repaired with the named error, or replanned for an error in the plan.

\subsubsection{Example problems}
\label{app:example}

\paragraph{Separate executable postcondition.}
\label{app:example-separate}

We illustrate the format with problem \textit{1955 Count Number of Special Subsequences} from \work-Test (Figures~\ref{fig:example-defs} and~\ref{fig:example-tests}). In this problem, we are given an array over $\{0,1,2\}$ and need to count, modulo $10^9+7$, the subsequences of the form $0^a 1^b 2^c$ with $a,b,c \geq 1$. The postcondition states the result as the size of the set of index sets whose selected elements take this form. The executable postcondition enumerates all subsets of the indices, a brute-force computation that finishes on the short test inputs. The implementation counts in 1 pass with 3 counters, for subsequences of 0s, of 0s followed by 1s, and of 0s, 1s and 2s, since an element either extends or skips the subsequences counted so far. Since neither specification mentions the counters, the correctness theorem relates the recurrence to the set of index sets.

For this problem, a natural postcondition characterizes the valid index sets $S$ as exactly those that select a subsequence of the form $0^+1^+2^+$ from the input. 
This characterization is not directly executable, since it ranges over arbitrary finite sets of natural numbers. 
However, every relevant index set is contained in the input indices. The executable postcondition therefore enumerates $2^{\{0,\ldots,\text{input.length}-1\}}$ and counts the subsets satisfying the same property. 
The two formulations describe the same relation but are not definitionally identical in Lean, so the postcondition equivalence theorem connects the logical characterization with the bounded enumeration. 

\begin{figure}[p]
\input{generated/fig_artifact.tex}
\caption{Declarations of the example problem, without namespace and docstrings (tests in Figure~\ref{fig:example-tests}).}
\label{fig:example-defs}
\end{figure}

\paragraph{Executable postcondition through type class resolution.}
\label{app:example-typeclass}

We illustrate the decidable form with problem \textit{523 Continuous Subarray Sum} from \work-Test (Figures~\ref{fig:example-typeclass-a}, \ref{fig:example-typeclass} and~\ref{fig:example-typeclass-tests}). In this problem, we are given an array of non-negative integers and a positive integer $k$ and need to decide whether some contiguous subarray of length at least 2 sums to a multiple of $k$. The postcondition states that the result is true exactly when there exist 2 indices and an integer multiplier $n$ such that the subarray between the indices sums to $n \cdot k$.

The executable postcondition applies \texttt{decide} to the postcondition, which compiles only when type class resolution finds a \texttt{Decidable} instance for the unfolded proposition. The existential over natural-number indices and an integer multiplier has no such instance as stated, since the quantifiers range over unbounded domains. The pipeline therefore synthesizes a type class instance that bounds both indices by the length of the array, replaces the multiplier by integer divisibility, and proves the bounded proposition equivalent to the original with \texttt{decidable\_of\_iff}. Type class resolution then assembles the executable postcondition from this instance and the library instances for bounded quantifiers and divisibility, so the executable postcondition agrees with the postcondition by construction (Appendix~\ref{app:equiv}). The implementation, omitted from the figure, scans prefix sums modulo $k$ with a map from each remainder to the earliest prefix index at which the remainder occurs, and the correctness theorem relates this scan to the existential of the postcondition.

\begin{figure}[p]
\input{generated/fig_artifact_tests.tex}
\caption{Test cases of the example problem, grouped by test layer.}
\label{fig:example-tests}

\vspace{1.5em}
\renewcommand{\thefigure}{\arabic{figure}a}
\input{generated/fig_artifact_typeclass_a.tex}
\caption{Declarations of the second example problem, part 1.}
\label{fig:example-typeclass-a}
\end{figure}

\begin{figure}[p]
\addtocounter{figure}{-1}
\renewcommand{\thefigure}{\arabic{figure}b}
\input{generated/fig_artifact_typeclass_b.tex}
\caption{Declarations of the second example problem, part 2, whose executable postcondition is \texttt{decide} of the postcondition, with the \texttt{Decidable} instance found by type class resolution from a synthesized instance. Namespace, docstrings, and the implementation with the correctness theorem are omitted (tests in Figure~\ref{fig:example-typeclass-tests}).}
\label{fig:example-typeclass}

\vspace{1.5em}
\renewcommand{\thefigure}{\arabic{figure}}
\input{generated/fig_artifact_typeclass_tests.tex}
\caption{Test cases of the second example problem, grouped by test layer.}
\label{fig:example-typeclass-tests}
\end{figure}

\clearpage

\subsubsection{Analysis}
\label{app:specanalysis}

The specification judges rarely reject, and the errors that pass all checks are in the meaning of the specification. Each verdict in Table~\ref{tab:stages} corresponds to a distinct component that compiles and passes the test layers of the component. The specification judges reject 1.7\% to 3.8\% of the verdicts, and all rejected problems are accepted after repair. The errors that proving or human review later found had all passed the test layers and the judge of the erroneous component. In 2 problems a correct rejection led to a repair that introduced the error found by review, and in 1 problem the judge accepted on a second reading the defect that the judge had named.

\begin{table}[htbp]
\caption{Verdicts of the judges on components that compile and pass the test layers of the component. A problem counts as rejected when at least 1 verdict on the problem is a rejection.}
\label{tab:stages}
\centering
\scriptsize
\setlength{\tabcolsep}{4pt}
\begin{tabular}{lrrrr}
\toprule
Judge & Verdicts & Rejections & Problems rejected & Accepted after rejection \\
\midrule
Faithfulness & 3{,}237 & 123 (3.8\%) & 112 & 112 \\
Non-tautology & 3{,}143 & 69 (2.2\%) & 61 & 61 \\
Completeness & 3{,}159 & 55 (1.7\%) & 42 & 42 \\
Implementation & 3{,}667 & 717 (19.6\%) & 414 & 281 \\
\bottomrule
\end{tabular}
\end{table}

\begin{figure}[htbp]
\centering
\includegraphics[width=\linewidth]{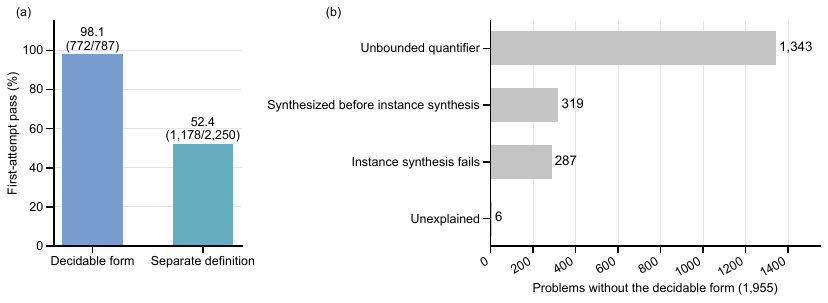}
\caption{(a) Executable postconditions that pass all checks at the first attempt, by route. (b) Reasons that the decidable form does not apply.}
\label{fig:execpost}
\end{figure}

The decidable form is reliable where the postcondition quantifies over bounded types (Figure~\ref{fig:execpost}). The decidable form serves 1{,}079 of the 3{,}037 problems (35.5\%). Of the other 1{,}958 problems, 1{,}343 have postconditions that quantify over unbounded types, 319 were synthesized before instance synthesis was added, 287 fail instance synthesis for 3 rounds, 6 remain unexplained, and 3 are routed to the decidable form but keep a separate definition in the final file. Where the decidable form applies, the executable postcondition passes all checks at the first attempt in 772 of 787 problems (98.1\%), compared with 1{,}178 of 2{,}250 (52.4\%) for separately generated executable postconditions. Missing instances arise from the decidable form. The pipeline first sets the executable postcondition to \texttt{decide} applied to the postcondition, which compiles only when Lean finds a decidability instance for the postcondition. A postcondition that quantifies over an unbounded type (e.g., over all natural numbers or all lists) has no computable instance as stated, and instance synthesis fails, for example on add-digits, where no decidability instance can be synthesized for a postcondition that quantifies over all digit roots $d$.

\subsection{Implementation}

Implementation planning is given the accepted specification as read-only text and selects an algorithm with cost argument, listing specification definitions to call and definitions to replace with reason. The plan is checked before generation for a termination argument based on a decreasing quantity of the input, a stated cost consistent with the approach and the input bounds, a replacement justified by cost or by a different question, and an algorithm that computes the answer, without search over candidates. Beyond the properties in Section~\ref{sec:pipeline}, the implementation judge reports computations already available in the specification.

\subsubsection{Analysis}
\label{app:stages}

\begin{figure}[H]
\centering
\includegraphics[width=\linewidth]{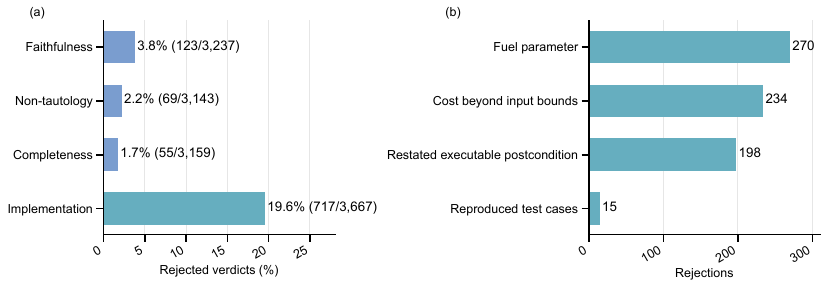}
\caption{(a) Rejections by each judge on compiled components that pass the test layers. (b) Implementation judge rejections by reason.}
\label{fig:judges}
\end{figure}

Test cases on small inputs admit false positives \citep{li2022alphacode}, and the implementation judge rejects 717 of the 3{,}667 implementations (19.6\%) that pass all test cases (Figure~\ref{fig:judges}).
\begin{itemize}[leftmargin=*, itemsep=0pt, topsep=2pt, parsep=0pt]
\item \textbf{Fuel parameters (270).} The implementation recurses on a fuel argument, which hides nontermination on inputs beyond the test cases.
\item \textbf{Cost beyond the input bounds (234).} The implementation is correct but exceeds the cost that the stated input bounds allow, known as a slow positive \citep{li2022alphacode}.
\item \textbf{Restated executable postconditions (198).} The implementation repeats a computation of the executable postcondition that already runs within the stated input bounds.
\item \textbf{Replayed test cases (15).} The implementation returns stored outputs for the tested inputs.
\end{itemize}
The implementation judge is waived after 2 retries, and 133 problems keep an implementation that the judge rejected. Of these, 89 are rejected for a fuel argument that may hide nontermination, 27 for cost beyond the input bounds, 12 for restating the executable postcondition and 5 for replaying test cases. 12 of the 133 are in \work-Test, and for the 9 among them rejected for fuel, the correctness proofs show that the fuel suffices on all valid inputs.

\subsection{Repair}

Compilation runs through the Kimina Lean server \citep{kiminaserver2025}, a REPL-based server \citep{leanrepl}. Repair revises rejected components separately for at most 18 rounds per run of the repair loop, which a problem re-enters after replanning. From the second round on, the repair input includes a digest of earlier attempts, and a root-cause analysis runs for errors that line numbers cannot attribute. Repair stops early when a revision repeats an earlier diff or does not change the component. Where a repair attributes the error to a test case, the attribution is reviewed by a judge, and a confirmed attribution replaces the test case with a held-back test case at most once per problem.

\subsubsection{Analysis}
\label{app:repairanalysis}

\begin{figure}[H]
\centering
\includegraphics[width=0.45\linewidth]{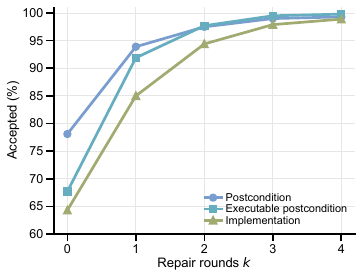}
\caption{Cumulative acceptance rate of each component over repair rounds on the 3{,}037 synthesized problems. The implementation depends most on repair, with 64.3\% accepted at the first attempt.}
\label{fig:repair}
\end{figure}

Repair recovers most first failures within 2 rounds (Figure~\ref{fig:repair}). The fraction of the 3{,}037 problems accepted within $k$ repair rounds rises from 78.1\% at $k=0$ to 93.9\% at $k=1$ and 97.5\% at $k=2$ for the postcondition, from 67.7\% to 91.9\% and 97.7\% for the executable postcondition, and from 64.3\% to 85.0\% and 94.4\% for the implementation.

\begin{figure}[H]
\centering
\includegraphics[width=0.75\linewidth]{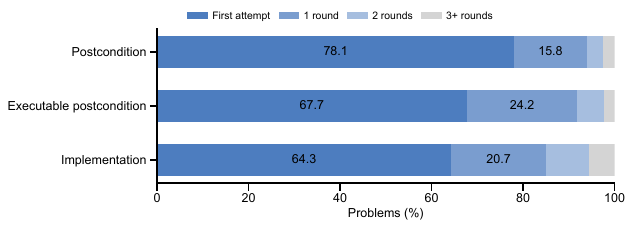}
\caption{Repair rounds needed by each component on the 3{,}037 synthesized problems.}
\label{fig:repaircomp}
\end{figure}

The remaining problems took 1{,}030 postcondition repair rounds, 1{,}321 executable postcondition rounds and 1{,}842 implementation rounds (Figure~\ref{fig:repaircomp}).

\begin{figure}[H]
\centering
\includegraphics[width=0.75\linewidth]{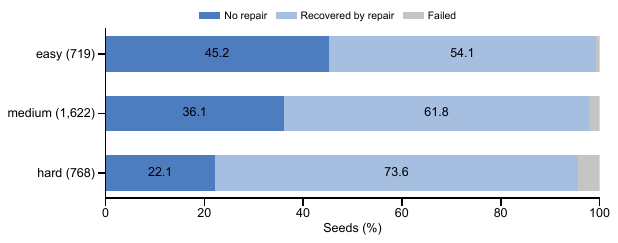}
\caption{Seeds of each difficulty accepted without repair, recovered by repair, and failed.}
\label{fig:repairdiff}
\end{figure}

1{,}081 problems (34.8\% of the 3{,}109 seeds) need no repair round for any component, and repair recovers most of the others, most on hard problems, where 22.1\% of the seeds pass without repair and 95.7\% with repair (Figure~\ref{fig:repairdiff}).

\begin{figure}[htbp]
\centering
\includegraphics[width=0.55\linewidth]{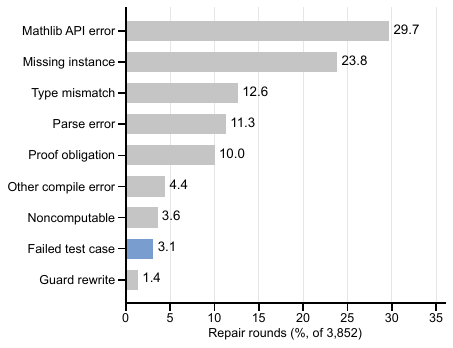}
\caption{Causes of 3{,}852 repair rounds. Proof obligations include termination, \texttt{omega} goals and index bounds.}
\label{fig:repaircauses}
\end{figure}

Repair never replaces a frozen test case. Repair attributed 113 failures on 96 problems to test cases, and the root-cause analyzer placed the fault in the model-generated \texttt{\#guard} command in 55 and in the plan in 52, never in a frozen test case. The most frequent case is a negative test case with a negative value for an input that the plan types as a natural number, which the judge resolves by changing the planned type to an integer. Repair rounds go to Lean usage rather than to failed test cases (Figure~\ref{fig:repaircauses}). Of 3{,}852 recorded rounds, Mathlib API errors take 29.7\%, missing instances 23.8\% (Appendix~\ref{app:specanalysis}), type mismatches 12.6\%, parse errors 11.3\%, proof obligations 10.0\%, and failed test cases 3.1\%. The fraction of Mathlib API errors matches the misuse of Mathlib declarations that motivates continued pretraining (Appendix~\ref{app:cpt}).

\subsection{Outcomes}
\label{app:accounts}

We break down the outcomes of Section~\ref{sec:outcomes} by difficulty (Table~\ref{tab:outcomes}).

\begin{table}[htbp]
\caption{Outcomes of the 3{,}109 seed problems. Percentages are over the seeds of each difficulty.}
\label{tab:outcomes}
\centering
\scriptsize
\setlength{\tabcolsep}{4pt}
\begin{tabular}{lrl}
\toprule
Outcome & Problems & Note \\
\midrule
Succeeded & 3{,}037 & Compiled, all frozen layers pass \\
\quad Easy & 714 & Of 719 seeds, 99.3\% \\
\quad Medium & 1{,}588 & Of 1{,}622 seeds, 97.9\% \\
\quad Hard & 735 & Of 768 seeds, 95.7\% \\
Failed & 72 & Terminal exits, stalled runs, and later withdrawals \\
\bottomrule
\end{tabular}
\end{table}

The decidable form of the postcondition is adopted in 1{,}079 problems (35.5\%) and a separate definition in 1{,}958 problems. The frozen \texttt{\#guard} commands of 90.0\% of the synthesized problems match the planned expressions verbatim, and the other 303 needed token-level matching or a judge.

\subsection{Training samples}
\label{app:traces}

The reasoning trace of a training sample is rephrased from the reasoning recorded while that problem was synthesized, which is the planning and generation reasoning for the generation tasks and the diagnostics, root cause, and repair reasoning for the repair tasks. A rephrased trace is admitted only when each sentence is grounded in the recorded reasoning, the answer is not stated before being derived, and no pipeline vocabulary appears.

\subsection{Cost}

Synthesis and supporting experiments cost \$2{,}604 over 226{,}351 model calls, divided across 9 accounts (Table~\ref{tab:cost}). Accounts follow record keeping, and reasoning-trace rephrasing, model comparisons, and discarded experiments are listed beside retrieval, test case construction, and formalization. Accounts cover different problems, and per-problem costs are not compared across accounts.

\begin{table}[htbp]
\caption{Spend by account. The last 2 accounts recorded no tokens, so their tokens and spend are estimated from the exact call counts with the mean usage of DeepSeek-V4-Flash judge calls at list price. The judging calls are an upper bound, since some were served from cache.}
\label{tab:cost}
\centering
\scriptsize
\setlength{\tabcolsep}{3pt}
\begin{tabular}{lrrrp{5.2cm}r}
\toprule
Account & Calls & Tokens In & Tokens Out & Models & Spend \\
\midrule
Retrieval & 7{,}421 & 32M & 71M & \raggedright DeepSeek-V4-Flash & \$24 \\
Test corpus & 43{,}880 & 109M & 341M & \raggedright DeepSeek-V4-Flash & \$98 \\
Formalization & 59{,}337 & 182M & 290M & \raggedright Kimi-K3 (generation, repair), DeepSeek-V4-Pro (planning), DeepSeek-V4-Flash (judging) & \$2{,}024 \\
Reasoning traces & 12{,}320 & 209M & 164M & \raggedright DeepSeek-V4-Flash & \$75 \\
Model comparisons & 3{,}183 & 20M & 20M & \raggedright Claude-Opus-5, DeepSeek-V4-Flash, DeepSeek-V4-Pro, GLM-5.2, Kimi-K3 & \$184 \\
Pipeline experiments & 435 & 3M & 2M & \raggedright Claude-Opus-5, DeepSeek-V4-Flash & \$36 \\
Bake-off & 39 & 0M & 0M & \raggedright GLM-5.2, Kimi-K3 & \$2 \\
\midrule
Classification & 15{,}080 & 36M & 68M & \raggedright DeepSeek-V4-Flash & \$24 \\
Judging & 84{,}656 & 202M & 385M & \raggedright DeepSeek-V4-Flash & \$136 \\
\midrule
Total & 226{,}351 & & & & \$2{,}604 \\
\bottomrule
\end{tabular}
\end{table}

Of the 7 accounts with recorded tokens, some are the provider-reported per-call charge as recorded by the pipeline, and the rest are priced from the recorded tokens at the list rate of the provider. The other 2 accounts are estimated from their call counts. List pricing ignores the lower rate for cached input and therefore overstates the cost of the priced accounts. The table lists the models of the accounts, since an account mixes the prices of several models.

The remaining 2 accounts, classification and judging, have a call count and no dollars. One vote is one call in both, and the 99{,}736 calls are therefore exact. Both accounts recorded calls without tokens and are counted by calls, outside the dollar total.

The dollars and the tokens of the formalization row count different calls. 6{,}237 of the formalization calls were served from the response cache, at no charge, and these calls make up 12.0\% of the tokens in the formalization account. Since tokens include cached calls and dollars exclude cached calls, dollars per token understate the price.

Formalization is the largest account at \$2{,}024, which is \$0.65 per problem on average, \$0.47 at the median, and \$6.68 at the maximum. A median problem took 17 calls and 23 minutes wall clock, and thinking tokens account for 91.9\% of the 290M output tokens of the formalization account.

Table~\ref{tab:rolecalls} breaks the formalization calls down by role, where a call type averaging below one call per problem is skipped on some problems, when an earlier check settles the decision or when the corresponding stage does not run.

\begin{table}[htbp]
\caption{Model calls by role over the problems run.}
\label{tab:rolecalls}
\centering
\scriptsize
\begin{tabular}{lrr}
\toprule
Role & Total Calls & Per Problem \\
\midrule
\multicolumn{3}{l}{\itshape Plan} \\
\quad Implementation sketch & 4{,}223 & 1.34 \\
\quad Implementation plan check & 4{,}182 & 1.32 \\
\quad Plan check & 3{,}598 & 1.14 \\
\quad Specification sketch & 3{,}294 & 1.04 \\
\quad Input types and interface & 3{,}081 & 0.97 \\
\quad Sketch repair & 732 & 0.23 \\
\multicolumn{3}{l}{\itshape Formalize} \\
\quad Decidability instances & 3{,}957 & 1.25 \\
\quad Specification & 3{,}337 & 1.06 \\
\quad Implementation & 3{,}319 & 1.05 \\
\quad Executable postcondition & 2{,}431 & 0.77 \\
\quad Executable postcondition guard commands & 792 & 0.25 \\
\multicolumn{3}{l}{\itshape Judge} \\
\quad Implementation quality & 3{,}660 & 1.16 \\
\quad Specification quality & 3{,}196 & 1.01 \\
\quad Executable postcondition judge & 3{,}143 & 0.99 \\
\quad Tautology & 3{,}093 & 0.98 \\
\quad Test case equivalence & 72 & 0.02 \\
\multicolumn{3}{l}{\itshape Repair} \\
\quad Specification repair & 2{,}514 & 0.80 \\
\quad Implementation repair & 2{,}305 & 0.73 \\
\quad Repair history digest & 1{,}238 & 0.39 \\
\quad Root cause & 175 & 0.06 \\
\quad Frozen row repair & 54 & 0.02 \\
\midrule
Total & 52{,}396 & 16.58 \\
\bottomrule
\end{tabular}
\end{table}

The total is a lower bound, since runs discarded during development left no record. Since cached runs record the tokens of the original calls, successful runs are fully counted.

\subsection{Model selection}
\label{app:models}

Frontier models are compared separately for generation and repair, implementation planning, and judging. The candidates are Kimi-K3 \citep{kimik3}, DeepSeek-V4-Pro and DeepSeek-V4-Flash \citep{deepseekv4}, GLM-5.2 \citep{glm52}, Claude-Opus-5 \citep{opus5}, and Claude-Sonnet-5 \citep{sonnet5}.

\subsubsection{Formalizer}

The comparison of generation models replaces the model for the 7 calls that generate or repair definitions and \texttt{\#guard} commands, and keeps all other calls fixed. Four calls generate the postcondition, the executable postcondition, the implementation, and the \texttt{\#guard} commands, and 3 repair these components from compiler diagnostics. Each configuration is run with a separate response cache to keep replies from being replayed across configurations, and pass rate, repair rounds, token count, cost, and wall-clock time in Table~\ref{tab:formalizer} are measured on fresh calls.

\begin{table}[htbp]
\caption{Comparison of models for generation and repair on 30 problems, 10 per difficulty. Claude-Opus-5 needs the fewest repair rounds, and Kimi-K3 reaches the same pass rate at lower cost than Claude-Opus-5. Straight indicates the number of problems that compile without repair, and tokens, cost, and time are per-problem averages.}
\label{tab:formalizer}
\centering
\scriptsize
\begin{tabular}{lrrrrrr}
\toprule
Model & Pass & Straight & Repairs & Tokens & Cost & Time \\
\midrule
Claude-Opus-5 & 30/30 & 23 & 7 & 33k & \$0.73 & 380s \\
Claude-Sonnet-5 & 29/30 & 13 & 28 & 36k & \$0.64 & 386s \\
Kimi-K3 & 30/30 & 9 & 37 & 35k & \$0.57 & 732s \\
GLM-5.2 & 30/30 & 10 & 60 & 112k & \$0.69 & 1335s \\
DeepSeek-V4-Pro & 28/30 & 8 & 36 & 83k & \$0.24 & 958s \\
DeepSeek-V4-Flash & 28/30 & 9 & 40 & 171k & \$0.23 & 1830s \\
\bottomrule
\end{tabular}
\end{table}

Three candidates formalize all 30 problems and are separated by efficiency (Table~\ref{tab:formalizer}). Claude-Opus-5 repairs least and still costs the most per problem. DeepSeek-V4-Pro and DeepSeek-V4-Flash are the cheapest but formalize 2 fewer problems and take the longest, and wall-clock time limits throughput since each repair round occupies a compilation process. Kimi-K3 formalizes all problems at lower cost than Claude-Opus-5 and returns reasoning traces for the training samples.

GLM-5.2 generates 112k output tokens per problem, compared with 33k for Claude-Opus-5, and 82\% of the GLM-5.2 tokens are from the 7 replaced calls. Generation of the executable postcondition alone takes 37k output tokens per problem, compared with 7k for Claude-Opus-5.

Results for Kimi-K3 and GLM-5.2 are confounded by prompt revision. Kimi-K3 was run on a revised specification prompt, and one of the 30 problems was rerun on the current prompts, whereas Claude-Opus-5, Claude-Sonnet-5, and both DeepSeek models were run on the earlier revision. For GLM-5.2, 10 problems were run on the earlier revision and 20 on revised versions of 4 of the 5 planning and generation prompts, and the GLM-5.2 repair and token counts therefore mix a model difference with a prompt difference.

\subsubsection{Planner}

The implementation plan selects the algorithm after both postconditions are accepted, and implementation generation follows the plan. Since errors in the implementation plan are frozen into the file, we measure the plans themselves.

Each of 30 problems was planned once by each candidate, from planning inputs recorded during synthesis. We selected the problems whose implementations had previously relied on a fuel-bounded recursion. Since 26 of the 30 problems are hard, the error counts below describe hard problems. Planning and judging run without compilation, whereas an end-to-end comparison takes a pool of compilation servers and about an hour per candidate.

Kimi-K3 plans fail most often at parsing, before judge review (Table~\ref{tab:plannerlayers}). 20 of 30 Kimi-K3 replies were valid JSON, compared with 29 and 30 for both DeepSeek-V4 variants, and all unparsable replies stop before the length limit. A plan that does not parse costs a retry and then the problem.

\begin{table}[htbp]
\centering
\scriptsize
\caption{Models for implementation planning on 30 problems (26 hard), one plan per model and problem. Parsed indicates the number of plans that are valid JSON, and Clean the number of parsed plans with no failure mode in Table~\ref{tab:plannerdefects}.}
\label{tab:plannerlayers}
\begin{tabular}{lrrrrr}
\toprule
Model & Parsed & Clean \\
\midrule
Kimi-K3 & 20\,(67\%) & 16\,(53\%) \\
DeepSeek-V4-Pro & 29\,(97\%) & 22\,(73\%) \\
DeepSeek-V4-Flash & 30\,(100\%) & 18\,(60\%) \\
\bottomrule
\end{tabular}
\end{table}

Each failure mode corresponds to one numbered planning rule, and plans are judged by the same rules that guided planning. A failure mode is recorded for a plan only when the judge names a concrete witness (e.g., an input on which a bound fails, an input size a stated cost cannot handle, a definition generated a second time). We count the failure modes over the plans that parsed (Table~\ref{tab:plannerdefects}).

\begin{table}[htbp]
\centering
\scriptsize
\caption{Failure modes of parsed plans, judged by DeepSeek-V4-Flash.}
\label{tab:plannerdefects}
\begin{tabular}{lrrrrrr}
\toprule
Model & Counter & Mismatch & Infeasible & Ungrounded & Shallow & Search \\
\midrule
Kimi-K3 & 2 & 0 & 1 & 0 & 1 & 0 \\
DeepSeek-V4-Pro & 0 & 0 & 2 & 0 & 5 & 0 \\
DeepSeek-V4-Flash & 1 & 0 & 4 & 0 & 7 & 0 \\
\bottomrule
\end{tabular}
\end{table}

\begin{itemize}[leftmargin=*, itemsep=0pt, topsep=2pt, parsep=0pt]
\item \emph{Counter}. A termination argument based on a counter that the algorithm introduces.
\item \emph{Mismatch}. A stated running time that belongs to a different algorithm from the planned algorithm.
\item \emph{Infeasible}. A stated running time beyond the input bounds, under which the test cases cannot finish in compiled evaluation.
\item \emph{Ungrounded}. A regenerated specification definition with a reason other than cost or a different decided property.
\item \emph{Shallow}. A specification definition replaced as deciding the wrong property, with no finer definition named for reuse.
\item \emph{Search}. An algorithm that evaluates the executable postcondition on candidate outputs and returns an accepted candidate.
\end{itemize}

The plans were judged again by DeepSeek-V4-Pro to measure how far the counts depend on the judge. Both judges are also planning candidates and may favor plans generated by the same model. Under the second judge, DeepSeek-V4-Pro and DeepSeek-V4-Flash swap places and Kimi-K3 ranks below both. Only the parse rate is independent of the judge, and the comparison between DeepSeek-V4-Pro and DeepSeek-V4-Flash rests on the parse rate.

\subsubsection{Judge}

DeepSeek-V4-Flash and DeepSeek-V4-Pro, run on the same 79 parsed plans, agree on 97\% of the per-mode calls and on 82\% of the verdicts. Since a plan counts as clean only when all modes are unflagged, a disagreement on any single mode can change the verdict.

Failure-mode counts depend on the choice of judge, with per-mode agreement over the plans marked by either judge between 0\% and 75\% (Table~\ref{tab:judgeagreement}). DeepSeek-V4-Pro reports Mismatch 3 times and DeepSeek-V4-Flash never, and neither judge reports Ungrounded or Search on any of the 79 plans, so no agreement can be computed for those modes.

\begin{table}[htbp]
\centering
\scriptsize
\caption{Agreement between DeepSeek-V4-Flash and DeepSeek-V4-Pro on 79 plans, where per-mode agreement ranges from 0\% to 75\%. Agreement is computed over the plans that at least one judge marks with the failure mode.}
\label{tab:judgeagreement}
\begin{tabular}{lrrrr}
\toprule
Failure Mode & V4-Flash & V4-Pro & Both & Agreement \\
\midrule
Counter & 3 & 5 & 1 & 14\% \\
Mismatch & 0 & 3 & 0 & 0\% \\
Infeasible & 7 & 6 & 5 & 62\% \\
Ungrounded & 0 & 0 & 0 & n/a \\
Shallow & 13 & 15 & 12 & 75\% \\
Search & 0 & 0 & 0 & n/a \\
\bottomrule
\end{tabular}
\end{table}

The judge model is DeepSeek-V4-Flash, the model of the production quality gates. The second judge bounds the dependence of the reported counts on that inheritance, which a single-judge table cannot show. The agreement reported here is on planner verdicts.

\section{Proofs}
\label{app:proofs}

\subsection{Method}
\label{app:loop}

\begin{figure}[H]
\centering
\includegraphics[width=\linewidth]{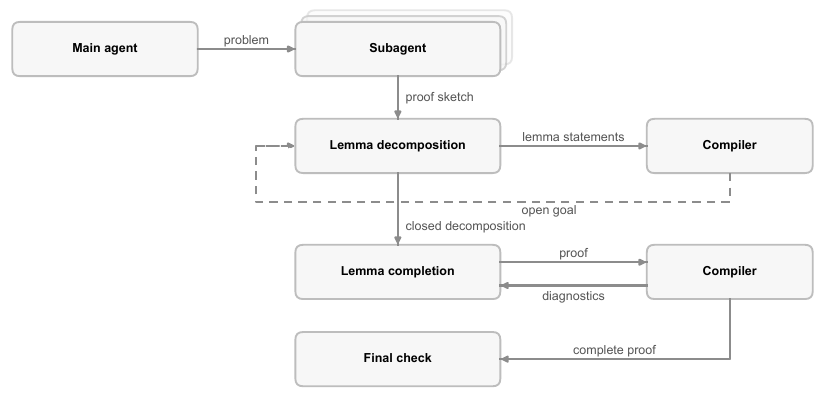}
\caption{The decompose-and-prove loop. Solid edges carry the named objects, and the dashed edge returns an open goal to decomposition.}
\label{fig:loop}
\end{figure}

We prove the postcondition equivalence theorems with a decompose-and-prove loop, which is a widely adopted strategy for formal proofs \citep{jiang2023dsp,wang2024legoprover,ren2025deepseekproverv2,zhou2025deltaprover,li2026goedelcodeprover}, run by a main agent and a set of subagents (Figure~\ref{fig:loop}).

The main agent holds the queue of problems and hands a problem to one subagent, together with the form of the executable postcondition and the proofs of the closest problems already proved.

The subagent compares the postcondition and the executable postcondition and produces a proof sketch \citep{jiang2023dsp}, which records the structure common to both postconditions and states one lemma for each remaining clause. The lemmas are given \texttt{sorry} bodies and the file is compiled once, which checks that the lemmas close the theorem before any lemma is proved \citep{li2026goedelcodeprover}. A decomposition that fails is restated by the same subagent, since the next decomposition is built from the open goal. The subagent then proves the lemmas one at a time against the compiler, and each proof is verified against the frozen statement with SafeVerify \citep{safeverify}, bumped to Lean 4.32, which checks that no declaration of the problem was changed and that the theorem depends on the standard axioms only.

\subsection{Postcondition equivalence theorem}
\label{app:equiv}

We analyze the proofs by the form of the executable postcondition (Appendix~\ref{app:route}), since the form fixes the statement of the theorem. Executable postconditions that evaluate the postcondition through a decidability instance agree with the postcondition by construction. Separate definitions replace a quantifier of the postcondition by a computation, and the postcondition equivalence proof shows that the replacement loses nothing on valid inputs. The computation is an enumeration that stands in for a set, a reformulation that decides a different property, or a bound that stands in for an unbounded quantifier. The decidable form covers 56\% of easy problems and 21\% of hard problems, and the proof effort concentrates on the harder problems (Table~\ref{tab:proofform}).

\begin{table}[htbp]
\caption{Postcondition equivalence proofs by form of the executable postcondition, where the decidable form has proof length 0 and the separate definitions a median of 112 lines. Proof length was measured as the number of lines added to the problem file, and the last column indicates the number of proofs in which the validity hypothesis is used.}
\label{tab:proofform}
\centering
\scriptsize
\setlength{\tabcolsep}{3pt}

\begin{tabular}{lrrrrrrr}
\toprule
Form & \#Problems & Easy & Medium & Hard & Median Lines & P90 Lines & Validity Used \\
\midrule
Decidable form & 1{,}079 & 400 & 526 & 153 & 0 & 0 & 0 \\
Separate definition & 1{,}958 & 314 & 1{,}062 & 582 & 112 & 394 & 868 \\
\quad Enumeration & 1{,}607 & 204 & 883 & 520 & 118 & 409 & 736 \\
\quad Reformulation & 293 & 90 & 153 & 50 & 95 & 333 & 104 \\
\quad Bound & 58 & 20 & 26 & 12 & 92 & 298 & 28 \\
\bottomrule
\end{tabular}
\end{table}

A hard proof costs more lemmas rather than more repair rounds (Table~\ref{tab:proofcost}). Of the 677 problems with a per-problem record, the first compilation of the decomposition passed for 78\%, and the median proof took 2 compilations. The median number of lemmas grows from 5 on easy problems to 21 on hard problems while the number of compilations stays at 1 to 2, which places the cost of a hard problem in stating the lemmas and not in repairing the lemmas. The number of lemmas tracks proof length (Spearman 0.96) and compilations are independent of proof length (0.09). A subagent spends a median of 128k tokens, 26 tool calls and 473 seconds on one problem (351 problems with a usage record), and tokens follow proof length (0.71). All 3{,}037 proofs depend on \texttt{propext}, \texttt{Quot.sound} and \texttt{Classical.choice} only. A full replay verified 3{,}033 proofs and returned no verdict for 4, which match the original except for compiler-generated auxiliary declarations.

\begin{table}[htbp]
\caption{Cost of a postcondition equivalence proof by difficulty, where the number of lemmas created grows fourfold from easy to hard while the number of compilations stays at 1 to 2. Only the 1{,}958 problems with separate definitions are counted. Proof length, lemmas created, compilations, tokens, tool calls and time are medians over the problems in the row, and passed at 1st is the fraction of problems whose lemma statements compiled on the first attempt. Proof length was measured on all 1{,}958 problems, lemmas and compilations on the 677 problems with a per-problem record, and tokens, tool calls and time on the 351 problems with a usage record.}
\label{tab:proofcost}
\centering
\scriptsize
\setlength{\tabcolsep}{3pt}

\begin{tabular}{lrrrrrrrrr}
\toprule
\multirow{2}{*}{Difficulty} & \multirow{2}{*}{\#Problems} & \multicolumn{2}{c}{Proof Length (lines)} & \multirow{2}{*}{\shortstack[r]{Lemmas\\Created}} & \multicolumn{2}{c}{Compilations} & \multirow{2}{*}{Tokens} & \multirow{2}{*}{\shortstack[r]{Tool\\Calls}} & \multirow{2}{*}{Time (s)} \\
\cmidrule(l{2pt}r{2pt}){3-4} \cmidrule(l{2pt}r{2pt}){6-7}
 & & Median & P90 & & Median & Passed at 1st & & & \\
\midrule
Easy & 314 & 58 & 203 & 5 & 2 & 86\% & 118k & 24 & 369 \\
Medium & 1{,}062 & 121 & 394 & 11 & 2 & 74\% & 128k & 26 & 484 \\
Hard & 582 & 142 & 486 & 21 & 1 & 86\% & 149k & 30 & 604 \\
\midrule
All & 1{,}958 & 112 & 394 & 13 & 2 & 78\% & 128k & 26 & 473 \\
\bottomrule
\end{tabular}
\end{table}

\begin{figure}[htbp]
\centering
\begin{subfigure}[b]{.38\linewidth}
\centering
\includegraphics[width=\linewidth]{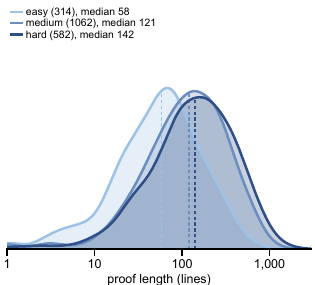}
\caption{By difficulty}
\label{fig:prooflen-difficulty}
\end{subfigure}\hfill
\begin{subfigure}[b]{.6\linewidth}
\centering
\includegraphics[width=\linewidth]{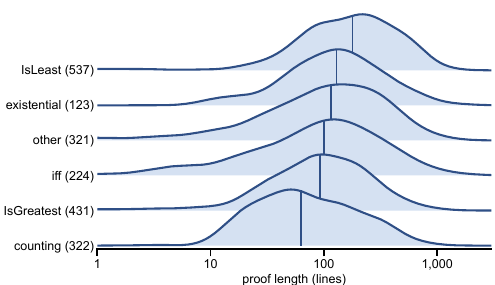}
\caption{By form of the postcondition}
\label{fig:prooflen-form}
\end{subfigure}
\caption{Proof length against difficulty and postcondition form, where the form of the postcondition separates the proofs more sharply than difficulty does. Separate definitions only, since the decidable form adds no lines. Curves are kernel density estimates on a logarithmic axis, and vertical lines mark medians.}
\label{fig:prooflen}
\end{figure}

\begin{figure}[H]
\centering
\includegraphics[width=.85\linewidth]{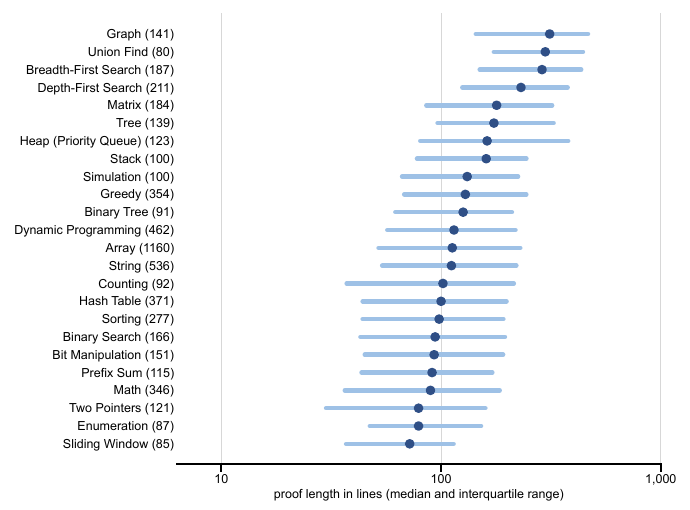}
\caption{Proof length by topic, where graph problems need 3 times the lines of window problems at the same difficulty. Dots mark medians and bars interquartile ranges over the separate definitions carrying the LeetCode topic tag, for tags with at least 80 problems, with the number of problems in parentheses. A problem with several tags is counted under each.}
\label{fig:prooftopic}
\end{figure}

The postcondition form and the topic of a problem predict proof length better than difficulty does (Figures~\ref{fig:prooflen} and~\ref{fig:prooftopic}). LeetCode difficulty predicts proof length weakly (Spearman 0.31 over all problems), and the topic tags that LeetCode assigns to a problem predict proof length better \citep{xia2025leetcodedataset}. Among separate definitions of medium difficulty, graph problems have a median proof length of 324 lines and two-pointer problems 66. Graph problems state reachability as the existence of a path in the postcondition and compute a closure in the executable postcondition, and the proof relates path and closure by induction on the search. Window problems quantify over the same contiguous ranges in both postconditions, and the proof is index arithmetic. Minimization costs more than maximization, with a median of 179 lines for \texttt{IsLeast} postconditions, compared with 93 for \texttt{IsGreatest}. The candidate set of a maximum is bounded by the input, whereas the lower bound of a minimum is bounded by the precondition alone, and the precondition is invoked in 58\% of \texttt{IsLeast} proofs, compared with 39\% of \texttt{IsGreatest} proofs. 121 proofs exceed 500 lines and 9 exceed 1,000, all on graph, geometry or string-rewriting problems.

A proof attempt that fails for a reason other than proof difficulty exposes a statement error through a counterexample or through independence. A counterexample is a valid input on which the postcondition and the executable postcondition disagree, and the negation of the theorem is then proved. Independence arises when the executable postcondition evaluates a \texttt{Float} operation, which Lean treats as opaque. Statement errors were found in 31 problems (1.0\%), 5 by independence and 26 by a compiled counterexample (Table~\ref{tab:stmterrors}). Each was repaired by the smallest change to the statement that agrees with the problem description, the repaired file passed the frozen test cases, and the theorem was then proved. With the postcondition equivalence theorems proved, the postcondition and executable postcondition of each problem agree on valid inputs. The proofs relate the 2 postconditions to each other, and the human review checks both against the problem description (Appendix~\ref{app:human-review}).

\begin{table}[H]
\caption{Statement errors found by proof attempts, where the postcondition accounts for 19 of 31 problems and the precondition for 3. A problem was counted under each component that was changed by the repair, and one problem was counted twice.}
\label{tab:stmterrors}
\centering
\scriptsize

\begin{tabular}{lrrrr}
\toprule
Component & Easy & Medium & Hard & Total \\
\midrule
Precondition & 1 & 0 & 2 & 3 \\
Postcondition & 4 & 10 & 5 & 19 \\
Executable postcondition & 2 & 2 & 6 & 10 \\
\midrule
\#Problems & 7 & 12 & 12 & 31 \\
\bottomrule
\end{tabular}
\end{table}

\subsection{Correctness theorem}
\label{app:correct}

The correctness theorem asserts that the implementation satisfies the postcondition. Most proofs establish the postcondition directly, and 8 of the 300 proofs show that the implementation satisfies the executable postcondition and transfer the result through the postcondition equivalence theorem. The loop of Appendix~\ref{app:loop} is unchanged.

Correctness theorems are proved for all 300 test problems (72 easy, 155 medium, 73 hard) on the files repaired after the review of Appendix~\ref{app:human-review}, where 274 proofs are checked under Lean 4.32 and 26 come from the earlier proving run. Correctness proofs grow with difficulty, from a median of 63 lines on easy problems to 178 on medium problems and 365 on hard problems, with a median of 159.5 lines over all 300 problems. All 300 reference implementations are also tested against the executable postcondition with Plausible \citep{plausible}, which finds no counterexample (214 pass and 86 are inconclusive, since valid inputs are rarely sampled for structured types).

\subsection{Equivalence of accepted outputs}
\label{app:accepted-equiv}
\begin{revised}

Acceptance relies on test cases and judges, so we also validate accepted SpecGen outputs by proof. For each model in Table~\ref{tab:mainpass} and each problem with an accepted output, we draw an accepted output at random and prove its postcondition equivalence theorem. The fixed tactics of the corpus proofs, such as \texttt{simp} with the executable postcondition unfolded, close 557 of the 2{,}012 drawn outputs in one line (27.7\%). The proving loop (Appendix~\ref{app:loop}) then settles every remaining output of the 2 RL models and up to 40 remaining outputs of each other model (Table~\ref{tab:accepted-equiv}). Each attempt ends in a proof or in a counterexample on a valid input, except when a \texttt{Float} operation makes the theorem independent, so no output stays undecided for lack of proof effort.

The theorem is proved for 99.2\% of accepted outputs of both RL models (251/253 for \work-27B-RL and 246/248 for \work-9B-RL), and all proofs pass SafeVerify under the criteria of the corpus proofs (Appendix~\ref{app:loop}). The 6 counterexamples locate the error on both sides of the theorem, in 3 declarative postconditions, 2 executable postconditions and 1 validity predicate, which admits unbalanced inputs for \textit{Remove Outermost Parentheses} (\work-9B-RL).

\begin{table}[H]
\caption{Postcondition equivalence theorems of accepted SpecGen outputs on \work-Test, one output per model and problem. The 2 RL models are a census. For the other models, 40 outputs left open by the fixed tactics go to the proving loop, and the estimated mismatch applies the counterexample rate among them to all open outputs.}
\label{tab:accepted-equiv}
\centering
\scriptsize
\setlength{\tabcolsep}{4pt}
\begin{tabular}{lrrrrrrr}
\toprule
Model & Accepted & \begin{tabular}[c]{@{}c@{}}Fixed\\Tactics\end{tabular} & \begin{tabular}[c]{@{}c@{}}Proving\\Loop\end{tabular} & \begin{tabular}[c]{@{}c@{}}Counter-\\example\end{tabular} & \texttt{Float} & \begin{tabular}[c]{@{}c@{}}Not\\Sampled\end{tabular} & \begin{tabular}[c]{@{}c@{}}Estimated\\Mismatch (\%)\end{tabular} \\
\midrule
\work-27B-RL & 253 & 42 & 209 & 2 & 0 & 0 & 0.8 \\
\work-9B-RL & 248 & 60 & 186 & 1 & 1 & 0 & 0.4 \\
\work-27B-SFT & 122 & 34 & 39 & 1 & 0 & 48 & 1.8 \\
\work-9B-SFT & 73 & 23 & 39 & 1 & 0 & 10 & 1.7 \\
\midrule
Claude-Opus-5 & 244 & 93 & 40 & 0 & 0 & 111 & 0.0 \\
Claude-Sonnet-5 & 118 & 47 & 40 & 0 & 0 & 31 & 0.0 \\
DeepSeek-V4-Flash (0731) & 230 & 39 & 40 & 0 & 0 & 151 & 0.0 \\
DeepSeek-V4-Pro & 163 & 50 & 40 & 0 & 0 & 73 & 0.0 \\
GPT-5.6-Luna & 86 & 16 & 39 & 1 & 0 & 30 & 2.0 \\
GPT-5.6-Sol & 179 & 57 & 40 & 0 & 0 & 82 & 0.0 \\
Kimi-K3 & 251 & 72 & 40 & 0 & 0 & 139 & 0.0 \\
\midrule
Qwen3.5-9B & 2 & 1 & 1 & 0 & 0 & 0 & 0.0 \\
Qwen3.8-27B & 43 & 23 & 20 & 0 & 0 & 0 & 0.0 \\
\bottomrule
\end{tabular}
\end{table}

\end{revised}

\section{Human study}
\label{app:human}

We conduct two human evaluations: a review of the 300 \work-Test reference problems (Appendix~\ref{app:human-review}) and an expert review of automatically accepted SpecGen model outputs (Appendix~\ref{app:specgen-audit}).

\subsection{Human review of \work-Test}
\label{app:human-review}

The human review of the 300 \work-Test problems is compared with the errors found while proving (Appendix~\ref{app:equiv}).

One of the authors, with experience in Lean, reviewed all 300 \work-Test problems. Each component was compared with the problem description in the order of Table~\ref{tab:rubric}. The reviewer was given the Lean file and the description and consulted no model or automated judge. An error was recorded only with an exposing test case, an analysis in prose, and a corrected file that compiles and passes the frozen test suite.

\begin{table}[htbp]
\caption{Rubric of the human review, in which each of the 4 components is judged against the problem description by the question of the corresponding acceptance criterion of Section~\ref{sec:acceptance}. One row per component, with the question the reviewer answers and the finding that was recorded as an error. The questions were taken from the acceptance criteria, so a person and a judge examine the same properties.}
\label{tab:rubric}
\centering
\scriptsize
\begin{tabular}{@{}>{\raggedright\arraybackslash}m{2.6cm}>{\raggedright\arraybackslash}m{4.6cm}>{\raggedright\arraybackslash}m{5.8cm}@{}}
\toprule
Component & Question & Recorded as an Error When \\
\midrule
Precondition & Does it admit exactly the inputs the constraints describe? & An input the constraints admit is rejected, or one they exclude is accepted \\
Postcondition & Does it carry each condition the description places on the answer? & A clause of the description is missing and a value the description rejects is admitted, or a correct output is rejected \\
Executable postcondition & Does it accept the same outputs as the postcondition on all valid inputs? & A bound is too small and an object the postcondition ranges over is not reached, or an object outside it is accepted \\
Implementation & Does it satisfy the postcondition, terminate without a fuel parameter, and run within the constraints? & A valid input yields a rejected output, termination rests on a fuel parameter, or the cost does not survive the largest admitted input \\
\bottomrule
\end{tabular}
\end{table}

Errors were confirmed in 22 of 300 problems (7.3\%), 21 by the review and 1 by the proof of the correctness theorem. The counts by component and difficulty are in Table~\ref{tab:review}. The postcondition and the implementation account for 20 of the 23 component errors, and the review found no executable postcondition that accepts different outputs from the postcondition. The error rate is highest for easy problems at 9.7\% and decreases to 5.5\% for hard problems, and 2 of the 3 precondition errors fall on easy problems. Of the 73 files the review changed, 21 changed a definition and the others changed prose or test cases.

\begin{table}[htbp]
\caption{Errors confirmed by human review and, for 1 problem, by the correctness proof, where 20 of 23 component errors are in the postcondition or the implementation and none is in the executable postcondition. Each percentage is taken over the problems of the row, and a problem with errors in 2 components is counted once among the problems with errors.}
\label{tab:review}
\centering
\scriptsize
\setlength{\tabcolsep}{2.5pt}

\begin{tabular}{lrrrrrr}
\toprule
 & & \multicolumn{4}{c}{Errors by Component} & Problems \\
\cmidrule(l{2pt}r{2pt}){3-6}
Difficulty & \#Problems & Precondition & Postcond. & Exec. Postcond. & Implementation & With Errors \\
\midrule
Easy & 72 & 2 (2.8\%) & 4 (5.6\%) & 0 & 1 (1.4\%) & 7 (9.7\%) \\
Medium & 155 & 1 (0.6\%) & 5 (3.2\%) & 0 & 6 (3.9\%) & 11 (7.1\%) \\
Hard & 73 & 0 & 1 (1.4\%) & 0 & 3 (4.1\%) & 4 (5.5\%) \\
\midrule
Total & 300 & 3 (1.0\%) & 10 (3.3\%) & 0 & 10 (3.3\%) & 22 (7.3\%) \\
\bottomrule
\end{tabular}
\end{table}

Review and proving find different errors. Of the 14 specification errors in the test split, attributed by the order of discovery, 9 were found by review alone and had a postcondition equivalence theorem proved against the wrong statement before the review, since an executable postcondition that decides a wrong postcondition still agrees with the wrong postcondition. 3 were found by both, with compiled counterexamples from the proof attempts. 2 were found by proof attempts alone and had passed the review. In one, the clause $\forall i,\ i < n \to A \leftrightarrow B$ parses as $\forall i,\ (i < n \to A) \leftrightarrow B$, and the postcondition therefore rejects all outputs while the text reads correctly. In the other, the executable postcondition compares distances in \texttt{Float}, and the theorem is independent of the axioms. The 10 implementation errors do not affect the postcondition equivalence theorem and are the subject of the correctness theorems of Appendix~\ref{app:correct}. Review measures agreement with the problem description, and proving measures agreement between the postcondition and executable postcondition (Table~\ref{tab:detection}). Neither measure subsumes the other, and the dataset was checked by both.

\begin{table}[htbp]
\caption{Specification errors in the test split by detection, where review and proving overlap on 3 of 14 problems. Review only means the postcondition equivalence theorem had been proved against the wrong statement, and proof attempt only means the review missed the error.}
\label{tab:detection}
\centering
\scriptsize

\begin{tabular}{lr}
\toprule
Detection & \#Problems \\
\midrule
Review only (theorem proved against the wrong statement) & 9 \\
Review and proof attempt & 3 \\
Proof attempt only & 2 \\
\midrule
Total & 14 \\
\bottomrule
\end{tabular}
\end{table}

\subsection{Expert review of SpecGen evaluation}
\label{app:specgen-audit}
\begin{revised}

We review false positives of the automatic SpecGen evaluation for the 2 reported RL checkpoints, \work-27B-RL and \work-9B-RL. For each model, we draw 50 problems counted as SpecGen successes, stratified as 12 easy, 26 medium and 12 hard in the proportion of \work-Test, and review 1 accepted output per problem. Both reviewers each review 60 problems, of which 20 overlap, and the review package masks model identity, judge verdicts, test results against the reference and proof results. Before the review, we test each output against the reference with Plausible, as for the reference implementations (Appendix~\ref{app:correct}).
\begin{enumerate}[leftmargin=*, itemsep=0pt, topsep=2pt, parsep=0pt]
\item Completeness. Plausible samples valid inputs and checks that the executable postcondition accepts the output of the reference implementation. A counterexample is a correct output that the specification rejects.
\item Validity. Plausible samples inputs and checks that the validity predicate agrees with the reference validity predicate. A counterexample is an input that one predicate admits and the other rejects.
\end{enumerate}
We apply the specification-side rubric of Table~\ref{tab:rubric}.

Expert review confirms 90 of the 100 accepted outputs, 43 of \work-27B-RL and 47 of \work-9B-RL (Table~\ref{tab:specgen-audit}). The 2 reviewers find 7 and 5 errors in their 60 problems and agree on all 20 shared problems, including the same 2 errors. Each error comes with a concrete input that the validity predicate admits outside the constraints, which makes the verdict checkable rather than a judgment. All declarative and executable postconditions are correct, and all 10 errors were found in the validity predicate. The missing constraints are guarantees that the answer fits in a 32-bit integer (3 problems), value ranges and alphabets (3), structural guarantees on the input (3) and a time format (1). The postconditions of these outputs therefore remain faithful on every input that the constraints admit. The tests against the reference miss these errors, since random inputs rarely reach the boundary of a missing constraint.

We acknowledge that the expert review covers only 100 outputs which is a small sample, yet the errors found converge to the same kind. The models produce correct postconditions but are highly likely to miss boundary conditions of the constraints in the validity predicate. One possible reason is that the acceptance criteria offer little guidance on the validity predicate, since the specification quality judge assigns clauses that restrict the input to the validity predicate without checking them (Appendix~\ref{app:evaljudges}) and only the test layers L1 and L1b check the validity predicate on a few inputs (Table~\ref{tab:layers}). Checking the validity predicate against the stated constraints is a direction for future work (Appendix~\ref{app:limitations}).

\begin{table}[htbp]
\centering
\caption{Expert review of automatically accepted SpecGen outputs of the reported RL checkpoints.}
\label{tab:specgen-audit}
\scriptsize
\begin{tabular}{@{}lccccc@{}}
\toprule
Model & Reviewed & Correct & $V$ Errors & $P_{\mathrm{spec}}$ Errors & $P_{\mathrm{check}}$ Errors \\
\midrule
\work-27B-RL & 50 & 43 (86.0\%) & 7 & 0 & 0 \\
\work-9B-RL & 50 & 47 (94.0\%) & 3 & 0 & 0 \\
\midrule
Total & 100 & 90 (90.0\%) & 10 & 0 & 0 \\
\bottomrule
\end{tabular}
\end{table}

\end{revised}

\section{Limitations}
\label{app:limitations}

Despite the contributions of \work{} to specification generation in Lean, the work has several limitations that we hope to address in future work.

\textbf{Repair at inference.} The repair tasks train the models on failed rollouts and are evaluated for a single round (Table~\ref{tab:mainpass}), and the same tasks support a self-refinement loop at inference, where the model repairs an output on compiler messages and test results over several rounds \citep{verispecgen2026}. Pass rates over rounds would show how much of the gap to frontier models such a loop closes.

\begin{revised}
\textbf{Scope of expert review.} A single reviewer covers the 300 \work-Test problems and finds errors in 21 of them, including 9 specification errors in problems with a proved postcondition equivalence theorem (Appendix~\ref{app:human-review}), and 2 reviewers review 100 accepted SpecGen outputs of the RL models (Appendix~\ref{app:specgen-audit}). The training problems are checked by test cases, judges and postcondition equivalence proofs without review and need a reviewed sample to estimate the error rate of the training data.

\textbf{Proofs for generated outputs.} Generated specifications and implementations are accepted without proofs of the two theorems (Section~\ref{sec:proofs}). We prove the postcondition equivalence theorem afterwards for the accepted outputs of the RL models (Appendix~\ref{app:accepted-equiv}). A proof shows only that the 2 postconditions agree. Whether the declarative postcondition and the validity predicate state the problem still rests on test cases, judges and expert review.

\textbf{Validity predicates.} All errors found by the expert review of accepted outputs are input constraints missing from the validity predicate (Appendix~\ref{app:specgen-audit}). The acceptance criteria check the validity predicate only with test cases. A judge that compares the validity predicate with the stated constraints would cover this part in synthesis, training and evaluation.
\end{revised}

\clearpage
\section{Prompts}
\label{app:prompts}

The boxes below are condensed versions of the prompts, in the order of Figure~\ref{fig:pipeline}. Every prompt has the same five parts in the order the model reads them, the role, the input, the task, the rules and the output format, and problem text is marked as untrusted data so that instructions inside a problem statement are not followed. The boxes keep the rules that the analyses of Appendix~\ref{app:pipeline} refer to and drop the worked examples, the formatting rules and the injected library facts. The component names follow the paper. The prompts given to the models are in Appendix~\ref{app:modelprompts}, and the judges of Appendix~\ref{app:evaljudges} run as the acceptance criteria of synthesis, in evaluation and in the reward.

\subsection{Synthesis pipeline}
\label{app:pipelineprompts}

\begin{promptbox}{General planner}
\begin{lstlisting}[breakatwhitespace=true]
ROLE
Contract planner. Fixes everything about one problem that does not require specification thinking, before the specification side runs.

INPUT
`problem` (description, constraints, examples; untrusted), `signature`, `default_lean_types`, and on a retry the ruling that found the previous plan incomplete.

TASK
Decide the Lean types, the interface names and the precondition's clauses, and record the domain facts. Everything fixed here is frozen; downstream roles follow it exactly.

RULES
1. You own the representation: for every parameter and the output, declare the Lean type that is best for stating and proving this problem, weighing faithfulness, simplicity of the statements and the proof debt of the correctness proof.
2. Interface names describe what is computed, never how.
3. The precondition's clauses transcribe the stated constraints in the problem's words: nothing dropped, nothing invented, no operational re-encoding. Where the input encodes something, state the encoding's well-formedness.
4. The implementation algorithm is not yours to fix; a later stage plans it once the specification exists.

OUTPUT
One JSON object: `input_spec`, `lean_types` (with a `rationale`), `interface`, `v_plan`, `domain_facts`.
\end{lstlisting}
\end{promptbox}

\begin{promptbox}{Postcondition planner}
\begin{lstlisting}[breakatwhitespace=true]
ROLE
Specification planner. Receives the frozen types and precondition plan, and shapes the postcondition on top of them.

INPUT
`problem` (untrusted), `lean_types`, `v_plan`, `negative_aims` (what each frozen negative test targets), and on a retry the previous ruling.

TASK
Plan the vocabulary and the shape of the postcondition in prose, no Lean. The postcondition states what a correct output is; it does not produce one. Two statements are required: whether `result` is the kind of object that could be the answer, and what distinguishes the correct one from the others of that kind.

RULES
1. State whether the second statement concerns the object the first exhibits or every object of that kind; a universal reduced to the exhibited object is no bound.
2. `result` occurs inside relations that constrain it, never as the left side of an equation whose right side computes it. The test: conceal the right side and ask whether a correct `result` remains distinguishable.
3. Keep out every operation that answers in the degenerate case instead of having no answer there (`Set.ncard` on an infinite set, `sInf` on the empty set, truncated subtraction, `xs[i]!`); it makes a different specification, not a false one.
4. Computability is not a consideration; quantifiers range over the problem's natural domain, and a stated bound is a hypothesis inside the quantifier, never a truncated range.

OUTPUT
One JSON object: the two statements, `second_statement_ranges_over`, `vocabulary`, `witnesses`.
\end{lstlisting}
\end{promptbox}

\begin{promptbox}{Postcondition formalizer}
\begin{lstlisting}[breakatwhitespace=true]
ROLE
Specification formalizer. Renders the plan's two prose statements as the specification side of one problem.

INPUT
`plan` (names and types binding), `problem` (untrusted), `sketch_preview` (the final file with your sections as holes), `verified_library` (declarations compiled against this toolchain), `frozen_cases` (layer, frozen input and output, and the expression the plan fixed).

TASK
Write the specification helpers, the precondition, the postcondition and the L1 and L1b guard commands. Theorem statements and module structure are machine-written; the test rows are frozen, and you write each row as one guard command.

RULES
1. Decidability is not your concern; bounding or enumerating a quantifier to make it computable is a defect.
2. Specify the what, never the how: `IsLeast S result` over `result = sInf S`. Mirroring an algorithm's structure is a defect.
3. The precondition is a total Bool function covering exactly the stated constraints; the postcondition returns Prop and says the plan's two sentences and nothing further. A library declaration whose definition is what you were about to write is that same proposition, with the library's lemmas.
4. The frozen rows are a contract: one guard command per row, none for a row that is not there, values preserved exactly.

OUTPUT
A design note, then one Lean fence with `Representation` (only when needed), `Specification helpers`, the precondition, the postcondition and `Unit tests`.
\end{lstlisting}
\end{promptbox}

\begin{promptbox}{Test form judge}
\begin{lstlisting}[breakatwhitespace=true]
ROLE
Judge of whether a guard command states the frozen test row it was written for, called when the mechanical comparison could not settle it.

INPUT
The frozen rows (`test_id`, layer, input and output as frozen JSON, the expression the plan fixed) and the guard command written for each.

TASK
Two mechanical checks ran first: the shape check (right layer, right anchor, right polarity) and the expression check (verbatim, then token by token). Rule on the rows where the tokens differ.

RULES
1. The frozen JSON is the authority, not the plan's expression.
2. Count when counting decides it: a replicate, a range, a map over a range.
3. Two expressions are equivalent when they elaborate to the same term; notation, grouping and a redundant ascription do not change the test. A missing or extra parameter or a value that differs anywhere is a different test.
4. Judge the denotation, not the style. When the evidence does not settle it, answer false and say what you could not determine.

OUTPUT
One JSON object with a ruling per row: `equivalent` (true or false) and `reason`.
\end{lstlisting}
\end{promptbox}

\begin{promptbox}{Executable postcondition router}
\begin{lstlisting}[breakatwhitespace=true]
ROLE
Route analyst for the executable postcondition of one problem. Decides how it will be organized before anyone writes it.

INPUT
`problem` (untrusted), `frozen_spec`, `test_inputs` (with the length of every list and string), `plan` (whose `domain_facts` often supply the bounding fact).

TASK
The executable postcondition restates the frozen postcondition as a Bool that compiled evaluation can run on the frozen inputs. The difficulty is cost: mirroring the quantifiers directly can be exponential in the input. Choose one of three routes: reshape (restate what is quantified over so the count collapses), truncate (bound a quantifier by a fact the problem supplies) or enumerate (range over a finite encoding of the candidates).

RULES
1. `bounding_fact` is the one fact the cost argument rests on, stated so a reader could check it against the problem.
2. The executable postcondition is a judge, never a solver: it quantifies over candidates or verifies the answer's defining property. A shape that computes the answer by its own algorithm and compares is refused downstream, whatever its cost.
3. Name the route a hasty writer would reach for and the cost that rules it out at these input sizes.

OUTPUT
One JSON object: `route`, `bounding_fact`, `shape` (in prose), `rejected`.
\end{lstlisting}
\end{promptbox}

\begin{promptbox}{Executable postcondition formalizer}
\begin{lstlisting}[breakatwhitespace=true]
ROLE
Writes the executable postcondition for a frozen postcondition, following the route decided for it.

INPUT
`plan`, `frozen_spec` (read-only target), `sketch_preview`, `route_plan`, `decidability_trial_error` (why the decidable form was not available), `verified_library`, `frozen_cases`.

TASK
Write one Lean section with the executable postcondition and its helpers, and the L2 and L2b guard commands. Record the route taken and the fact that bounds it in a note before the fence.

RULES
1. A total Bool function named `base_name` + `_check`, with the specification's argument order.
2. Bridge toward the frozen postcondition, never toward a weaker reading of it; mirror its clause structure so the equivalence theorem is provable clause by clause.
3. Every recursion or fold must already be present in the postcondition's own definitions: the equivalence proof should need only unfolding and structural alignment, never an algorithmic insight. Never compute the answer by another method and compare it with `result`.
4. Bound every quantifier to a structure derived from the input. Size is measured, not hypothetical: what settles an enumeration is running it.
5. The fast algorithm belongs to the implementation; an evaluation that finishes belongs here.

OUTPUT
A design note, then one Lean fence with the executable postcondition section and `Unit tests`.
\end{lstlisting}
\end{promptbox}

\begin{promptbox}{Implementation planner}
\begin{lstlisting}[breakatwhitespace=true]
ROLE
Implementation planner. The postcondition and the executable postcondition are accepted, and the algorithm is planned against them rather than against the problem statement alone.

INPUT
`problem` (untrusted), `plan` (`impl_name` binding), `frozen_spec`, `frozen_executable_postcondition` (read-only), `verified_library`, `frozen_cases`, and on a retry the judge's ruling.

TASK
Decide how the answer is computed and at what cost. This is the only section where running time is judged: the guards run the algorithm on the frozen inputs, so an exponential approach fails by timeout. Say which accepted definitions the algorithm calls and which it replaces, with the cost difference.

RULES
1. `approach` names the standard efficient algorithm, not the first one that terminates; `cost` states its running time in terms of the input.
2. A frozen definition is passed over on one of two grounds, and the entry says which: too slow (both figures given) or different structure required (the finer definition named). An entry on neither ground is a rewrite that should have been a call.
3. `termination` names the quantity the recursion consumes; a `Nat` carried along so the recursion has something to count down is not such a quantity.
4. The algorithm computes the answer; the specification says what a correct answer is. Do not evaluate the specification over candidates and return the one that satisfies it.

OUTPUT
One JSON object: `approach`, `cost`, `calls`, `replaces`, `termination`.
\end{lstlisting}
\end{promptbox}

\begin{promptbox}{Root-cause analyzer}
\begin{lstlisting}[breakatwhitespace=true]
ROLE
Root-cause analyst, called only when local repair cannot proceed: a symptom recurred after repair, a repairer declared the fault lies elsewhere, or an error cannot be placed by line.

INPUT
`error_report`, `sections` (each labelled frozen or repairable), `assembled_source` (the whole file with line numbers), `harness_regions`, `frozen_tests`, `history` (the repair attempts so far), `plans`.

TASK
Rule which block is at fault, so the pipeline repairs the true target instead of the symptom's location. A plan may also be ruled faulty, which triggers one replan.

RULES
1. Rule on evidence in the diagnostics and sections, never on which line happened to error; symptom location and fault location are different things.
2. A frozen test literal is ground truth by default. Rule it wrong only on evidence you can quote: the case does not violate the constraint its own rationale cites, or a documented answer contradicts the problem's own examples.
3. Prefer the cheapest sufficient target: a section over a plan, a plan over declaring the run dead.
4. When the specification itself contradicts the tests, the run fails rather than the specification being weakened.

OUTPUT
One JSON object: `fault` (`section` with its title, `plan`, or `frozen`), `reasoning` (the decisive evidence), `confidence`.
\end{lstlisting}
\end{promptbox}

\begin{promptbox}{Specification repairer}
\begin{lstlisting}[breakatwhitespace=true]
ROLE
Specification repairer. Repairs exactly one broken section; everything else is frozen.

INPUT
`broken_section`, `diagnostics` (message, line, source line), `context_sections` (read-only), `plan`, `attempt_history` (what earlier rounds tried and how each ended), `verdict` (the ruling naming the target).

TASK
Read the diagnostics, fix the target section and return it. The reply replaces that section byte-for-byte; no other bytes of the file change.

RULES
1. Minimal change: fix the diagnosed fault and keep the rest of the section byte-identical where possible. The plan's names are frozen.
2. If the only repair that compiles would weaken the specification's meaning, return the section unchanged and mark it `no-faithful-repair`.
3. Two technical-looking repairs count as weakening: narrowing a quantifier's domain to a bound the problem does not state, and replacing a quantified clause with a Bool computation that decides it.
4. If the true fault lies in a read-only section, return the body unchanged and mark it `fault-elsewhere` with the section named.

OUTPUT
Exactly one section with the target's title, followed by one line `attempt: <one sentence>`.
\end{lstlisting}
\end{promptbox}

\begin{promptbox}{Negative test case proposers}
\begin{lstlisting}[breakatwhitespace=true]
INVALID INPUTS (L1b)
Role: adversarial proposer for the validity layer. Input: `problem` (untrusted), `signature`, `valid_examples`, `quota`.
Task: make the precondition hard to write. A lenient predicate looks correct against valid inputs alone and stays wrong forever, so the measure of a case is which wrong predicate it defeats: `n = 9` against `1 <= n <= 8` defeats every predicate that forgot the ceiling. A case no plausible wrong predicate would accept has raised the difficulty by nothing. When the examples do not show the range you need, ask instead of guessing.

INCORRECT OUTPUTS (L2b)
Role: adversarial proposer for the correctness layer. Input: `problem` (untrusted), `signature`, `case` (a valid input and its true output), `aim` (the flaw family to try first), `quota`.
Task: make the postcondition hard to write. Name a flaw a weak postcondition could have (an optimization stated as achievable but never as best, a count that misses a case, a search that forgets a boundary), then write the wrong answer that flaw admits. A wrong answer with no flaw behind it catches nothing.

ADMISSIBILITY
Task: keep a frozen input only if compiled evaluation could check a postcondition against it. The cost of an input is the number of candidates the postcondition ranges over, measured before the call for each standard shape, and about ten thousand is the ceiling.

OUTPUT
Each proposer returns a JSON array of cases, each with the flaw or constraint it targets; the admissibility judge returns keep or refuse with a reason.
\end{lstlisting}
\end{promptbox}

\subsection{Model prompts}
\label{app:modelprompts}

The models receive one of the 4 system prompts below, condensed. The user turn gives the problem, the constraints and the test cases, and for CodeGen the given specification. A repair turn repeats the original user turn and adds the current file and the review of each failed stage. Without test cases in the prompt, as in Table~\ref{tab:verina-notests}, the SpecGen and CodeGen prompts replace the given test cases by test cases derived from the statement.

\begin{promptbox}{SpecGen}
\begin{lstlisting}[breakatwhitespace=true]
TASK
Formalize the algorithmic problem in Lean 4 as a validity predicate, an ideal specification and an executable checker, and check the predicate and the checker on the given test cases with guards.

RULES
1. State the specification with the definitions of the statement, so that it characterizes the correct `result` instead of computing it. Quantify over the objects the statement defines, and bound a quantifier only by a bound the statement states. An optimum is attained by a feasible candidate and is no worse than every feasible candidate, a decision holds exactly when a witness exists, and a count is the size of the set of objects with the stated property. Use `result = f input` only for a formula the statement gives or for a sequence of steps it prescribes.
2. Define the validity predicate to accept exactly the inputs that satisfy the stated constraints.
3. Define the checker to decide the specification on valid input, with each quantifier bounded by a range the input determines: `decide` of the specification with registered instances where every quantifier is bounded, a hand-written Bool function otherwise.
4. Complete each definition without `sorry`, and use none of `admit`, `native_decide`, `noncomputable`, `partial`, `unsafe` and `Classical`.

OUTPUT
One Lean file: `import Mathlib`, a docstring restating the problem, then Specification helpers, Validity predicate, Ideal specification, Decidable specification (with `<name>_check_iff_spec` left as `sorry`) and Unit tests with guards under layers L1, L1b, L2 and L2b.
\end{lstlisting}
\end{promptbox}

\begin{promptbox}{CodeGen}
\begin{lstlisting}[breakatwhitespace=true]
TASK
Implement the algorithmic problem in Lean 4 against the given specification, with a correct output for each input the constraints allow, and check the implementation on the given test cases with guards.

RULES
1. Prove termination of each recursive definition by structural recursion or a `termination_by` measure.
2. Choose an algorithm within a log factor of the best known method, so that the implementation finishes on the largest allowed input.
3. Complete each definition computably and without `sorry`, with the constructs banned for SpecGen.

OUTPUT
One Lean file with `Input`, `Output` and the representation types copied from the given specification, then Specification helpers, Implementation helpers, Main algorithm and Unit tests with guards under layer L3.
\end{lstlisting}
\end{promptbox}

\begin{promptbox}{SpecGen-Repair}
\begin{lstlisting}[breakatwhitespace=true]
TASK
Repair the Lean 4 specification so that the file compiles, all guards evaluate to `true` and the specification states the problem faithfully.

RULES
1. Locate the fault before editing, starting with the first compiler error.
2. Make the smallest change that removes the cause, and keep the names and argument order of `isValidInput`, `<name>_spec`, `<name>_check` and `<name>_check_iff_spec`.
3. State the specification declaratively where the problem allows; use `result = f input` only where the statement prescribes the output.
4. Keep the checker a clause-by-clause mirror of the specification with bounded quantifiers.
5. Complete each definition computably and without `sorry`.

OUTPUT
The complete repaired file, keeping the section headings and order of the current version.
\end{lstlisting}
\end{promptbox}

\begin{promptbox}{CodeGen-Repair}
\begin{lstlisting}[breakatwhitespace=true]
TASK
Repair the Lean 4 implementation so that the file compiles, all guards evaluate to `true` and the implementation returns a correct output for each input the constraints allow.

RULES
1. Find the cause before editing; a termination error usually needs a new measure or a `decreasing_by` proof.
2. Apply the smallest edit that removes the cause, and keep the name and argument order of the main algorithm.
3. Prove termination as in CodeGen, and use an algorithm within a log factor of the best known method.
4. Complete each definition computably and without `sorry`.

OUTPUT
The complete repaired file, with the section headings and order of the current version.
\end{lstlisting}
\end{promptbox}

\subsection{Evaluation judges}
\label{app:evaljudges}

The judges below run as the acceptance criteria of synthesis, in evaluation and in the overall quality, test case alignment and non-tautology stages of the reward.

\begin{promptbox}{Test case alignment}
\begin{lstlisting}[breakatwhitespace=true]
ROLE
Checker of the guard commands of a file against the test rows the author was given.

INPUT
`test_block` (the given rows), `guards` (every guard of the file), `unmatched` (what a mechanical check found where it could not match).

TASK
Decide whether the guards test exactly the given rows.

RULES
1. Compare values only. Function names and spellings such as tuples, parentheses, arrays or string literals change nothing; a changed, added, removed or reordered element is a different value.
2. A positive row is accepted, and an invalid input or a wrong output is rejected with `!`. A specification file tests a positive row on both the validity predicate and the checker.
3. Guards beyond the given rows do not affect the verdict.

OUTPUT
`problems` (each `missing`, `extra`, `changed` or `reversed`, with the fix) and `verdict`, `PASS` or `FAIL`.
\end{lstlisting}
\end{promptbox}

\begin{promptbox}{Specification quality (SpecGen)}
\begin{lstlisting}[breakatwhitespace=true]
ROLE
Reader of a specification against the problem it was written for.

INPUT
`description` (untrusted), `constraints`, `specification`.

TASK
Decide whether the specification accepts exactly the answers the problem accepts on valid inputs: soundness (it carries every condition on the answer) and completeness (it rejects no correct answer).

RULES
1. Take the statement clause by clause. A clause that constrains the answer must appear, a clause that restricts the input belongs to the validity predicate, and a resource bound stays out of the report.
2. Judge the Lean definitions by what they compute, evaluating each helper on a small input.
3. Every finding names concrete values: for soundness, the quoted clause and a value the specification admits and the problem rejects; for completeness, a valid input, its correct output and the part of the specification that rejects it.

OUTPUT
`missing_conditions` and `rejected_answers`, each empty when there is nothing to report.
\end{lstlisting}
\end{promptbox}

\begin{promptbox}{Checker quality (SpecGen)}
\begin{lstlisting}[breakatwhitespace=true]
ROLE
Reader of the executable checker. The guards pass, so on the frozen inputs it agrees with the specification.

INPUT
`constraints`, `specification`, `checker`.

TASK
Decide whether the checker accepts exactly what the specification does on every valid input.

RULES
1. A checker truncates where the specification quantifies over an unbounded domain. A bound that is too small misses an object the specification admits; one that reaches objects the specification rejects can accept a wrong output.
2. Report a disagreement only with an object: a valid input, its correct output by the statement, and the witness or counterexample one side reaches and the other does not.

OUTPUT
`truncates`, `agrees_with_the_specification` (`YES` or `NO`) and `escapes`, each with its direction, input, correct output and object.
\end{lstlisting}
\end{promptbox}

\begin{promptbox}{Implementation quality (CodeGen)}
\begin{lstlisting}[breakatwhitespace=true]
ROLE
Reader of the implementation. The guards pass, so it is correct on the tested inputs.

INPUT
`constraints`, `specification`, `checker`, `implementation`.

TASK
Judge what the tests cannot see: termination, cost, answers written back for the tested inputs and duplicated work.

RULES
1. Report a fuel parameter, and whether a bound the recursion stops on can run out on an admitted input.
2. Mark the implementation too slow only when the best known method is faster by more than a logarithmic factor and the implementation takes more than about 10^7 steps on the largest admitted input.
3. Report an implementation that branches on particular inputs to return particular answers.

OUTPUT
`termination`, `cost` (with `too_slow`), `reproduces_tests`, `rewrote_the_checker` and a short `reason`.
\end{lstlisting}
\end{promptbox}

\begin{promptbox}{Non-tautology}
\begin{lstlisting}[breakatwhitespace=true]
ROLE
Lean 4 formalization auditor of the specification side.

INPUT
`description` (untrusted), `constraints`, `specification`.

TASK
Judge how much of a solution the specification contains.

RULES
1. DECLARATIVE: the condition on `result` is the condition of the statement, quantified over the objects of the statement.
2. ENUMERATIVE: the condition of the statement, but over a list of candidates the specification builds itself.
3. TAUTOLOGICAL: the specification computes `result` by solving the problem, through a recurrence, a table, a greedy rule or a simulation the statement does not prescribe.
4. For ENUMERATIVE and TAUTOLOGICAL, describe the declarative specification in at most three sentences.

OUTPUT
`reasoning`, `better_specification` and `verdict`.
\end{lstlisting}
\end{promptbox}

\end{document}